%% file: gauge.tex
\documentstyle[epsf]{article}

\newcommand{\newsection}{ \setcounter{equation}{0} \section}
\newcommand{\beq}{\begin{equation}} \newcommand{\eeq}{\end{equation}}
\newcommand{\bea}{\begin{eqnarray}} \newcommand{\eea}{\end{eqnarray}}

  \newcommand
{\Romannumeral}[1]{\uppercase\expandafter{\romannumeral#1}}

\newcommand{\be}{\begin{enumerate}} \newcommand{\ee}{\end{enumerate}}
\newcommand{\bi}{\begin{itemize}} \newcommand{\ei}{\end{itemize}}
\newcommand{\ba}{\begin{array}} \newcommand{\ea}{\end{array}}
\newcommand{\bc}{\begin{center}} \newcommand{\ec}{\end{center}}
\newcommand{\bt}{\begin{tabular}} \newcommand{\et}{\end{tabular}}

\def\lsim{\mathrel{\rlap{\lower4pt\hbox{\hskip1pt$\sim$}}
    \raise1pt\hbox{$<$}}}           % less than or approx. symbol
\def\gsim{\mathrel{\rlap{\lower4pt\hbox{\hskip1pt$\sim$}}
    \raise1pt\hbox{$>$}}}           % greater than or approx. symbol

\newcommand{\Tr}{\mathop{\rm Tr}}           % Trace
\newcommand{\half}{\textstyle {1\over2} \displaystyle}    % One half
\newcommand{\third}{\textstyle {1\over3} \displaystyle}   % One third
\newcommand{\quarter}{\textstyle {1\over4} \displaystyle} % One quarter
\newcommand{\sixth}{\textstyle {1\over6} \displaystyle}   % One sixth
\newcommand{\eigth}{\textstyle {1\over8} \displaystyle}   % One ieight
\newcommand{\Dslash}{{\hbox{D}\kern-0.6em\raise0.15ex\hbox{/}}} % D slash

\begin{document}

\setlength{\oddsidemargin}{0cm} \setlength{\baselineskip}{7mm}

\input epsf

% \topmargin 0pt
% \oddsidemargin 5mm
% \headheight 0pt
% \topskip 0mm

% \addtolength{\baselineskip}{0.20\baselineskip}

% \renewcommand{\thefootnote}{\fnsymbol{footnote}}

\begin{normalsize}\begin{flushright}
%   UCI-96-xx \\
    DAMTP-96-68 \\
    July 1996 \\
\end{flushright}\end{normalsize}

\begin{center}
  
\vspace{25pt}
  
{\Large \bf GAUGE INVARIANCE IN SIMPLICIAL GRAVITY }

\vspace{40pt}
  
{\sl Herbert W. Hamber}
$^{}$\footnote{e-mail address : hamber@cern.ch; permanent address:
University of California, Irvine Ca 92717 USA}
and {\sl Ruth M. Williams}
$^{}$\footnote{e-mail address : rmw7@damtp.cam.ac.uk; permanent address:
DAMTP, Silver Street, Cambridge CB3 9EW, England}
\\

\vspace{20pt}

Theoretical Physics Division, CERN \\
CH-1211 Geneva 23, Switzerland \\

\end{center}

\vspace{40pt}

\begin{center} {\bf ABSTRACT } \end{center}
\vspace{12pt}
\noindent

The issue of local gauge invariance in the simplicial lattice
formulation of gravity is examined.  We exhibit explicitly, both in
the weak field expansion about flat space, and subsequently for
arbitrarily triangulated background manifolds, the exact local gauge
invariance of the gravitational action, which includes in general both
cosmological constant and curvature squared terms.
We show that the local invariance of the discrete action and the
ensuing zero modes correspond precisely to the diffeomorphism
invariance in the continuum, by carefully relating the fundamental
variables in the discrete theory (the edge lengths) to the induced
metric components in the continuum.
We discuss mostly the two dimensional case, but argue that our
results have general validity.
The previous analysis is then extended to the coupling with a scalar
field, and the invariance properties of the scalar field action under
lattice diffeomorphisms are exhibited.  The 
construction of the lattice conformal gauge is then described,
as well as the separation of
lattice metric perturbations into orthogonal conformal and
diffeomorphism part.
The local gauge invariance properties of the lattice
action show that no Fadeev-Popov determinant is required in the
gravitational measure, unless lattice perturbation theory is performed
with a gauge-fixed action, such as the one arising in the lattice
analog of the conformal or harmonic gauges.

\vspace{24pt}

\vfill

\newpage

\vskip 10pt
\newsection{Introduction}
\hspace*{\parindent}

In the quantization of gravitational interactions one expects
non-perturbative effects to play an important role
~\cite{haw}.
One formulation
available for studying such effects is Regge's simplicial lattice
theory of gravity ~\cite{regge}.  It is the only lattice model with a
local gauge invariance ~\cite{lesh}, and the only model known to
contain gravitons in four dimensions ~\cite{rowi}.  One would hope
that a number of
fundamental issues in quantum gravity, such as the existence of a
non-trivial ultraviolet fixed point of the renormalization group in
four dimensions and the recovery of general relativity at large
distances, could in principle be addressed in such a model.  The
presence of a local gauge invariance, which is analogous to the
diffeomorphism group in the continuum, makes the model attractive as a
regulated theory of gravity ~\cite{hartle}, while the existence of a
phase transition in three ~\cite{hw3d} and four
dimensions ~\cite{lesh,hw84,berg,phases,beirl} (but not in two ~\cite{hw2d})
suggests the existence of a (somewhat unusual) lattice continuum limit.
The two {\it phases} of quantized gravity found in ~\cite{phases},
can loosely be described as having in one
phase ($G<G_c$, rough, polymer-like phase)
\beq
\langle g_{\mu\nu} \rangle \; = \; 0 \;\; ,
\eeq
and in the other phase ($G>G_c$, smooth phase),
\beq
\langle g_{\mu\nu} \rangle \; \approx \; c \; \eta_{\mu\nu} \;\; ,
\eeq
with a small negative average
curvature (anti-DeSitter space) in the vicinity of the critical point
at $G_c$.
A physically similar two-phase structure was later proposed also
in ~\cite{polylesh}; see also the ideas found in ~\cite{fubini}.
A discussion of the properties of the
two phases characterizing four-dimensional gravity, and of the
associated critical exponents, can be found in ~\cite{phases}. For
additional recent numerical results we refer the reader
to ~\cite{beirl}, while for some
earlier attempts we refer to the work in ~\cite{lesh,hw84,berg}.  Recently
calculations have progressed to the point that a first calculation of
the Newtonian potential from the correlation of heavy particle world
lines, following the suggestive proposal of ~\cite{moda}, seems
feasible ~\cite{lines}.
The results so far indicate that in the lattice quantum
theory of gravity the potential between heavy spinless bodies is
attractive, and has roughly the correct heavy mass dependence.
In the same work a
general scaling theory for gravitational correlations, valid in the
vicinity of the fixed point, was put forward.  We also refer the
reader to ~\cite{wt}, where a more complete set of references to
earlier work on Regge gravity can be found.
For results with an alternative and 
complementary approach based on dynamical triangulations, we refer
the reader to the references in ~\cite{smit}.

In view of this recent progress it would seem desirable to further
elucidate the correspondence between continuum and lattice theories.
The weak field expansion is available to systematically develop this
correspondence, and it is well known that such an expansion can be
carried out in both formulations.  Not unexpectedly, it is technically
somewhat more complex in the lattice theory due to the presence of
additional vertices, as happens in ordinary lattice gauge theories.
In the past most perturbative studies of lattice gravity have focused
on the lowest order terms, and in particular the lattice graviton
propagators ~\cite{rowi,hw2d,hw3d}.  Recently it has been extended to
include the vertex functions, and the results have been used to
compute the one-loop amplitudes relevant for the conformal anomaly in two
dimensions ~\cite{fey}.

One central issue in a regularized theory of quantum gravity is the
{\it nature of its invariance properties}. Although some discussions of
these issues have appeared before, no systematic
and coherent exposition has been presented yet in the literature.
In this paper we address the question of what exactly the local gauge
invariance built into Regge's simplicial gravity looks like.
Its existence is intimately tied in with the appearance of gravitons
(in four dimensions) in the lattice weak field expansion about
a flat background.
It need not be emphasized here that local gauge invariance plays a central
role in both the classical and
quantum formulation of gravity, and its preservation in the lattice
theory must therefore be considered of paramount importance.
Physically, it expresses the fact that
the same physical geometry can be described by equivalent metrics.
Classically, it leads for example to the invariance of the infinitesimal line
element and the Bianchi identities for the curvature. In the quantum
theory it is known to give rise to the Slavnov-Taylor identities for
the gravitational Green's functions. One would therefore expect that local
gauge transformations should play a central role in the lattice theory
as well. This aspect will be therefore the focus of the first part of
the paper, where the analog of local gauge transformations on the
lattice will be constructed.  The requirement of gauge invariance will
have implications for both the gravitational measure and the coupling
to a scalar field, and we will present in this paper a detailed analysis of
its consequences. The second part of this paper will be devoted
to a number of relevant applications.

The plan of the paper is as follows.
In Section 2, we introduce our notation, describe the choice of
lattice structure and the relevant degrees of freedom in the lattice
theory, the squared edge lengths. We discuss the discrete actions
for the gravitational degrees of freedom, and the relationship
to their well-known continuum counterparts.
In Section 3 we move on to the lattice weak field expansion, and
discuss in detail the two-dimensional case (with cosmological and
curvature squared terms). We exhibit explicitly the gauge zero modes
and their corresponding eigenvectors, which are shown to correspond
precisely to local gauge transformations in the continuum.
We then compute explicitly and analytically the zero modes for
fluctuations about
a non-flat background (the tetrahedral, octahedral and icosahedral 
tessellations of the two-sphere), and show that the counting of the
zero modes is consistent with the expectation from the continuum
theory. We then give further arguments supporting the identification
of the zero modes with the diffeomorphisms in the continuum, which
we argue is valid in any dimension.
In Section 4 we extend the previous analysis to arbitrary curved backgrounds
and show explicitly the persistence of a local gauge invariance for
the area, curvature and curvature squared terms.
In section 5 we introduce a scalar field coupled invariantly to the
gravitational
degrees of freedom. We again exhibit its invariance properties
under local gauge variations of the squared edge lengths, at least for
sufficiently smooth scalar field configurations, by working out the
concrete case of background lattices which are close to either
equilateral or square. We then discuss the more general case
of arbitrary background lattices, and the construction of the
energy-momentum tensor for the scalar field.
Section 6 discusses the implications of the preceding results 
for the lattice gravitational measure, and we give arguments that the
lattice measure is essentially unique, up to local volume factors.
We will argue therefore that the lattice measure is essentially no
less unique than the original continuum (DeWitt) measure.
In Section 7 we consider the possibility of introducing a gauge fixing
term in the lattice action, in order to remove the gauge zero modes of
the gravitational action and subsequently perform perturbative
calculations, and in close analogy with the procedure followed in
the usual continuum perturbation theory.
As an example, we discuss the explicit construction of
the lattice conformal gauge, starting from an arbitrary configuration
of squared edge lengths.
Finally, Section 8 contains some concluding remarks.

\vskip 10pt
\newsection{The Discretized Theory}
\hspace*{\parindent}

In this section we will briefly review the construction of the action
describing the gravitational field on the lattice, and define the
necessary notation used later in the paper.  In concrete examples we
will often refer, because of its simplicity, to the two-dimensional
case, where a number of results can be derived easily and
transparently. But in a number of instances here, and throughout the
paper, important aspects of the discussion and of the conclusions will
be quite general, and not restricted to specific aspects of the
two-dimensional case.

In simplicial gravity the elementary building blocks for
$d$-dimensional space-time are simplices $ \sigma^d $ of dimension
$d$.  A 0-simplex is a point, a 1-simplex is an edge and a 2-simplex
is a triangle.  A $d$-simplex is a $d$-dimensional object with $d+1$
vertices and $d(d+1)/2$ edges connecting them. Each simplex in turn
contains $ \left ( { d+1 \atop k+1 } \right ) $ sub-simplices
$\sigma^k$ of dimension $k$.  Thus in two dimensions we shall consider
here a fixed closed simplicial two-manifold consisting of $N_0$
vertices, $N_1$ edges and $N_2$ triangles, joined in such a way that
each point has a neighborhood homeomorphic to the interior of a
two-dimensional sphere.  A simplicial geometry is then specified by
the assignment of squared edge lengths $l_i^2$, $i=1\dots N_1$, and a
flat Riemannian metric can be assigned to the interior regions of the
simplices in a way that is consistent with the edge length values.
Further restrictions arise from the fact that the triangle
inequalities (and their higher dimensional analogs in $d$ dimensions)
have to be satisfied.

The correspondence between squared edge lengths and an assigned
continuum metric field can be made more precise, with the
identification
\beq
l_{ab} \; = \; \int_{\tau(a)}^{\tau(b)} d \tau
\sqrt{ \textstyle g_{\mu\nu} ( x (\tau) ) {d x^{\mu} \over d \tau}
{d x^{\nu} \over d \tau} \displaystyle } \; = \; \sqrt{ \textstyle
g_{\mu\nu} l_{ab}^{\mu} l_{ab}^{\nu} \displaystyle } \;\; ,
\label{eq:distance}
\eeq
where $l_{ab}$ is the length of the edge connecting neighboring
points $a$ and $b$. For a given set of edge lengths, the metric $
g_{\mu\nu} (x) $ has initially support on the edges only.
\footnote{The above identification parallels an analogous correspondence
used sometimes in ordinary lattice gauge theories,
where the $SU(n)$ matrix-valued
lattice field $A_{n \mu}$ has support only on the {\it links} of a
hypercubic lattice, $ U_{n \mu} \equiv e^{ i a A_{n \mu}} $$
= P \exp \left ( i a \int_n^{n+\mu} d x^\mu A_\mu (x) \right ) $.
This definition is a convenient starting point for performing
perturbation theory and defining the lattice Feynman rules.  For the
same construction in perturbative simplicial gravity see
\cite{fey}.}
For a metric that is constant inside each simplex,
$l_i^2 = g_{\mu\nu} l_i^{\mu} l_i^{\nu}$, where $i$ labels the edge
from $a$ to $b$ and the $l_i^{\mu}$'s are the components of the edge
lengths.

\vskip 10pt
\subsection{Lattice Structure}
\hspace*{\parindent}

In two dimensions quantum gravity can be defined on a
two-di\-men\-sio\-nal surface consisting of a network of flat
triangles.  The underlying lattice may be constructed in a number of
ways.  Points may be distributed randomly on the surface and then
joined to form triangles according to some algorithm.  In such
lattices the coordination number at each vertex can be kept fixed
(quenched random lattice), or allowed to vary (annealed random
lattice), by considering it as an additional, dynamical variable of the
model.  An alternative procedure is to start with a regular lattice,
like a regular tessellation of the two sphere, or a lattice of squares
divided into triangles by drawing in parallel sets of diagonals, and
then allow the edge lengths to vary, which will give rise to curvature
localized on the vertices.  It should be emphasized that for arbitrary
assignments of edge lengths, consistent with the imposition of the
triangle inequalities constraints, such a lattice is in general far
from regular, and resembles more a random lattice.

The incidence matrix, which provides the information on which edges
are adjacent, and fixes therefore the local coordination number $q_i$,
describes the topology of the manifold. It can be chosen to correspond
to a fixed regular or to a fixed random lattice.  But one word of
caution should be spent here on the terminology.  Since the edge
lengths are dynamical variables, the lattice is in fact random in
either case: contrary to a fixed regular lattice (such as the square
or triangular one in two dimensions), there are a priori no preferred
directions even for a lattice with fixed coordination number, as
neighboring points can have any relative orientation as long as they
are consistent with the triangle inequalities and their higher
dimensional analogs.  Universality arguments would then suggest that
the choice of local coordination number should not affect the large
distance limit of the model, and a number of explicit calculations on
random lattices have shown to some extent that this is indeed the case
~\cite{itzcar,r2rand}.

In the following we will often narrow down the discussion and be even
more specific, and usually think of the ``regular'' lattice as
consisting of a network of triangles with a fixed coordination number
of six, $q_i=6$, although many of the results in this work are quite
general and do not rely on the specific choice of local coordination
numbers.

Quenched random lattices, where the local coordination number $q_i$
(which is the number of edges meeting at $i$) is random but fixed,
were considered in \cite{lee,cfl,itzcar,itzd}. For such Poissonian random
lattices, the average coordination number is also $q=6$ in two
dimensions, independent of the topology. This follows from the
expression for the Euler characteristic
$ \chi \; = \; N_0 - N_1 + N_2 $ with
$ 2 N_1 \; = \; 3 N_2 \; = \; \sum_i q_i \;\; $
in two dimensions, which gives for large $N_0$
\beq
q \equiv
\lim_{ N_0 \rightarrow \infty } { \sum_i q_i \over \sum_i 1 } \; = \;
6 \;\; ,
\eeq
irrespective of the value of $\chi$, as well as $N_1 = 3
N_0$ and $N_2 = 2 N_0$.  In general on such random lattices one does
not have, strictly speaking, translational or rotational invariance
for a {\it fixed} assignment of edge lengths. The latter only hold on
the average. Explicit calculations confirm that this is indeed the
case, at least in two dimensions \cite{itzcar,gid,rlattis}.

When the edge lengths are allowed to fluctuate one would expect the
situation to be different, since now locally there are no preferred
directions any more, as the lattice structure fluctuates from edge
length configuration to edge length configuration. Lastly, one can
allow the local coordination number to change (annealed random
lattice) by re-linking neighboring vertices, although there is no
unique algorithm to do so which preserves the geometry. In this case
the coordination number fluctuation $\delta q_i = q_i - 6 $ becomes an
additional dynamical variable, and is indeed the only dynamical
variable in the so-called dynamical triangulations.  Randomness can be
shown to be a relevant perturbation in two dimensions, changing the
universality class already for flat surfaces. We shall not consider
dynamical random lattices here any further, as we are interested in
discretizations for gravity coupled to matter which maintain the
crucial property of reducing to the ordinary, known flat space field
theories in the limit of zero local curvatures.  A review of the
properties of random lattices and their relation to matrix models in
two dimensions can be found in \cite{david}.

\vskip 10pt
\subsection{Degrees of Freedom}
\hspace*{\parindent}

The elementary degrees of freedom on the lattice are the edge lengths,
with the correspondence between continuum and lattice degrees of
freedom given locally by
\beq
\left \{ \; g_{\mu\nu} (x) \; \right \}_{x \epsilon {\cal M}}
\rightarrow
\left \{ \; l_i^2 \; \right \}_{ i = 1 \dots N_1 } \;\; ,
\eeq
where the index $i$ ranges over all $N$
edges in the lattice.  In general the dynamical lattice will give rise
to some average lattice spacing $a_0 = [ \langle l^2 \rangle ]^{\half}
$, which in turn will naturally supply the ultraviolet cutoff that is
needed to define the quantum theory.  An important difference with
ordinary lattice field theories lies in the fact that the momentum
cutoff $\Lambda = 1/l_0$ is not determined a priori, but follows
instead from the dynamics (i.e. from the lattice action and lattice
measure).  The dynamical cutoff turns out to be determined mostly by
the cosmological constant term and the measure factor ~\cite{phases}.

Furthermore for finite volumes the lattice theory will have a finite
number of degrees of freedom $N$, and will therefore inherit an
infrared cutoff of the order of $1/L$, where $L$ is the physical
linear extent of the lattice.

In the discrete case all the metric information on the piecewise
linear space is contained in the values of the edge lengths.  As
already emphasized by Regge, and in accordance with the usual view of
lattice discretization of continuum field theories, the discrete
manifold $L$ is thought of as an approximation to some continuum
manifold $S$ (this is illustrated in Figure 1).  In the limit as the
average lattice spacing $a_0$ is sent to zero, the original continuum
theory is recovered.  In four dimensions it has been rigorously
proven, for the Einstein-Regge action, that if a piecewise flat space
approximates a smooth space in a suitable sense, then the
corresponding curvatures are close in the sense of
measures ~\cite{cms}; see also the results of ~\cite{frlee}.
In general the expectation is that the lattice and
continuum theory will differ by higher order corrections, with the two
actions related to each other by
\beq
I_L (l^2) \; = \; I_C (
g_{\mu\nu} ) + a_0 \; \delta I + a_0^2 \; \delta^2 I + \cdots \;\; .
\eeq
All corrections can in principle systematically be evaluated by
the standard procedure of replacing the finite differences which
appear in the lattice action by derivatives, for example according to
the formula
\beq
{ g(n+a_0) - g(n-a_0) \over 2 a_0 } \; = \; g' (n) + {1 \over 6}
\, a_0 \, g''' (n) + O (a_0^4 \; g^{(5)} (n) ) \;\; .
\eeq
It should be noted
that higher order corrections are expected to involve higher
derivatives of the metric.  The above expansion procedure can be
thought of being equivalent to introducing a continuum metric on the
piecewise linear manifold, and expand in the difference between the
continuum and the piecewise linear metric.

\begin{center}
  \leavevmode
  \epsfysize=5cm \epsfbox{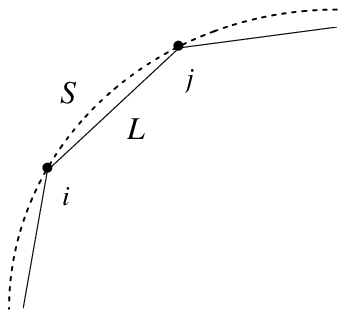}
\end{center}
\noindent
{\small{\it Fig.\ 1.  Piecewise linear space $L$ as an approximation
    to a smooth $d$-dimensional enveloping surface $S$.  \medskip}}

This interpretation is analogous to the situation in
ordinary lattice gauge theories, where the lattice gauge fields
$U_{n\mu}$ are defined on the links only; the continuum fields
$A_{\mu}(x)$ can then be reconstructed by some suitable interpolation
to the interior regions of the lattice.
It is of course possible to endow the piecewise linear space with a 
continuum metric $g_{\mu\nu}(x)$ which is defined everywhere, including
the interiors of the simplices. In this case a continuum curvature
$R_{\mu\nu\rho\sigma}(x)$ can be defined as well, but since the interior of
the simplices is flat, the curvature acquires delta-function singularities
on the hinges where the discrete curvature resides.
While such a description can be useful in certain circumstances, it has also
some drawbacks, which have led to considerable confusion in some
of the literature. The obvious ones are that the resulting model is
no longer an ultraviolet regulator for the continuum theory, as space-time
has become continuous again.
Furthermore the fields are singular, due to the
delta-function singularities in the curvature,
and the number of degrees of freedom is no
longer finite due to the re-introduction of a continuum metric.
In this formalism new divergences appear, which have to be regulated
by some ad-hoc procedure such as the smoothing out of conical singularities,
and lead to difficulties in defining higher order invariant operators
such as the ones containing curvature squared terms \cite{hw84}.

This point of view, while certainly legitimate in discussing some
classical aspects of the theory, is therefore in our opinion not
useful in describing a {\it regulated} theory of quantum gravity.
It leads instead to a string of paradoxical results when lattice and continuum
language are mixed together, and can be especially misleading when
discussing such subtle issues as the gravitational functional
integration measure.

It should be emphasized here that
in the following we shall restrict our attention almost exclusively
to the {\it lattice} theory,
which is defined in terms of its lattice degrees of freedom {\it only}.
Since it is our purpose to describe an ultraviolet regulated theory of
quantum gravity, we shall follow the usual procedure followed
in discussing lattice field theories, and discuss the model exclusively
in terms of its primary, lattice degrees of freedom: the squared edge lengths.
As such, the theory will not require any additional ad-hoc regulators.
Below we shall discuss further at length a number of issues related to the
precise correspondence between the lattice degrees of freedom and the continuum
ones, the local gauge invariance of the lattice action (which gives
rise in the quantum theory the lattice analogs of the Taylor-Slavnov
identities) and the need for (or lack of) gauge fixing.

\vskip 10pt
\subsection{Curvature and Discretized Action}
\hspace*{\parindent}

The construction of the lattice action starts from the definition of
the elementary building blocks for spacetime, the $n$-dimensional simplices.
Consider an $n$-dimensional simplex with vertices 1, 2, 3, ... n+1 and
square edge lengths $l_{12}^2 = l_{21}^2$, ... .
Its vertices are specified by a set of vectors $\vec e_0 =0$,
$\vec e_1$, ... $\vec e_n$ in flat Euclidean space.
The matrix
\beq
g_{ i j } \; = \; \vec e_i \cdot \vec e_j \;\; ,
\eeq
with $1 \leq i,j \leq n $, is positive definite. In terms of the edge lengths
$l_{ij} = | \vec e_i - \vec e_j | $ (see Figure 2) it is given by
\beq
g_{ij} (l^2) \; = \; \half \;
\Bigl [ l_{0i}^2 + l_{0j}^2 - l_{ij}^2 \Bigr ] \;\; .
\label{eq:gij_simplex}
\eeq
The volume of a general $n$-simplex is then given by an $n$-dimensional
generalization of the well known formula for the volume of a tetrahedron,
\beq
V_n (l^2) \; = \; {1 \over n ! } \sqrt { \det g_{ij}(l^2) } \;\; .
\eeq
Conversely, in order to obtain a simplex for an arbitrary assignment of
edge lengths, the generalization to higher dimensions of the
triangle inequalities require that $V_n^{(i)} (l^2) \geq 0 $,
with $n=1\dots d$ and $i=1\dots N_n$ be satisfied for every edge,
triangle, tetrahedron etc. in the lattice. This can be stated
equivalently by requiring
\beq
\det g_{ij} (l^2) \; > \; 0
\eeq
for every sub-determinant of the highest dimension 
$ \det g_{ij} $.
In $d$ dimensions the matrix $g_{ij}$ has $d(d+1)/2$ components,
just as there are $d(d+1)/2$ components for the metric $g_{\mu\nu}(x)$
per space-time point in the continuum.
\begin{center}
\leavevmode
\epsfysize=7cm
\epsfbox{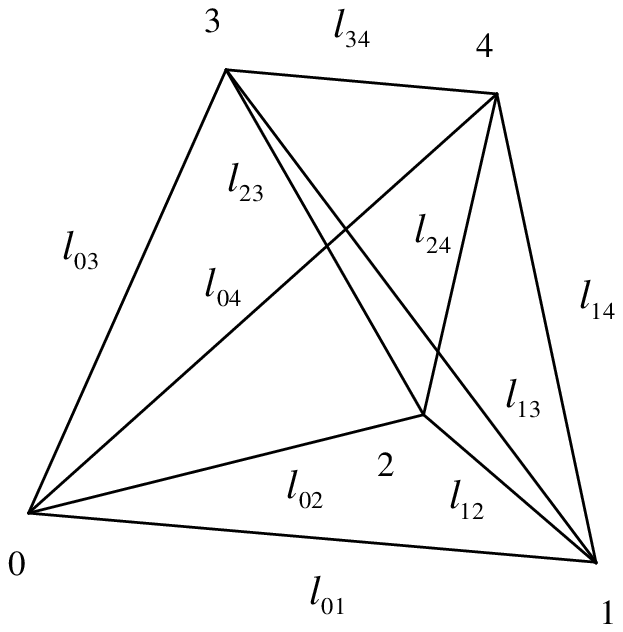}
\end{center}
\noindent
{\small{\it Fig.\ 2.
Assignments of edge lengths for a four-dimensional simplex.
\medskip}}
In this paper we shall often refer to the two-dimensional case.
In two dimensions one has simply
\beq
g_{ij} (l^2) \, = \, \left( \begin{array}{cc}
l_{01}^2 & \half (l_{01}^2 + l_{02}^2 - l_{12}^2 ) \\
\half (l_{01}^2 + l_{02}^2 - l_{12}^2 ) & l_{02}^2 \\
\end{array} \right) \;\; ,
\label{eq:gij_triangle} 
\eeq
and therefore
\beq
\det g_{ij} (l^2) \, = \, \quarter \left [
2 ( l_{01}^2 l_{02}^2 + l_{02}^2 l_{12}^2 + l_{12}^2 l_{01}^2 ) -
l_{01}^4 - l_{02}^4 - l_{12}^4 \right ]
\eeq
and 
\beq
\sqrt{ \det g_{ij} (l^2) } \, = \, 2 A_T (l^2) \;\; ,
\label{eq:detgij} 
\eeq
where $A_T (l^2)$ is the area of the given triangle (see Figure 3).
\begin{center}
\leavevmode
\epsfysize=5cm
\epsfbox{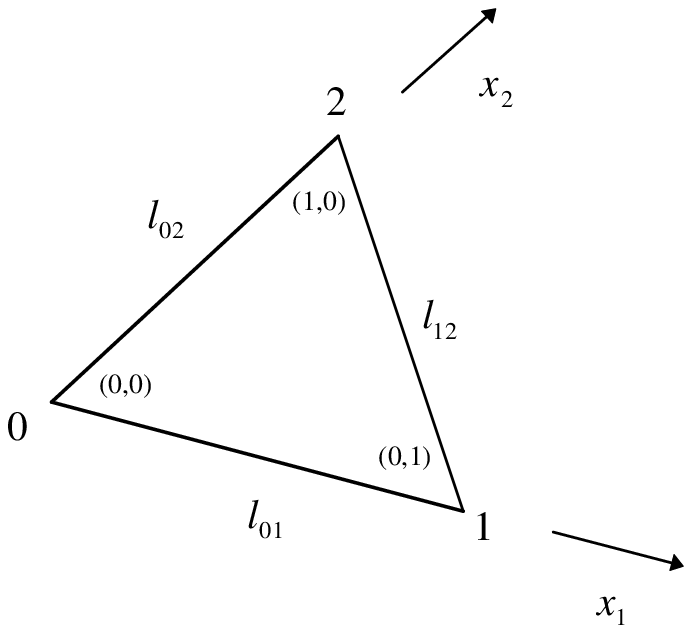}
\end{center}
\noindent
{\small{\it Fig.\ 3.
Assignments of edge lengths and natural coordinates for a triangle.
\medskip}}

In simplicial gravity the curvature is concentrated on the hinges, which
are subspaces of dimensions $d-2$, and is entirely determined from the
assignment of the edge lengths.
In two dimensions the hinges correspond to the vertices
and $\delta_h$, the deficit angle at a hinge, is defined by
\beq
\delta_h \; = \; 
2\pi - \sum_{\rm triangles~t \atop meeting~at~h} \theta_t \;\; ,
\label{eq:deficit} 
\eeq
where $\theta_t$ is the dihedral angle associated with the triangle
$t$ at the vertex $h$ (see Figure 4).
In $d$ dimensions several $d$-simplices meet on a ($d-2$)-dimensional
hinge, and the deficit angle is defined by
\beq
\delta_h (l^2) 
\; = \; 2 \pi - \sum_{ {\rm d-simplices \atop meeting \, on \, h }}
\theta_d (l^2) \;\; ,
\eeq
where $\theta_d$ is the dihedral angle in $d$ dimensions.
The sine of the dihedral angle can be computed from the well known formula
\beq
\sin \theta_d (l^2) \; = \;
{ d \over d-1 } { V_d V_{d-2} \over V_{d-1} V_{d-1} ' } \;\; ,
\eeq
where $ V_{d-2} $ is the volume of the hinge, $ V_d $ is the volume
of the $d$-simplex, and $ V_{d-1}$, $ V_{d-1} '$ the volumes of the two
($d-1$)-dimensional faces that meet on the hinge.
A general derivation of these formulae can be found in ~\cite{hartle},
with some additional results in ~\cite{hw84}.
Since the sine does not uniquely determine the angle, it can be useful
to obtain an expression for the cosine of the dihedral angle, which can be
found in ~\cite{hw84}.
In two dimensions the dihedral angle is given by
\beq
\cos \theta_d \; = \;
{ l_{01}^2 + l_{02}^2 - l_{12}^2 \over 2 l_{01} l_{02} } \;\; .
\label{eq:dihedral}
\eeq

\begin{center}
\leavevmode
\epsfysize=5cm
\epsfbox{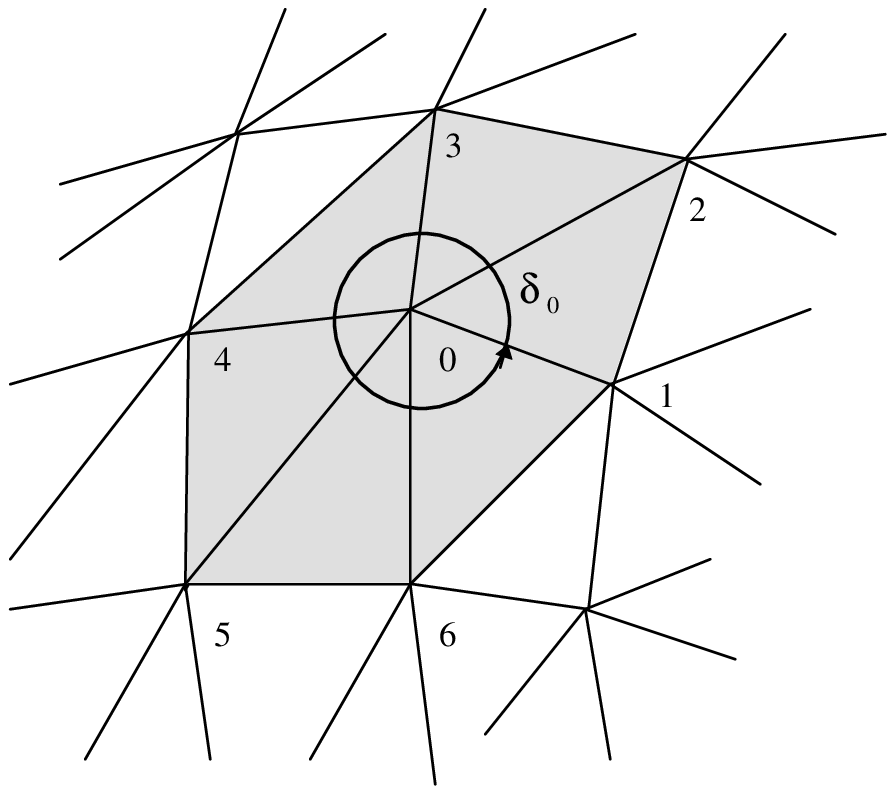}
\end{center}
\noindent
{\small{\it Fig.\ 4.
In two dimensions the computation of the
deficit angle $\delta_0$ at the vertex $0$ involves the values for the edge
lengths associated with the shaded triangles.
\medskip}}

It is useful to introduce a dual lattice following, for example,
the Dirichlet-Voronoi cell construction, which consists in
introducing the perpendicular bisectors of the edges in each triangle and
joining the resulting vertices.
This provides for a natural subdivision of the original lattice
in a set of non-overlapping exhaustive cells, and has a natural
generalization to higher dimensions.
It is easy to see that
the vertices of the original lattice then reside on circumscribed
circles, centered on the vertices of the dual lattice.
For the vertex $0$ the dihedral dual volume contribution,
shown in Figure 5, is given by
\beq
A_d (l^2) \; = \; {1 \over 32 A } 
\Bigl [ l_{12}^2 ( l_{01}^2 + l_{02}^2 ) - ( l_{01}^2 - l_{02}^2 )^2 \Bigr ]
\;\; .
\label{eq:voronoi} 
\eeq
It is clear that the above subdivision is not unique.
Alternatively, one can introduce a baricenter for each triangle, defined
as the point equidistant from all three vertices, and again
join the resulting vertices.
The vertices of the original lattice then reside on inscribed
circles, centered on the vertices of the dual lattice.
The baricentric dihedral volume is simply given by 
\beq
A_d (l^2) \; = \; A/3 \;\; .
\label{eq:baricentric} 
\eeq
In general, if the original lattice has local coordination number $q_i$
at the site $i$, then the dual cell centered on $i$ will have $q_i$ faces.
A fairly complete set of formulae for dual volumes relevant for lattice gravity
and their derivation can be found in \cite{hw84}.
In the following we shall refer to the Voronoi cell construction as the
``dual subdivision'', while we will call the baricentric
cell construction the ``baricentric subdivision''.

\begin{center}
\leavevmode
\epsfysize=6cm
\epsfbox{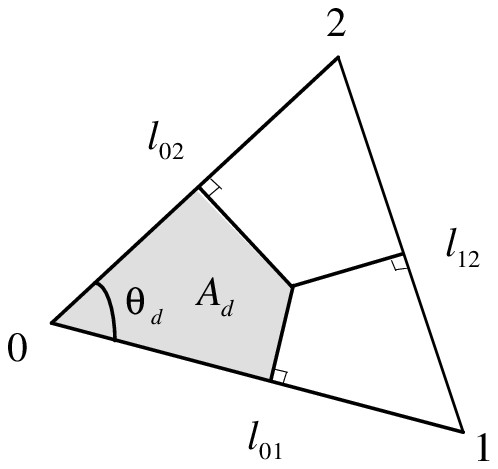}
\end{center}
\noindent
{\small{\it Fig.\ 5.
Dual area $A_d$ associated with vertex $0$, and the corresponding dihedral
angle $\theta_d$.
\medskip}}

Two-dimensional Einstein gravity 
is trivial because the Einstein action is constant and the Ricci 
tensor vanishes identically.  When a cosmological constant term 
and a curvature-squared term are included in the action,
\beq
I \; = \; \int d^2 x \, \sqrt g \, \Bigl [ \lambda - k R + a R^2 \Bigr ] \;\; ,
\eeq
the classical solutions have constant curvature with 
$R= \pm \sqrt{\lambda /a}$ (there being no real solutions for 
$\lambda < 0)$.
Thus the theory with the Einstein action and a cosmological 
constant is metrically trivial, having neither dynamical degrees 
of freedom nor field equations. On the other hand the functional 
measure can lead to a non-trivial effective action.
However, for a system with fixed topology, the only non-classical 
aspects of $1+1$ di\-men\-sio\-nal gravity are fluctuations in 
the local volumes $\sqrt{g(x)}$.

The Einstein action for a two-dimensional simplicial lattice is 
given by \cite{regge}
\beq
\int d^2 x \sqrt g \, R \, \longrightarrow
\, 2 \sum_{\rm hinges~h} \delta_h \;\; .
\label{eq:reggea} 
\eeq
According to the Gauss-Bonnet theorem the Einstein action in 
two-dimensions is equal to $4\pi$ times the Euler characteristic
of the surface.
The same result is true on the lattice, with
$ \sum_h \delta_h = 2 \pi \chi $, where $\chi$ is the Euler characteristic.
It is a constant provided we consider, as we shall do below, surfaces with
a fixed topology.

A cosmological constant term can be included in the action in the form 
\beq
\lambda \int d^2 x \sqrt g \, \longrightarrow \, 
\lambda \sum_{\rm triangles~t} A_t \;\; ,
\label{eq:cosma} 
\eeq
where $A_t$ is the area of triangle $t$.  Equivalently one may 
subdivide the triangles into areas associated with each hinge $A_h$ 
and use the expression
\beq
\lambda \sum_{\rm hinges~h} A_h \;\; .
\eeq
For the baricentric subdivision one has simply
\beq
A_h \; = \; \third \sum_{\rm triangles~t\atop meeting~at~h} A_t \;\; .
\label{eq:duala} 
\eeq
$A_h$ can also be taken to be the area of the cell surrounding 
$h$ in the dual lattice (see Figure 6), with
\beq
A_h \; = \; \sum_{\rm triangles~t\atop meeting~at~h} A_d \;\; ,
\label{eq:baria} 
\eeq
with the dual area contribution for each triangle $A_d$ given in
Eq.~(\ref{eq:voronoi}).
 
\begin{center}
\leavevmode
\epsfysize=7cm
\epsfbox{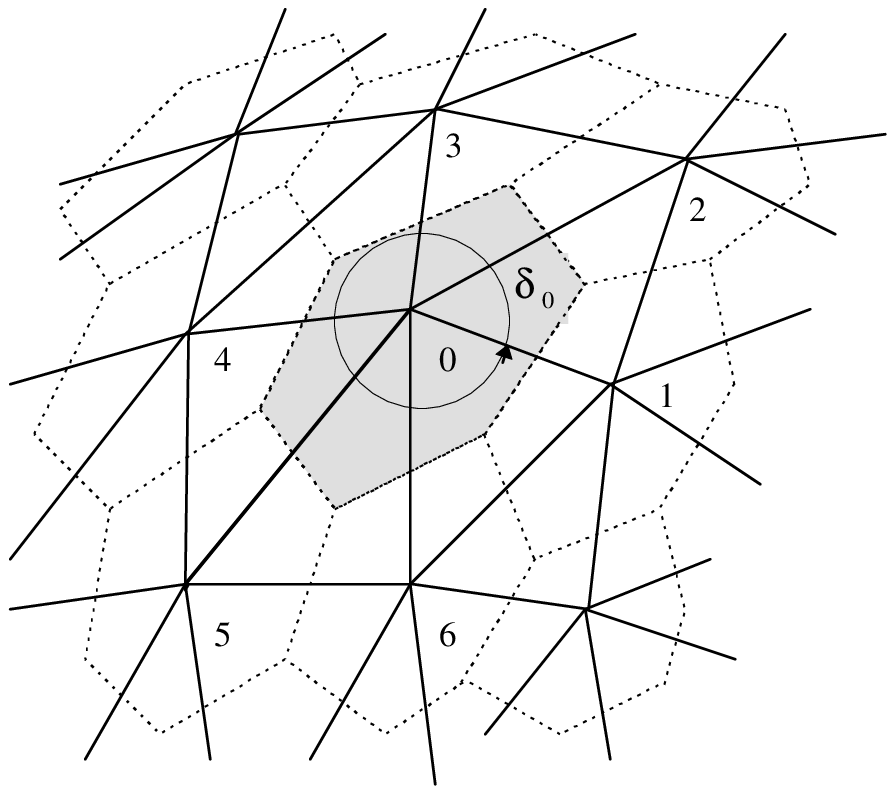}
\end{center}
\noindent
{\small{\it Fig.\ 6.
Original simplicial lattice (continuous lines) and dual lattice (dotted lines)
in two dimensions.
The shaded region corresponds to the dual area associated with vertex $0$.
\medskip}}

In two dimensions the Weyl tensor vanishes identically,
while the other curvature-squared terms are all proportional 
to each other,
\beq
R_{\mu \nu \rho \sigma} R^{\mu \nu \rho \sigma} \; = \; \half \,
R_{\mu \nu}R^{\mu \nu} \; = \; R^2 \;\; .
\eeq
One therefore needs only one term quadratic in the curvature for
the lattice action.
Using the requirements that it be a sum over 
hinges (the only places where the curvature is non-zero), that it 
be quadratic in the deficit angle, and that it have the correct 
dimension (length)$^{-2}$, one is led to the unique expression
\beq
\int d^2 x \sqrt g \, R^2 \, \longrightarrow \,
4 \sum {\delta^2_h \over A_h} \;\; .
\label{eq:curv2a} 
\eeq
It can be shown that this formula is exact for all 
regular tessellations of the two-sphere, in the sense that the
discrete lattice expression does not depend on the how fine the
tessellation is, once the area of the surface is kept fixed ~\cite{hw84}.

The lattice action corresponding to pure gravity is then
\beq
I (l^2) \; = \; \sum_h \, \Bigl [ \lambda \, A_h - 2 k \, \delta_h 
+ 4 a \, { \delta_h^2 \over A_h } \Bigr ] \;\; ,
\label{eq:pure} 
\eeq
which can be written equivalently as
\beq
I (l^2) \; = \; \sum_h \, V_h \, \Bigl [ \lambda - k \, R_h 
+ a \, R_h^2 \Bigr ] \;\; ,
\eeq
with the two-dimensional volume element $V_h=A_h$,
and the local curvature given by $R_h = 2 \; \delta_h / A_h $.
In the limit of small fluctuations around a smooth background, $I(l^2)$
corresponds to the continuum action
\beq
I [g] \; = \; \int d^2 x \,
\sqrt g \, \Bigl [ \lambda - k R + a R^2 \Bigr ] \;\; .
\eeq
For a manifold of fixed topology the term proportional to $k$ can be
dropped, since $ \sum_h \delta_h = 2 \pi \chi $,
where $\chi$ is the Euler characteristic.
The curvature-squared leads to non-trivial interactions in two dimensions,
although the resulting theory is not unitary. In the next section we shall
discuss properties of the above action in the weak field expansion about
flat space, and later about an arbitrary lattice manifold.

Arguments based on perturbation theory about two dimensions (where the
gravitational coupling is dimensionless and the
Einstein theory becomes renormalizable) suggest that there should be no
non-trivial ultraviolet fixed point of the renormalization group
in two dimensions. Explicit calculations in the lattice theory
have shown conclusively that this is indeed the case in the absence of matter
\cite{hw2d,gh,elba,tabor,holm}.
The equations of motion for pure gravity in two dimensions
then follow from the variation
\beq
\delta I[g] \; = \; \half \int d^2 x \, \sqrt g \, \left [ 
\lambda - a R^2 \right ] g^{\mu\nu} \delta g_{\mu\nu} \; = \; 0 \;\; ,
\eeq
and read
\beq
{ a \over 2 } \, R^2 \, g_{\mu\nu} - {\lambda \over 2} \, g_{\mu\nu}
\; = \; 0 \;\; ,
\eeq
or, in contracted form, $ R^2 = {\lambda \over a} $.
For an arbitrary {\it gauge} variation of the metric,
\beq
\delta g_{\mu\nu} (x) \, = \, 
- g_{\mu\lambda} (x) \, \partial_\nu \chi^\lambda (x)
- g_{\lambda\nu} (x) \, \partial_\mu \chi^\lambda (x)
- \partial_{\lambda} g_{\mu\nu} (x) \, \chi^\lambda (x) \;\; ,
\eeq
one obtains after an integration by parts, and using the fact that
the gauge function $\chi^{\lambda}$ is arbitrary
(and that $\left ( g^{\mu\nu} \right )_{; \nu} = 0 $),
\beq
\left ( R^2 g^{\mu\nu} \right )_{; \nu} \; = \; 0 \;\; .
\eeq
This is the two-dimensional analog of the (contracted) Bianchi identity.
Since the squared edge lengths are the primary degrees of freedom,
the corresponding lattice field equations of motion are obtained, in any
dimension, from
\beq
{ \partial \; I [ l^2 ] \over \partial \; l^2_i } \; = \; 0 \;\; .
\eeq
Already in the two-dimensional case they are rather unwieldy when written out
explicitly, and will not be recorded here.

A candidate for the discrete analog of the
two-dimensional Bianchi identity is simply
\beq
\sum_{h(i)} \delta_h \left ( l_i^2 + \delta l_i^2 \right )
\; - \;
\sum_{h(i)} \delta_h \left ( l_i^2 \right ) \; = \; 0 \;\; ,
\eeq
where the sum includes the four hinges $h$ belonging to
the two triangles bordering the edge $i$, and $\delta l_i^2 $
represents a variation of the edges meeting at the vertex $h$.
By considering gauge variations of the edge lengths in higher dimensions,
the corresponding exact lattice Bianchi identities can in
principle be written down. Some further discussion of the Bianchi
identities in higher dimensions can be found in the second reference
in ~\cite{rowi}.

\vskip 10pt
\newsection{Lattice Weak Field Expansion and Zero Modes}
\hspace*{\parindent}

One of the simplest problems which can be studied analytically in the continuum
as well as on the lattice
is the analysis of small fluctuations about some classical background solution.
In the continuum, the weak field expansion is often performed by expanding
the metric and the action about flat Euclidean space
\beq
g_{\mu\nu} (x) \; = \;
\delta_{\mu\nu} \; + \; \kappa \; h_{\mu\nu} (x) \;\; .
\eeq
In four dimensions $\kappa=\sqrt{32 \pi G}$, which shows that the weak
field expansion there corresponds to an expansion in powers of $G$.
In two dimensions this is no longer the case and the relation between
$\kappa$ and $G$ is lost; instead one should regard $\kappa$ as
a dimensionless expansion parameter which is eventually set to one,
$\kappa = 1 $, at the end of the calculation. The procedure will be
sensible as long as wildly fluctuating geometries are not important
in two dimensions (on the lattice or in the continuum).
The influence of the latter configurations 
can only be studied by numerical simulations of the full
path integral ~\cite{hw2d,gh}.

In the lattice case the weak field calculations can be carried out in
three ~\cite{hw3d} and four ~\cite{rowi} dimensional flat background
space with the Regge-Einstein action.
One finds that the Regge gravity propagator 
indeed agrees exactly with the continuum result ~\cite{veltman}
in the weak-field limit.
As a result, the existence of gravitational waves and gravitons in the
discrete lattice theory has been established (indeed it is the only
lattice theory of gravity for which such a result has been obtained
\footnote{A discretization of the edge lengths, and therefore of the
curvatures, as advocated in some models for lattice
gravity, can be considered, the dynamical triangulations being
one specific example. This procedure leads obviously to a loss
of the graviton excitation, at least in the weak field expansion.
In ordinary non-abelian lattice gauge
theories, models based on discrete subgroups of $SU(N)$ have an
artificial freezing transition at finite coupling and no lattice
continuum limit, and do not seem to represent a useful discretization
of the original continuum theory ~\cite{rebbi}.}
).

The weak field expansion about flat space is relevant for the continuum
limit of the lattice quantum theory. Consider a simplicial lattice
approximation to a given continuum manifold.
For an arbitrary continuum manifold, one can envision a triangulation
which is successively refined by making the simplices and the corresponding
edge lengths smaller and smaller.
As the average lattice spacing is reduced, the curvature on the scale
of the lattice spacing becomes eventually sufficiently small that the
simplicial manifold can be regarded as being locally close to flat.
In this limit the curvature is small on the scale of the local volume,
and in two dimensions one has
\beq
\vert curvature \vert_h \; \equiv \;
\Bigl \vert {\delta_h \over A_h} \Bigr \vert \; \ll \;
( volume )_h^{-1} \; \equiv \; { 1 \over A_h } \;\;\;\;\; {\rm or} \;\;\;\;\;
\vert\delta_h\vert \; \ll \; 1 \;\; .
\eeq
In such regions, which become larger and larger in size as the lattice
spacing is reduced, one can meaningfully apply the weak field expansion
about flat space, which becomes only an approximation when it
is truncated to any finite order.

In the following we shall consider in detail only the
two-dimensional case, although
similar calculations can in principle be performed in higher dimensions, with
considerable more algebraic effort.
In the pure gravity case the Einstein-Regge action is a topological
invariant in two dimensions, and one has to consider the next non-trivial
invariant contribution to the action.
We shall therefore consider a two-dimensional lattice with the higher
derivative action of Eq.~(\ref{eq:pure}) and $\lambda=0$,
\beq
I (l^2) \; = \; 4 a \sum_{{\rm hinges \, h}} { \delta^2_h \over A_h } \;\; .
\eeq
The weak field expansion for such a term has largely been done in
~\cite{hw2d}, and we will first recall here the main results.
Since flat space is a classical solution for such an $R^2-$type action, one
can take as a background space a network of unit squares 
divided into triangles by drawing in parallel sets of diagonals 
(see Figure 7). This is one of an infinite number of possible choices
for the background lattice, and a rather convenient one. Physical
results should in the end be insensitive to the choice of
the background lattice used as a starting point for the weak field
expansion.
Opposite edges of the network are supposed to be identified so that the 
lattice acquires the topology of a torus.
(In the following we will be concerned
with local properties of the action, and the detailed nature of the
boundary conditions will play only a marginal role).

\begin{center}
\leavevmode
\epsfysize=5cm
\epsfbox{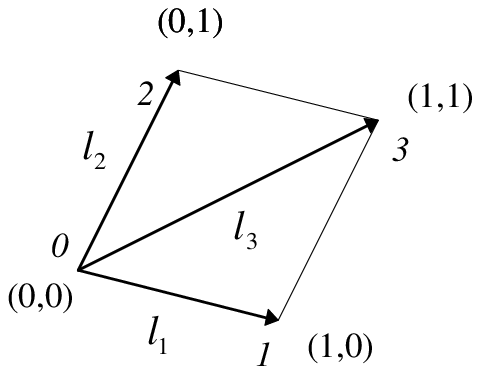}
\end{center}
\noindent
{\small{\it Fig.\ 7.
Notation for the weak-field expansion about the rigid square lattice.
\medskip}}

It is also convenient to use the binary notation for vertices described in
references ~\cite{rowi}.
As discussed in the previous section, the edge lengths
on the lattice correspond to the metric degrees of freedom in the
continuum. 
The edge lengths are thus allowed to fluctuate around their flat 
space values,
\beq
l_i \; = \; l_i^0 \; (1 + \epsilon_i) \;\; ,
\label{eq:epsilon} 
\eeq
with $l_1^0 = l_2^0 =1$ and $l_3^0 = \sqrt{2}$ for our choice of lattice.
The second variation of the 
action is then expressed as a quadratic form in the $ \epsilon $'s,
\beq
\delta^2 I \; = \; 4a \sum_{ij} \, \epsilon_i \; M_{ij} \; \epsilon_j \;\; .
\eeq
The properties of $M_{ij}$ are best studied by going to momentum space.
One assumes that the fluctuation $\epsilon_i$ 
at the point $i$, $j$ steps in one coordinate direction and $k$ steps in the 
other coordinate direction from the origin, is related to the 
corresponding $\epsilon_i$ at the origin by 
\beq
\epsilon_i^{(j+k)} \, = \, \omega_1^j \, \omega_2^k \, \epsilon_i^{(0)} \;\; ,
\eeq
where $\omega_i=e^{- i k_{i} }$ and $k_i$ is the momentum in the direction $i$.
The matrix $M$ then reduces to a 
$3 \times 3$ matrix $M_\omega $ with components given by ~\cite{hw2d} 
\bea
(M_\omega )_{11} & = & 2+ \omega_1 - 2 \omega_2 - 2 \omega_1 \omega_2 + 
\omega_1 \omega_2^2 + c.c.
\nonumber \\
(M_\omega )_{12} & = & 2 - \omega_1 - \bar \omega_2 - \omega_1 \omega_2 - 
\bar \omega_1 \bar \omega_2 - \omega_1^2 - \bar \omega_2^2+ \omega_1^2 
\omega_2+ \bar \omega_1 \bar \omega_2^2 + 2 \omega_1 \bar \omega_2
\nonumber \\
(M_\omega )_{13} & = & 2(-1+2 \omega_1 - \bar \omega_1+ \omega_2 -
\bar \omega_2
- \omega_1 \omega_2 + 2 \bar \omega_1 \bar \omega_2+ \bar \omega_2^2- 
\bar \omega_1 \bar \omega_2^2- \omega_1 \bar \omega_2)
\nonumber \\
(M_\omega )_{33} & = & 4(2-2 \omega_1-2 \omega_2 + \omega_1 \omega_2 + 
\bar \omega_1 \omega_2 + c.c.) \nonumber \\
\label{eq:momega}
\eea
with the other components easily obtained by symmetry.
For small momenta $M_{\omega}$ takes the form
\beq
M_{\omega} \; = \; l^4 \left( \begin{array}{ccc}
k_2^2(k_1+k_2)^2&k_1k_2(k_1+k_2)^2&-2k_1k_2^2(k_1+k_2)\\
k_1k_2(k_1+k_2)^2&k_1^2(k_1+k_2)^2&-2k_1^2k_2(k_1+k_2)\\
-2k_1k_2^2(k_1+k_2)&-2k_1^2k_2(k_1+k_2)&4k_1^2k_2^2\\
\end{array} \right) \,\, + O(k^5) \;\; .
\eeq
The change of variables
\beq
\epsilon_1' \; = \; \epsilon_1 \,\,\,\,
\epsilon_2' \; = \; \epsilon_2 \,\,\,\,
\epsilon_3' \; = \; \half ( \epsilon_1+ \epsilon_2)+ \epsilon_3 \;\; .
\label{eq:change} 
\eeq
leads for small momenta to the matrix $M_{\omega}' $ given by 
\beq
M_{\omega}' \; = \; l^4 \left( \begin{array}{ccc}
k_2^4&k_1^2k_2^2&-2k_1k_2^3\\
k_1^2k_2^2&k_1^4&-2k_1^3k_2\\
-2k_1k_2^3&-2k_1^3k_2&4k_1^2k_2^2\\
\end{array} \right) \,\, + O(k^5) \;\; .
\eeq
This expression is identical to what one obtains from the corresponding
weak-field limit in the continuum theory.
To see this, define as usual the small fluctuation field
$h_{\mu\nu} $ about flat space by setting
\beq
g_{ \mu \nu } \; = \; \delta_{ \mu \nu } \; + \; h_{ \mu \nu } \;\; .
\eeq
In two dimensions one has
\beq
R \; = \; h_{11,22} + h_{22,11} - 2h_{12,12} + O(h^2) \;\; ,
\eeq
and also
\beq
\sqrt{g} \; = \; 1 + \half (h_{11} + h_{22}) + O(h^2) \;\; ,
\eeq
which gives
\beq
\sqrt{g} \, R^2 \; = \; (h_{11,22} + h_{22,11} - 2h_{12,12})^2 + O(h^3)
\;\; .
\eeq
In momentum space, each derivative $ \partial_\nu $ produces a 
factor of $ k_\nu $, and so one obtains
\beq
\sqrt{g} \, R^2 \; = \; h_{ \mu \nu } \, V_{ \mu \nu , \rho \sigma } \,
h_{ \rho \sigma } \;\; ,
\eeq
where $ V_{ \mu \nu , \rho \sigma } $ coincides with $ M' $ 
above (when the metric components are re-labeled according to 
$ 11 \rightarrow 1, \; 22 \rightarrow 2, \; 12 \rightarrow 3) $.

One might wonder what the origin of the change of variables in
Eq.~(\ref{eq:change}) is.
Given the three edges in Figure 7, one can write for the metric at the origin
\beq
g_{ij} \; = \; \left( \begin{array}{cc}
l_1^2 & \half (l_3^2 - l_1^2 - l_2^2 ) \\
\half (l_3^2 - l_1^2 - l_2^2 ) & l_2^2 \\
\end{array} \right) \;\; .
\label{eq:gij_square}
\eeq
The apparent contradiction with the earlier expression for $g_{ij}$
given in Eq.~(\ref{eq:gij_triangle}) arises from the different
choice of coordinates in the triangles (compare Figure 3 with Figure 7).
Inserting $l_i \, = \, l_i^0 \, ( 1 + \epsilon_i )$,
with $l_i^0 = 1 $ for the body principals
($i=1,2$) and $l_i^0 = \sqrt{2} $ for the diagonal ($i=3$), one obtains
\bea
l_1^2 \; = \; ( 1 + \epsilon_1 )^2 & \, = \, & 1 + h_{11}
\nonumber \\ 
l_2^2 \; = \; ( 1 + \epsilon_2 )^2 & \, = \, & 1 + h_{22}
\nonumber \\ 
\half l_3^2 \; = \; ( 1 + \epsilon_3 )^2 & \, = \, & 1 +
\half ( h_{11} + h_{22} ) + h_{12}
\nonumber \\ 
\label{eq:htoeps}
\eea
which can be inverted to give
\bea
\epsilon_1 & \, = \, & \half h_{11} - \eigth h_{11}^2 + O(h_{11}^3)
\nonumber \\ 
\epsilon_2 & \, = \, & \half h_{22} - \eigth h_{22}^2 + O(h_{22}^3)
\nonumber \\ 
\epsilon_3 & \, = \, & \quarter ( h_{11} + h_{22} + 2 h_{12} )
- {\textstyle {1\over32} \displaystyle} 
( h_{11} + h_{22} + 2 h_{12} )^2 + O(h^3)
\nonumber \\ 
\label{eq:epstoh}
\eea
and which was then used in Eq.~(\ref{eq:change}).
Thus the matrix
$M_{\omega}$ was brought into the continuum form after performing
a suitable local rotation from the local edge lengths to the local metric
components.

The weak field expansion for the purely gravitational part can be carried
out to higher order, and the Feynman rules for the vertices of order
$h^3$, $h^4$, $\dots$ in the $R^2$-action of
Eq.~(\ref{eq:pure}) can be derived. Since their expressions are rather
complicated, they will not be recorded here.

\vskip 10pt
\subsection{Lattice Diffeomorphisms}
\hspace*{\parindent}

It is easy to determine the eigenvalues and eigenvectors of the matrix
$M_\omega$ of Eq.~(\ref{eq:momega}).
The eigenvalues of the matrix $M_\omega$ are given by
\bea
\lambda_1 & = & 0 
\nonumber \\ 
\lambda_2 & = & 0
\nonumber \\ 
\lambda_3 & = &
24 - 9 ( \omega_1 + \bar \omega_1 + \omega_2 + \bar \omega_2 )
+ 4 ( \omega_1 \bar \omega_2 + \bar \omega_1 \omega_2 )
\nonumber \\ 
&& + \omega_1 \omega_2^2 + \omega_1^2 \omega_2
+ \bar \omega_1 \bar \omega_2^2 + \bar \omega_1^2 \bar \omega_2
\nonumber \\
\eea
and there are thus two exact zero modes in the weak field limit.
It should be emphasized that the exact zero modes appear for arbitrary
$\omega_i$, and not just for small momenta. We shall see later that
their presence reflects an exact local continuous invariance of the
gravitational action.

If one were interested in doing lattice perturbation theory,
one would have to add a lattice gauge fixing term to remove the zero modes,
such as the lattice analog of the term
\beq
{ 1 \over \kappa^2 } 
\left ( \partial_\mu {\textstyle \sqrt{g(x)} \displaystyle} \;
g^{\mu\nu} \right )^2 \;\; ,
\eeq
and add the necessary Fadeev-Popov non-local ghost determinant.
A similar term would have to be included as well if one were to pick the
lattice analog of the conformal gauge ~\cite{poly},
to which we shall return later.
If one is {\it not} doing perturbation theory, then of course the contribution
of the zero modes will cancel out between the numerator and denominator
in the Feynman path integral representation for operator averages, and
such a term should not be included, as in ordinary lattice
formulations of Yang-Mills gauge theories.

The eigenvectors corresponding to the two zero modes can be written as
\beq
\left( \begin{array}{c}
\epsilon_1 ( \omega ) \\
\epsilon_2 ( \omega ) \\
\epsilon_3 ( \omega ) \\
\end{array} \right)
\, = \,
\left( \begin{array}{cc}
1 - \omega_1 & 0          \\
0            & 1 - \omega_2 \\
\half ( 1 - \omega_1 \omega_2 ) &
\half ( 1 - \omega_1 \omega_2 ) \\
\end{array} \right)
\left( \begin{array}{c}
\chi_1 ( \omega ) \\
\chi_2 ( \omega ) \\
\end{array} \right) \;\; ,
\label{eq:gauge_wfe_o}
\eeq
where $\chi_1 (\omega) $ and $\chi_2 (\omega) $ are arbitrary.
One might worry that the above result is restricted to two
dimensions. This is not the case.
Completely analogous zero modes are found for the Regge
action in three ~\cite{hw3d} and four ~\cite{rowi} dimensions,
leading to expressions rather similar to Eq.~(\ref{eq:gauge_wfe_o}), with
as expected $d$ zero modes in $d$ dimensions.
As we shall see, this is not a coincidence.
We give here for comparison the corresponding expression in three
dimensions \cite{hw3d}, obtained from the weak field expansion of
the Regge action, 
\beq
\left( \begin{array}{c}
\epsilon_1  ( \omega ) \\ \epsilon_2  ( \omega ) \\ 
\epsilon_4  ( \omega ) \\ \epsilon_3  ( \omega ) \\
\epsilon_5  ( \omega ) \\ \epsilon_6  ( \omega ) \\
\epsilon_7  ( \omega ) \\
\end{array} \right)
\, = \,
\left( \begin{array}{ccc}
1- \omega_1 & 0 & 0 \\
0 & 1 - \omega_2 & 0 \\
0 & 0 & 1 - \omega_4 \\
\half ( 1 - \omega_1 \omega_2 ) & 
\half ( 1 - \omega_1 \omega_2 ) & 0 \\
\half ( 1 - \omega_1 \omega_4 ) &  0 &
\half ( 1 - \omega_1 \omega_4 ) \\
0 & \half ( 1 - \omega_2 \omega_4 ) & \half ( 1 - \omega_2 \omega_4 ) \\ 
\third ( 1 - \omega_1 \omega_2 \omega_4 ) & 
\third ( 1 - \omega_1 \omega_2 \omega_4 ) & 
\third ( 1 - \omega_1 \omega_2 \omega_4 ) \\
\end{array} \right)
\left( \begin{array}{c}
\chi_1  ( \omega ) \\ \chi_2 ( \omega ) \\ \chi_3 ( \omega ) \\
\end{array} \right) \;\; ,
\eeq
again with $\chi_1 (\omega) $, $\chi_2 (\omega) $
and $\chi_3 (\omega) $ arbitrary gauge functions
(in the binary notation for the edges, the indices 1,2,4 correspond to
the body principals, the indices 3,5,6 to the face diagonals, and 7 to
the body diagonal).

\begin{center}
\leavevmode
\epsfysize=5cm
\epsfbox{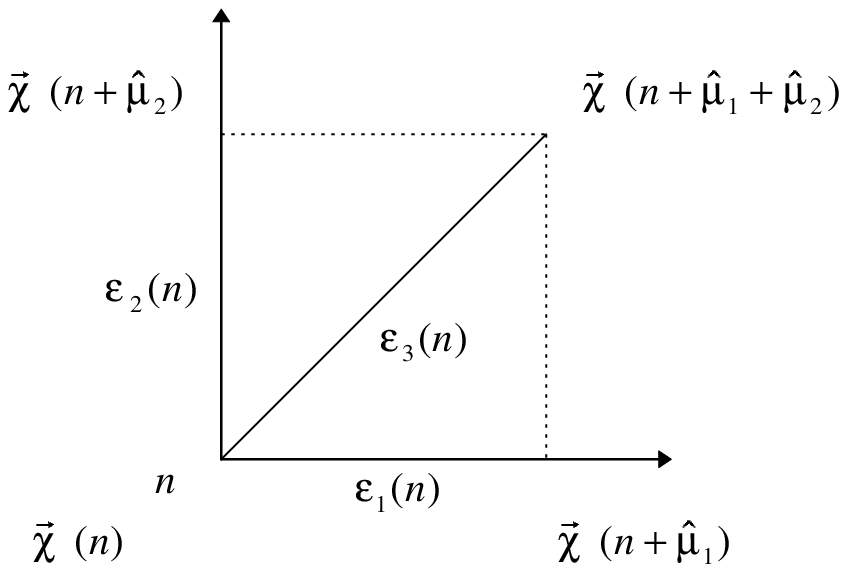}
\end{center}
\noindent
{\small{\it Fig.\ 8.
Edge length gauge deformations $\epsilon_i(n)$, and corresponding 
gauge transformation vector field $\vec \chi (n)$, defined on the sites.
\medskip}}

It is useful to look at the above relations, and in particular
Eq.~(\ref{eq:gauge_wfe_o}, in real space.
The replacement
$ e^{ i k_i } \rightarrow e^{d \over dx} $ and
$ e^{ \vec a \cdot \nabla } f(\vec x) = f(\vec x + \vec a) $ yields
\bea
\epsilon_1 (n) & \, = \, & \chi_1 (n) - \chi_1 ( n + \hat \mu_1 )
\nonumber \\ 
\epsilon_2 (n) & \, = \, & \chi_2 (n) - \chi_2 ( n + \hat \mu_2 )
\nonumber \\ 
\epsilon_3 (n) & \, = \, & \half \chi_1 (n) + \half \chi_2 (n) -
\half \chi_1 ( n + \hat \mu_1 + \hat \mu_2 ) -
\half \chi_2 ( n + \hat \mu_1 + \hat \mu_2 ) \;\; .
\nonumber \\
\label{eq:gauge_wfe}
\eea
For our notation, we refer to the drawing in Figure 8.
Note how the arbitrary gauge variations act on the two ends of an edge.
This is true in any dimension, where elementary local gauge transformations
are always defined on the {\it vertices} of the lattice.
The above gauge transformation law is in fact remarkably simple
for those edges that lie in the direction of the chosen coordinates,
namely
\beq 
\delta l^2_{ij} \; = \; {\chi'}_i \; - \; {\chi'}_j \;\; ,
\label{eq:gauge_like1d}
\eeq
where $i$ and $j$ labels the end-points of the edge, and
we have rescaled the arbitrary functions $\chi$ by $l_{ij}^2$
so that the quantities ${\chi'}_i$ now have dimensions of length
squared.

It is easy to see that the above equations solve the constraint
$\sum_i \epsilon_i (n) = 0 $ at the vertex $n$ where $\vec \chi_n \neq 0 $,
\bea
\sum_{i=1}^6 \; \epsilon_i (n)  & = & 
\chi_1 (n) + \half \chi_1 (n) + \half \chi_2 (n)
+ \chi_2 (n)
- \chi_1 (n) - \half \chi_1 (n) - \half \chi_2 (n)
- \chi_2 (n) \,
\nonumber \\
& = & 0 \;\; .
\nonumber \\
\eea
It can be written equivalently in terms of variations of the squared edge
lengths meeting at the vertex $n$, labeled clockwise around the vertex
$n$ starting with the edge in the positive 1 direction,
\beq
a \; \left ( \delta l^2_{n1} + \delta l^2_{n3} + \delta l^2_{n4} +
\delta l^2_{n6} \right ) \; + \;
b \;  \left ( \delta l^2_{n2} + \delta l^2_{n5} \right )
\; = \; 0 \;\; ,
\label{eq:gauge_r}
\eeq
with $a$ and $b$ arbitrary constants at this point.
In other words, gauge variations of the squared edge lengths are
recognized as special variations, where all edges meeting at a
point (in two dimensions) are considered, and which either have
the explicit form given in Eq.~(\ref{eq:gauge_wfe}) for the weak field
case, or equivalently (and more generally) satisfy a set of
defining constraints such as the one in Eq.~(\ref{eq:gauge_r}).

Incidentally, it should be noted here that conformal transformations,
which in the continuum take the form
$ \delta g_{\mu\nu} (x) = g_{\mu\nu} (x) \; \delta \varphi (x) $,
have a natural lattice analog. They contract
(or expand) locally all the edges meeting on a given vertex $n$ by the
same amount,
\beq
\delta l^2_{i} (n) \; = \; l^2_{i} (n) \;\; \delta \varphi (n) \;\; ,
\label{eq:confdef}
\eeq
and therefore {\it do} change locally the curvature at $n$.
When constructed in this way, they
can be considered orthogonal to the $\chi$ transformations of
Eqs.~(\ref{eq:gauge_wfe_o}) and (\ref{eq:gauge_wfe}).

The cosmological constant term can also be shown to be invariant under
the same set of continuous local transformations,
since, using the same notation for the expansion about the square lattice,
one obtains
\beq
\sum_h A_h \, = \, \sum_n \left [ \; 1 + \third ( 
\epsilon_1 (n) + \epsilon_2 (n) ) + O(\epsilon^2) \; \right ] \;\; ,
\eeq
and at the vertex $P_n$ where $\vec \chi_n \neq 0 $ one has again
\beq
\sum_{i=1}^4 \; \epsilon_i (n) \, = \,
\chi_1 (n) + \chi_2 (n) - \chi_1 (n) - \chi_2 (n) \, = \, 0 \;\; ,
\eeq
where the sum is over the four edges pointing in the four
principal directions.
Written in terms of the variations of the squared edge lengths, one has
\beq
\delta l^2_{n1} + \delta l^2_{n3} + \delta l^2_{n4} +
\delta l^2_{n6} \; = \; 0 \;\; .
\label{eq:gauge_v}
\eeq
In general one has in two dimensions six edges meeting at a vertex,
and therefore four allowed constraints on the gauge edge length variations.
In conclusion we have exhibited an exact local gauge invariance of the
gravitational action in the weak field limit. Later on we shall show that
it corresponds precisely to the lattice analog of the diffeomorphisms.

It is important to notice that the appearance of zero modes in
the weak field expansion is not specific to the expansion about flat space.
One can look at the same procedure for variations about 
spaces which are classical solutions for the 
gravitational action with a cosmological constant term
as in Eq.~(\ref{eq:pure}), such as the 
regular tessellations of the two-sphere \cite{hw84}.
In the following we will consider edge length
fluctuations about the regular tetrahedron
(with 6 edges), octahedron (12 edges), and icosahedron (30 edges).

\begin{center}
\leavevmode
\epsfysize=5cm
\epsfbox{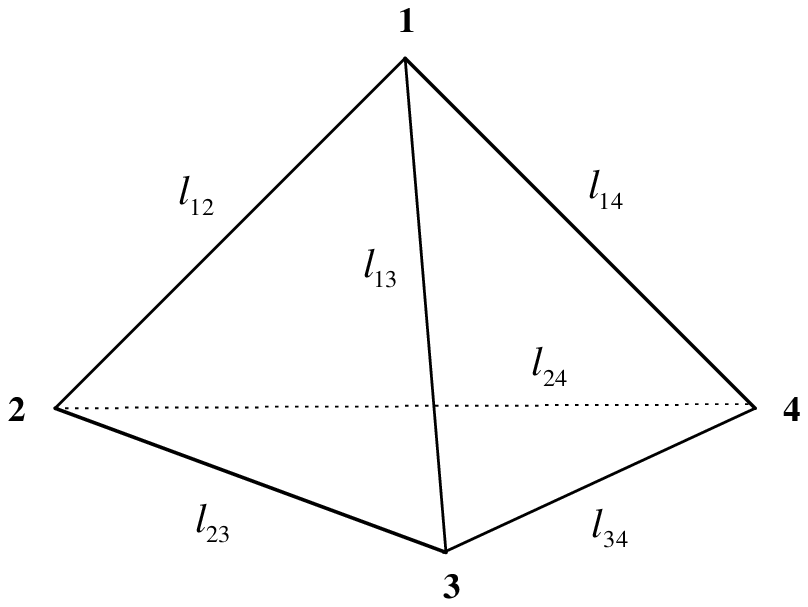}
\end{center}
\noindent
{\small{\it Fig.\ 9.
Tetrahedral tessellation of the two-sphere, with arbitrary edge
length assignments.
\medskip}}

After expanding about the equilateral configuration,
the action at the stationary point reduces to
\beq
I \; = \; \lambda \; 8 \pi \sqrt{a / \lambda} +
a \; 8 \pi / \sqrt{a / \lambda} \; = \; 16 \pi \sqrt{a \; \lambda} \;\;\; ,
\eeq
in fact independently of the tessellation considered.
Vanishing of the linear terms in the small fluctuation
expansion gives for the average edge length
\beq
l_0 \; = \; \left [ c \pi^2 (4 a / \lambda) \right ]^{1/4} \;\;\; ,
\eeq
with $c=16/3,4/3,16/75$ for the tetrahedron, octahedron and
icosahedron, respectively.
For fluctuations about the classical solution for a tetrahedral
tessellation of $S^2$ (see Figure 9) the small
edge length fluctuation matrix gives
rise to the following coefficients
\bea
\epsilon_{12}^2 & \rightarrow &
16 \sqrt{a \lambda} \; (54 - 6 \sqrt 3 \pi + 5 \pi^2) / 81 \pi
\nonumber \\
\epsilon_{12} \; \epsilon_{13} & \rightarrow &
16 \sqrt{a \lambda} \; \pi / 9
\nonumber \\
\epsilon_{12} \; \epsilon_{15} & \rightarrow &
64 \sqrt{a \lambda} \; (-27 + 3 \sqrt 3 \pi + 2 \pi^2) / 81 \pi
\nonumber \\
\eea
with the remaining coefficients being determined by symmetry.
The small fluctuation matrix is therefore given by
\beq
{8 \pi \sqrt{a \lambda} \over 9 } \,
\left( \begin{array}{cccccc}
\mu&1&1&1&1&2- \mu\\
1&\mu&1&2- \mu&1&1\\
1&1&\mu&1&2- \mu&1\\
1&2- \mu&1&\mu&1&1\\
1&1&2- \mu&1&\mu&1\\
2- \mu&1&1&1&1&\mu\\
\end{array} \right) \;\; ,
\eeq
where $\mu = 2 (5 \pi^2-6 \sqrt 3 \pi + 54)/9 \pi^2 \approx 1.5919$.
(the $\lambda/a$ dependence has disappeared since the couplings $a$ and
$\lambda$ only appear in the dimensionless combination $\sqrt{a \lambda}$.
The eigenvalues of the above matrix (neglecting the constants in front
of it) are $0$ (with multiplicity 2), $2(\mu -1)$ (with multiplicity 3)
and 6 (with multiplicity 1).
The zero modes correspond to flat directions for which deformations
of the edge lengths leave the lattice geometry unchanged.
The multiplicities of the eigenvalues are agree with the dimensions
of the irreducible representations of the symmetry group of the tetrahedron.

\begin{center}
\leavevmode
\epsfysize=7cm
\epsfbox{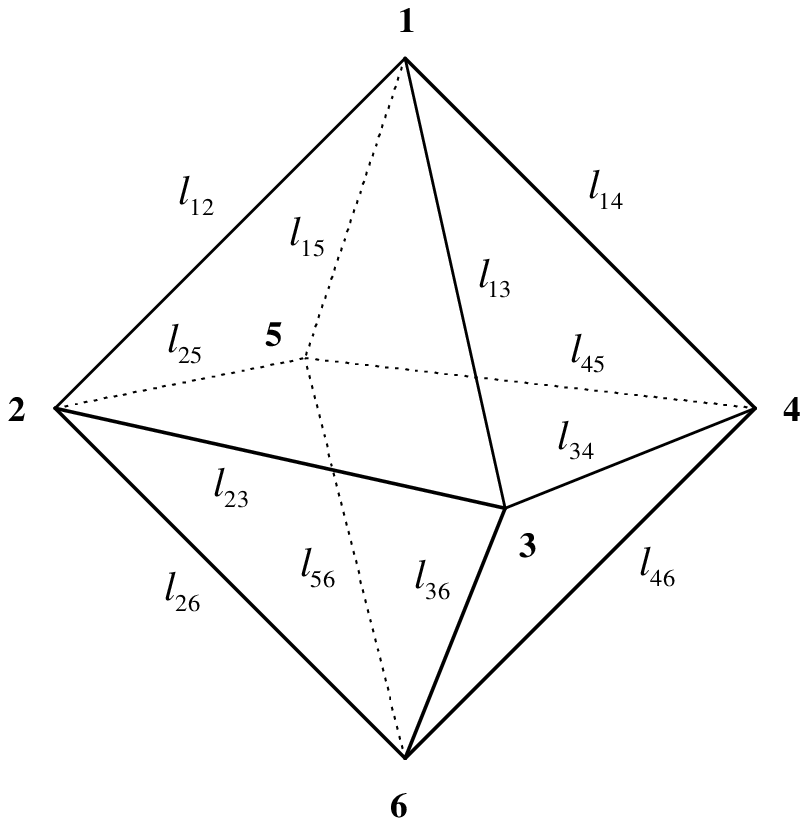}
\end{center}
\noindent
{\small{\it Fig.\ 10.
Octahedral tessellation of the two-sphere, with arbitrary edge
length assignments.
\medskip}}

For the octahedron (see Figure 10) one obtains instead the following
coefficients of the small fluctuation matrix
\bea
\epsilon_{12}^2 & \rightarrow &
2 \sqrt{a \lambda} \; (216 - 12 \sqrt 3 \pi + 5 \pi^2) / 27 \pi
\nonumber \\
\epsilon_{12} \; \epsilon_{13} & \rightarrow &
8 \sqrt{a \lambda} \; (-27 - 3 \sqrt 3 \pi + 2 \pi^2) / 27 \pi
\nonumber \\
\epsilon_{12} \; \epsilon_{14} & \rightarrow &
4 \sqrt{a \lambda} \; (54 + \pi^2) / 9 \pi
\nonumber \\
\epsilon_{12} \; \epsilon_{34} & \rightarrow &
8 \sqrt{a \lambda} \; (-54 + 3 \sqrt 3 \pi + \pi^2) / 27 \pi
\nonumber \\
\epsilon_{12} \; \epsilon_{46} & \rightarrow &
4 \sqrt{a \lambda} \; (108 + 12 \sqrt 3 \pi + \pi^2) / 27 \pi
\nonumber \\
\eea
again with the remaining coefficients being determined by symmetry.
Up to a common factor of 
$ 2 \sqrt{a \lambda} / 27 \pi $, the eigenvalues of the
$12 \times 12$ small fluctuation matrix are given by $36 \pi^2$ (with
multiplicity 1), $972$ (with multiplicity 2), and
$8 (3 \sqrt{3}-\pi)^2 $ (with multiplicity 3), and zero (with
multiplicity 6).

\begin{center}
\leavevmode
\epsfysize=9cm
\epsfbox{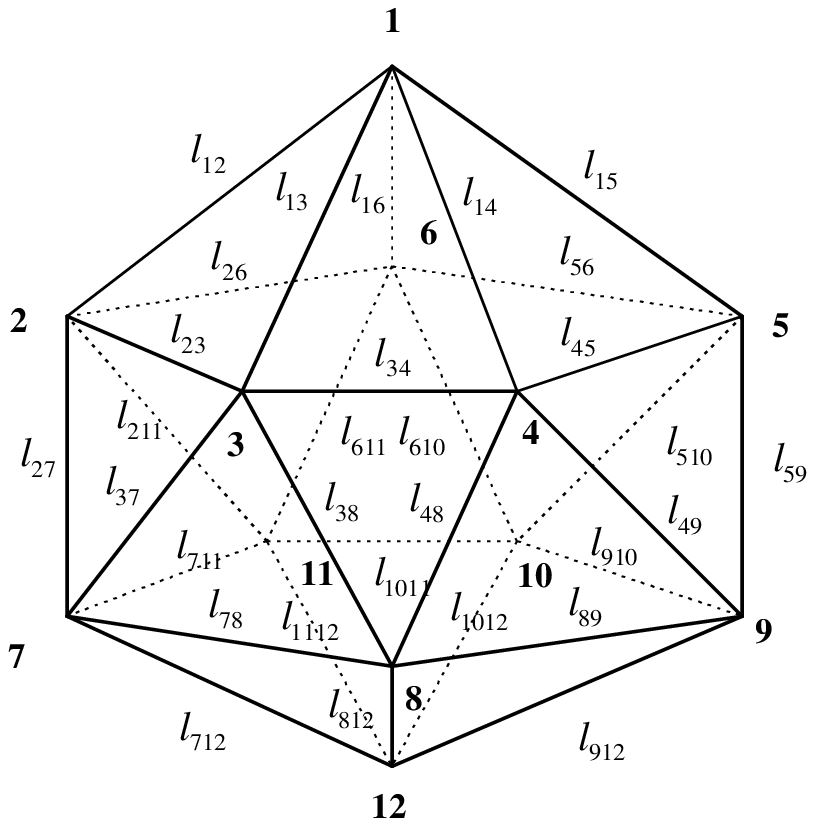}
\end{center}
\noindent
{\small{\it Fig.\ 11.
Icosahedral tessellation of the two-sphere, with arbitrary edge
length assignments.
\medskip}}

Finally, for the icosahedron (shown in Figure 11) one 
computes the following coefficients of the small fluctuation matrix
\bea
\epsilon_{12}^2 & \rightarrow &
16 \sqrt{a \lambda} \; (270 - 6 \sqrt 3 \pi + \pi^2) / 135 \pi
\nonumber \\
\epsilon_{12} \; \epsilon_{13} & \rightarrow &
16 \sqrt{a \lambda} \; (-675 - 30 \sqrt 3 \pi + 8 \pi^2) / 675 \pi
\nonumber \\
\epsilon_{12} \; \epsilon_{14} & \rightarrow &
16 \sqrt{a \lambda} \; (270 - 6 \sqrt{3} + \pi^2) / 135 \pi
\nonumber \\
\epsilon_{12} \; \epsilon_{34} & \rightarrow &
32 \sqrt{a \lambda} \; (-675 + 15 \sqrt{3} \pi + 2 \pi^2) / 675 \pi
\nonumber \\
\epsilon_{12} \; \epsilon_{45} & \rightarrow &
16 \sqrt{a \lambda} \; (-675 + 15 \sqrt{3} \pi + 2 \pi^2) / 675 \pi
\nonumber \\
\epsilon_{12} \; \epsilon_{38} & \rightarrow &
16 \sqrt{a \lambda} \; (-675 + 15 \sqrt{3} \pi + 2 \pi^2) / 675 \pi
\nonumber \\
\epsilon_{12} \; \epsilon_{48} & \rightarrow &
16 \sqrt{a \lambda} \; (675 + 30 \sqrt{3} \pi + \pi^2) / 675 \pi
\nonumber \\
\eea
with the remaining coefficients being determined by symmetry.
Up to a common factor of 
$ 8 \sqrt{a \lambda} / 675 \pi $, the eigenvalues of the
$30 \times 30$ small edge length
fluctuation matrix are given (numerically) by $12340.173$ (with
multiplicity 3), $7238.984$ (with multiplicity 5),
$888.264=90\pi^2$ (with multiplicity 1), $20.887$ (with multiplicity 3),
and zero (with multiplicity 18).

The presence of the zero modes is related
to the gauge (diffeomorphism) invariance of the gravitational action.
The previous results can in fact be summarized as
\bea
{\rm Tetrahedron} \; (N_0 = 4) : && {\rm 2 \; zero \; modes}
\nonumber \\
{\rm Octahedron} (N_0 = 6) : && {\rm 6 \; zero \; modes}
\nonumber \\
{\rm Icosahedron} (N_0 = 12 ) : && {\rm 18 \; zero \; modes}
\nonumber \\
\eea
If the number of zero modes for each triangulation of the sphere is
denoted by $N_{z.m.}$, then the results can be re-expressed as
\beq
N_{z.m.} \; = \; 2 N_0 \; - \; 6 \;\; ,
\eeq
which agrees with the expectation that in the continuum limit,
$N_0 \rightarrow \infty$, $N_{z.m.}/N_0$ should approach the constant
value $d$ in $d$ space-time dimensions, which is
the number of local parameters for a diffeomorphism.
On the lattice the diffeomorphisms correspond to local deformations
of the edge lengths about a vertex, which leave the local geometry
physically unchanged, the latter being described by the values of local
lattice operators corresponding to local volumes, and curvatures.
The lesson is that the correct count of zero modes will in general only be
recovered asymptotically for large triangulations, where $N_0 $ is
roughly much larger than the number of neighbors to a point in $d$ dimensions.
It should be possible to find a similar pattern in higher dimensions.

\vskip 10pt
\subsection{Edge Lengths as Metric Components}
\hspace*{\parindent}

Returning to the weak field expansion about flat space,
it is easy to see that the above lattice gauge transformation corresponds
to the diffeomorphisms in the continuum. Using the relationship
between the metric perturbations and the edge length variations,
obtained by choosing coordinates axes along the edges (as in
Eq.~(\ref{eq:gij_square}) and Figure 7),
\beq
\delta g_{ij} (n) \; = \; \left( \begin{array}{cc}
\delta l_1^2 (n) & \half (\delta l_3^2 (n) -
\delta l_1^2 (n) - \delta l_2^2 (n) ) \\
\half (\delta l_3^2 (n) - \delta l_1^2 (n) - 
\delta l_2^2 (n) ) & \delta l_2^2 (n) \\
\end{array} \right) \;\; ,
\label{eq:gpert}
\eeq
one obtains from Eq.~(\ref{eq:gauge_wfe}) the result
\bea
\delta g_{11} & = & \delta l_1^2 \; = \;
2 \, \chi_1 (n) - 2 \, \chi_1 ( n + \hat \mu_1 ) \approx
- 2 \, \partial_1 \chi_1
\nonumber \\ 
\delta g_{22} & = & \delta l_2^2 \; = \;
2 \, \chi_2 (n) - 2 \, \chi_2 ( n + \hat \mu_2 ) \approx
- 2 \, \partial_2 \chi_2
\nonumber \\ 
\delta g_{12} & = & \half ( \delta l_3^2 - \delta l_1^2 - \delta l_2^2 )
\nonumber \\ 
& = & \chi_1 ( n + \hat \mu_1 ) - \chi_1 ( n + \hat \mu_1 + \hat \mu_2 )
+ \chi_2 ( n + \hat \mu_2 ) - \chi_2 ( n + \hat \mu_1 + \hat \mu_2 )
\nonumber \\ 
& \approx & - \partial_2 \chi_1 - \partial_1 \chi_2 \;\; ,
\nonumber \\
\eea
which can then be combined into the single familiar expression
\beq
\delta g_{\mu\nu} \; = \; 
- \partial_\mu \chi_\nu - \partial_\nu \chi_\mu \;\; ,
\eeq
and which is indeed the correct gauge variation in the weak field limit.

Conversely, the above form of the lattice gauge transformation,
Eq.~(\ref{eq:gauge_wfe}), can be obtained from the form of infinitesimal
diffeomorphisms in the continuum.
In order to see this result, start from the definitions of
diffeomorphisms
\beq
g'_{\mu\nu} (x') \, = \, { \partial x^\rho \over \partial x_\mu '}
{ \partial x^\sigma \over \partial x_\nu '} \,
g_{\rho\sigma} (x) \;\; ,
\eeq
as transformations which leave the infinitesimal line element, as well
as any other coordinate invariant quantity, unchanged
\beq
ds'^2 \, \equiv \, g'_{\mu\nu} (x') \, dx'^\mu dx'^\nu
\, = \, ds^2 \, \equiv \, g_{\rho\sigma} (x) \, dx^\rho dx^\sigma
\eeq
under an arbitrary change of coordinates
\beq
x'^\mu \, = \, x^\mu + \chi^\mu (x) \;\; .
\eeq
For infinitesimal variations one obtains
\beq
g'_{\mu\nu} (x') \, = \,
g_{\mu\nu} (x) + \delta g_{\mu\nu} (x) \, = \, g_{\mu\nu} (x) 
- g_{\mu\lambda} (x) \, \partial_\nu \chi^\lambda (x)
- g_{\lambda\nu} (x) \, \partial_\mu \chi^\lambda (x) + O ( \chi^2 ) \;\; .
\label{eq:diffeo}
\eeq
The above relationships express the well-known fact that metrics related by
a coordinate transformation describe the same physical manifold. 
In the discrete case it reflects the invariance of the lattice action under
local deformations of the simplicial manifold which leave the local curvatures
unchanged ~\cite{lesh}.
Since the continuum metric degrees of freedom correspond on the lattice
to the values of edge lengths squared, one would expect to find 
analogous deformations of the edge lengths that leave the lattice geometry
invariant, the latter being specified by the local {\it lattice}
areas and curvatures, in accordance with the principle of discussing the
geometric properties of the lattice theory in terms of lattice quantities only.
Clearly the distance between lattice vertices
will change under such a transformation, in accordance with the fact
that only distances between {\it fixed} points will remain the same.
This invariance is spoiled by the presence of the triangle inequalities,
which places a constraint on how far the individual edge lengths
can be deformed.
In the perturbative, weak field expansion about
a fixed background the triangle inequalities are not seen to any order
in perturbation theory, they represent non-perturbative constraints.

These considerations are further illustrated by the following
elementary example \cite{oned}.
In one dimension (zero space, one time dimension)
one can discretize the line by introducing $N$ points,
joined by segments of lengths $l_n$. The only invariant term in one
dimension is obviously the length of the curve,
\beq
L (l) \; = \; \sum_{n=1}^N \; l_n \;\; .
\label{eq:length}
\eeq
From the expression for the invariant line element, $ds^2 = g dx^2$,
one naturally associates $g(x)$ with $l_n^2$, and the coordinate increment
with the lattice spacing, $dx=1$.
One can take the view that distances can only be assigned
between vertices which appear on some lattice in the ensemble, although
this is not strictly necessary as distances can also be defined for
locations that do not coincide with any specific vertex.

The above one-dimensional action has an exact local invariance
(compare with Eq.~(\ref{eq:gauge_wfe}) in two dimensions)
\beq
\delta l_n \; = \; \chi_{n+1} - \chi_n \;\; ,
\label{eq:gauge_1d} 
\eeq
where the $\chi_n$'s represent continuous gauge transformations defined on
the lattice vertices (actually,
in order for the edge lengths to remain positive,
one should also require $\chi_{n} - \chi_{n+1} < l_n $, which
is satisfied for sufficiently small $\chi$'s).
These transformations are in fact remarkably close in structure to the ones
found in two dimensions (see Eq.~(\ref{eq:gauge_like1d})).
Physically, the local invariance
reflects the re-parameterization, or coordinate invariance, of the original
continuum action $L = \int dx \sqrt{ g(x) }$.
It is the discrete form of the change $\delta g(x) = 2 g \partial \chi $.
Variations of the edges
which satisfy Eq.~(\ref{eq:gauge_1d}) leave the physical length of
the curve unchanged. In addition, given any two points on the curve,
independent local gauge transformations can be performed on any of the
vertices situated between the
two points, while at the same time maintaining the same physical distance
between them (which, for any assignments of edge lengths,
is simply obtained by adding up the intervening edge lengths).
It justifies the name {\it lattice diffeomorphisms} for the transformations
of Eq.~(\ref{eq:gauge_1d}). 
\footnote{There is here an (incomplete) analogy with ordinary lattice gauge
theories, in the sense that if one defines $U_n = e^{l_n}$ and
$V_n = e^{\chi_n}$, then the gauge transformation law of
Eq.~(\ref{eq:gauge_1d}) can be re-written as
$ U_n \rightarrow V^{-1}_{n} U_n V_{n+1}$, which parallels the gauge 
transformation law for the $SU(N)$ gauge fields $U_{n\mu}$ in $d$ dimensions,
$ U_{n\mu} \rightarrow V^{-1}_{n} U_{n\mu} V_{n+\mu}$, where
$V_{n\mu}$ are arbitrary $N \times N$ $SU(N)$ matrices.
The ``update'' $U_n \rightarrow V \cdot U_n$
corresponds here to $l_n \rightarrow l_n + \delta l_n$.  }

In two dimensions one starts from the relationship of Eq.~(\ref{eq:gij_square})
between the squared edge lengths and the metric, and uses
the expression for metric perturbations given in
Eq.~(\ref{eq:gpert}).
From Eq.~(\ref{eq:diffeo}) one obtains
\bea
\delta l_1^2 & = & \delta g_{11} \; = \;
- 2 \; l_1^2 \, \partial_1 \chi^1 +
( l_1^2 + l_2^2 - l_3^2 ) \, \partial_1 \chi^2
\nonumber \\ 
\delta l_2^2 & = & \delta g_{22} \; = \;
- 2 \; l_2^2 \, \partial_2 \chi^2 +
( l_1^2 + l_2^2 - l_3^2 ) \, \partial_2 \chi^1
\nonumber \\ 
\delta l_3^2 & = & \delta g_{11} +
\delta g_{22} + 2 \delta g_{12} 
\nonumber \\ 
& = &(-l_1^2 + l_2^2 - l_3^2 ) \, ( \partial_1 \chi^1 + \partial_2 \chi^1 ) +
( l_1^2 - l_2^2 - l_3^2 ) \, ( \partial_1 \chi^2 + \partial_2 \chi^2 )
\nonumber \\
\eea
After introducing appropriate finite differences for the fields $\chi^{\mu}$
one then has
\bea
\delta l_1^2 & = &
- 2 \; l_1^2 \, ( \chi_1^1 - \chi_0^1 ) +
( l_1^2 + l_2^2 - l_3^2 ) \, ( \chi_1^2 - \chi_0^2 )
\nonumber \\ 
\delta l_2^2 & = &
- 2 \; l_2^2 \, ( \chi_2^2 - \chi_0^2 ) +
( l_1^2 + l_2^2 - l_3^2 ) \, ( \chi_2^1 - \chi_0^1 )
\nonumber \\ 
\delta l_3^2 & = &
(-l_1^2 + l_2^2 - l_3^2 ) \, ( \chi_3^1 - \chi_0^1 ) +
( l_1^2 - l_2^2 - l_3^2 ) \, ( \chi_3^2 - \chi_0^2 )
\nonumber \\
\label{eq:gauge_guess}
\eea
Here upper indices on $\chi$ label the components, while lower
indices indicate the position.
Taking $\vec \chi_1 = \vec \chi_2 = \vec \chi_3 = 0 $,
$\vec \chi_0 \neq 0 $, as well as $l_1 = l_2 =1 $, $l_3 = \sqrt{2} $
(as appropriate for the square lattice), one finally obtains the simple result
\bea
\delta l_1^2 & = & 2 \, \chi_0^1
\nonumber \\ 
\delta l_2^2 & = & 2 \, \chi_0^2
\nonumber \\ 
\delta l_3^2 & = & 2 \, ( \chi_0^1 + \chi_0^2 )
\nonumber \\ 
\eea
which is indeed equivalent to Eq.~(\ref{eq:gauge_wfe}).
This shows that the zero modes described in Eq.~(\ref{eq:gauge_wfe})
correspond to lattice diffeomorphisms. (The term {\it lattice coordinate
transformations} would appear to be equally suitable, provided
one identifies the directions associated with the edges with a preferred
coordinate system, and identifies changes in these coordinates as
corresponding to variations in the squared edge lengths which leave the local
curvatures unchanged).
The case of flat space is obviously the simplest. By moving the
location of the vertices around in flat space, one can find a different
assignment of edge lengths which represents the same flat geometry.
This leads to a $ d \cdot N_0$-parameter family of transformations for the
edge lengths in flat space, and to a set of equivalent metrics which
are all related by lattice diffeomorphisms, i.e. deformations of the
edge lengths which leave the local curvature invariants unchanged, and respect
the triangle inequalities.

In conclusion, the previous analysis shows a direct correspondence
between gauge transformations on the simplicial lattice
\beq
l_i^2 \; \longrightarrow \; l_i^2 \; + \; \delta l_i^2 (\chi) \;\; ,
\eeq
and the analogous diffeomorphisms in the continuum
\beq
g_{\mu\nu} (x) \; \longrightarrow \; g_{\mu\nu}(x) \;
+ \; \delta g_{\mu\nu} (\chi(x)) \;\; .
\eeq
These transformations, in which suitable deformations of the edge
lengths are shown to correspond to the local gauge transformations, should
be contrasted with the set of what can be called {\it trivial} coordinate
transformations.
For a given assignment of edge lengths, introduce an arbitrary coordinate
system, and a corresponding flat metric, within each triangle (or simplex
in higher dimensions).
Coordinate changes can then be performed within any triangle such that
\bea
{l'}_{01}^2 (n) & = & l_{01}^2 (n)
\nonumber \\ 
{l'}_{02}^2 (n) & = & l_{02}^2 (n)
\nonumber \\ 
{l'}_{12}^2 (n) & = & l_{12}^2 (n)
\nonumber \\ 
\eea
within each triangle. These diffeomorphisms
are trivial, in the sense that they correspond to a change in an
arbitrary coordinate system, which was {\it not} part of the
theory to begin with,
as Regge's lattice theory is formulated exclusively in terms of
coordinate-independent squared edge lengths,
and not piecewise flat continuum metrics, which are highly degenerate.

The confusion between the two types of invariance is, in our
opinion, at the root of the erroneous conclusions drawn in
Ref. ~\cite{jnm},
where it is argued that Regge gravity always needs a non-local gauge
fixing term, to compensate for the fact that in the functional
integral
for gravity the integration is over ``invariants'', the edge lengths.
As the edge lengths correspond to components of the metric
(see Eqs.~(\ref{eq:gij_simplex}), (\ref{eq:epstoh}),
(\ref{eq:gauge_wfe_o}), (\ref{eq:gauge_wfe})), this cannot be true.
The above discussion shows in detail that the situation
is more subtle, and that a non-local additional, and in our
opinion ad-hoc, gauge fixing term
will most likely lead to an incorrect weighting, as already pointed out 
in ~\cite{cargese}.

\vskip 10pt
\newsection{Arbitrary Curved Backgrounds}
\hspace*{\parindent}

The previous discussion dealt with the case of an expansion of the
gravitational action about flat space, or a regular tessellation of the
sphere, a manifold of constant curvature.
To complete our discussion, we now turn to the more complex task of
exhibiting explicitly the local gauge invariance of the simplicial
theory, for an arbitrary background simplicial complex.
To this end we write
\beq
l_i^2 \; = \; l_{0i}^2 + q_i + \delta l^2_i \;\; ,
\eeq
where $q_i$ describes an arbitrary but small
deviation from a regular lattice, and
$\delta l^2_i$ is a gauge fluctuation, whose form needs to be determined.
We shall keep terms $ O(q^2)$ and $O(q \; \delta l^2 )$, but neglect terms
$O(\delta l^4)$.

The squared volumes $V_n^2 (\sigma)$ of n-dimensional simplices $\sigma$
are given by homogeneous polynomials of order $ (l^2)^n $. 
In particular for the area of a triangle $A_\Delta$ with arbitrary edges
$l_1,l_2,l_3$ one has
\beq
\delta A_{\Delta}^2 \; = \;
\eigth ( -l_1^2 + l_2^2 + l_3^2 ) \, \delta l_1^2 +
\eigth ( l_1^2 - l_2^2 + l_3^2 ) \, \delta l_2^2 +
\eigth ( l_1^2 + l_2^2 - l_3^2 ) \, \delta l_3^2 \;\; ,
\eeq
and similarly for the other quantities in Eqs.~(\ref{eq:dihedral}),
(\ref{eq:voronoi}) and (\ref{eq:baricentric}) which are needed in 
order to construct the action.

For our notation in two dimensions we will refer to Figure 12.
The subsequent Figures 13 and 14 illustrate the difference
between a {\it gauge} deformation of the surface which leaves
the area and curvature at the point labeled by $0$ invariant,
and a {\it physical} deformation which corresponds to a
re-assignment of edge lengths meeting at the vertex $0$ such that it alters
the area and curvature at $0$. In the following we will
characterize unambiguously what we mean by the two different
operations.

\begin{center}
\leavevmode
\epsfysize=5cm
\epsfbox{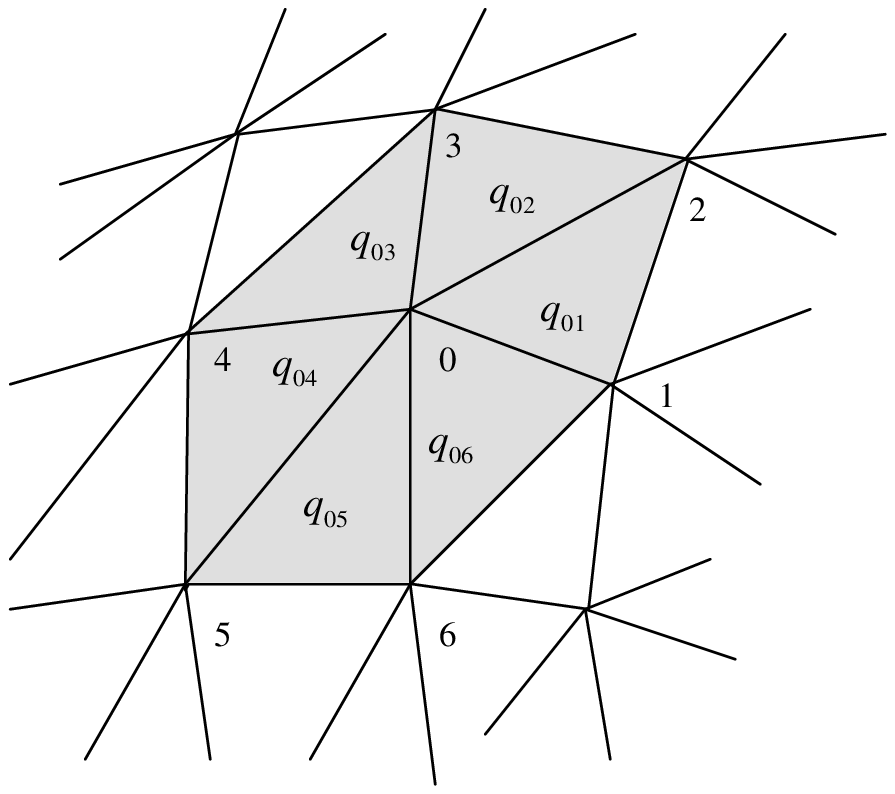}
\end{center}
\noindent
{\small{\it Fig.\ 12.
Notation for an arbitrary simplicial lattice, where the edge lengths
meeting at the vertex $0$ have
been deformed away from a regular lattice by a small amount $q_i$
(minimally deformed equilateral lattice).
\medskip}}

\begin{center}
\leavevmode
\epsfysize=5cm
\epsfbox{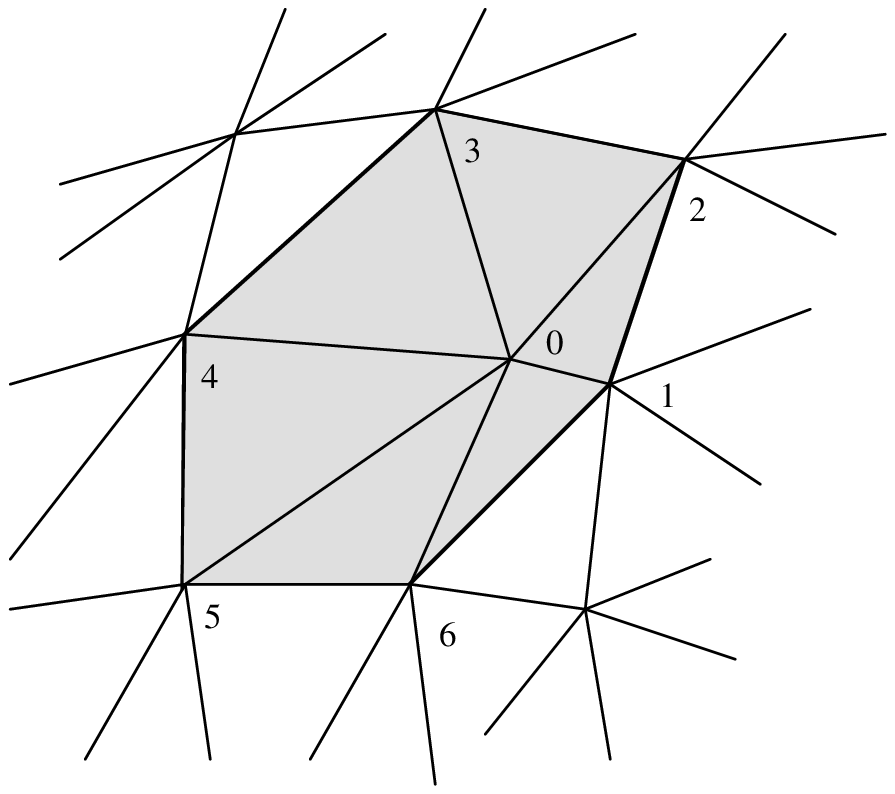}
\end{center}
\noindent
{\small{\it Fig.\ 13.
Local gauge deformations of the lattice act on the edge lengths
meeting at the vertex $0$, and are performed in such a way that the area and
curvature at the vertex $0$ are left unchanged.
\medskip}}

\begin{center}
\leavevmode
\epsfysize=5cm
\epsfbox{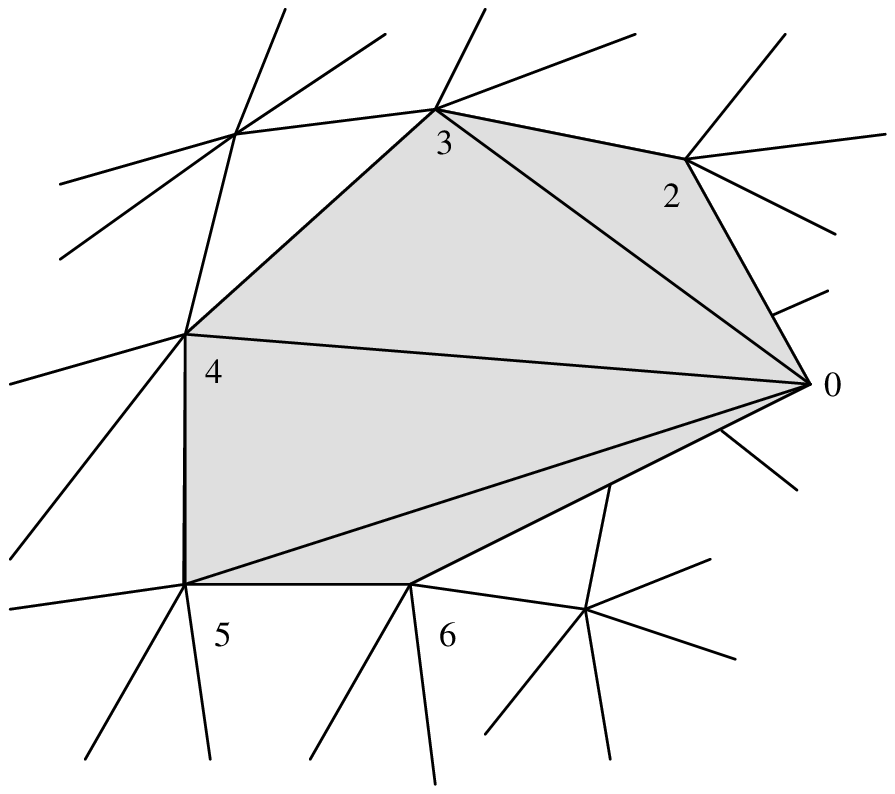}
\end{center}
\noindent
{\small{\it Fig.\ 14.
Physical deformations change the area and curvature at the vertex $0$,
thus changing the lattice geometry.
\medskip}}

\vskip 10pt
\subsection{Equilateral Lattices}
\hspace*{\parindent}

First consider the expansion about a deformed equilateral lattice, for
which $l_{0i} = 1$ to start with. A motivation for this choice is
provided by the fact that in the numerical studies of two-dimensional
gravity the averages of the squared edge lengths in the three
principal directions turn out to be equal,
$\langle l_1^2 \rangle = \langle l_2^2 \rangle = \langle l_3^2 \rangle $.
The baricentric area associated with vertex $0$ is then given by
\bea
A \, = \, A_0 (q) + {1 \over 2 \cdot 3^{5/2} } \Bigl [
& \delta l^2_{01} &
\left ( 3 + q_{06} - 4 q_{01} + q_{02} + q_{16} + q_{12} \right )
\nonumber \\
+ & \delta l^2_{02} &
\left ( 3 + q_{01} - 4 q_{02} + q_{03} + q_{12} + q_{23} \right )
\nonumber \\
+ & \delta l^2_{03} &
\left ( 3 + q_{02} - 4 q_{03} + q_{04} + q_{23} + q_{34} \right )
\nonumber \\
+ & \delta l^2_{04} &
\left ( 3 + q_{03} - 4 q_{04} + q_{05} + q_{34} + q_{45} \right )
\nonumber \\
+ & \delta l^2_{05} &
\left ( 3 + q_{04} - 4 q_{05} + q_{06} + q_{45} + q_{56} \right )
\nonumber \\
+ & \delta l^2_{06} &
\left ( 3 + q_{05} - 4 q_{06} + q_{01} + q_{56} + q_{16} \right )
\Bigr ]
\nonumber \\
+ && O(\delta l^4 ) \;\; .
\nonumber \\
\eea
Our normalization is such that $A_0 = { \sqrt{3} \over 2 } $ for $q_i=0$.
Equivalently one can write, in more compact notation, at the vertex $0$
\beq
A \, = \, A_0 (q) 
+ \third \vec v_A (q) \cdot \vec {\delta l^2} + O(\delta l^4 ) \;\; ,
\eeq
with $\vec {\delta l^2} = ( \delta l^2_{01}, \dots, \delta l^2_{06})$.
After adding the contributions from the neighboring vertices one 
obtains
\beq
\sum_{P_0 \dots P_6} A \, = \,
\sum_{P_0 \dots P_6} A_0 (q) +
\vec v_{A} (q) \cdot \vec {\delta l^2} + O(\delta l^4 ) \;\; .
\eeq
Therefore the area associated with the vertex $0$ will remain
unchanged provided the variations in the squared edge lengths meeting
at $0$ satisfy the constraint
\beq
\vec v_{A} (q) \cdot \vec {\delta l^2} \; = \; 0 \;\; .
\eeq
This is nothing but the curved space equivalent of the condition
of Eq.~(\ref{eq:gauge_v}), which for the flat equilateral lattice
takes the
form
\beq
\sum_{i=1}^6 \; \delta l^2_i (n) \; = \; 0 \;\; .
\eeq
Alternatively, if one considers a dual subdivision, one has to
consider the dual area associated with vertex $0$. In this case
one has
\bea
A \, = \, A_0 (q) + {1 \over 4 \cdot 3^{3/2} } \Bigl [
& \delta l^2_{01} &
\left ( 1 + 2 q_{06} - 4 q_{01} + 2 q_{02} \right )
\nonumber \\
+ & \delta l^2_{02} &
\left ( 1 + 2 q_{01} - 4 q_{02} + 2 q_{03} \right )
\nonumber \\
+ & \delta l^2_{03} &
\left ( 1 + 2 q_{02} - 4 q_{03} + 2 q_{04} \right )
\nonumber \\
+ & \delta l^2_{04} &
\left ( 1 + 2 q_{03} - 4 q_{04} + 2 q_{05} \right )
\nonumber \\
+ & \delta l^2_{05} &
\left ( 1 + 2 q_{04} - 4 q_{05} + 2 q_{06} \right )
\nonumber \\
+ & \delta l^2_{06} &
\left ( 1 + 2 q_{05} - 4 q_{06} + 2 q_{01} \right )
\Bigr ]
\nonumber \\
+ && O(\delta l^4 ) \;\; .
\nonumber \\
\eea
leading to a result formally similar (in fact in this
case identical) to the baricentric case.

A similar calculation can be done for the curvature associated
with vertex $0$. One has for the deficit angle at $0$
\bea
\delta \, = \, \delta_0 (q) + {1 \over 3^{3/2} } \, \Bigl [
& \delta l^2_{01} &
\left ( 3 - 2 q_{06} - q_{01} - 2 q_{02} + q_{16} + q_{12} \right )
\nonumber \\
+ & \delta l^2_{02} &
\left ( 3 - 2 q_{01} - q_{02} - 2 q_{03} + q_{12} + q_{23} \right )
\nonumber \\
+ & \cdots & \Bigr ] + O(\delta l^4 ) \;\; .
\nonumber \\
\eea
and therefore for the variation of the sum of the deficit angles
surrounding $0$
\beq
\Delta \left ( \sum_h \delta_h \right ) \; = \;
\sum_{P_0 \dots P_6} \Delta \delta \;\; ,
\eeq
and 
\beq
\sum_{P_0 \dots P_6} \delta \, = \, 
\sum_{P_0 \dots P_6} \delta_0 (q) +
\vec v_{R} (q) \cdot \vec {\delta l^2} + O(\delta l^4 ) \;\; ,
\eeq
with in this case, as expected, $ \vec v_{R} (q) \equiv 0 $.

Finally for the curvature squared associated with vertex $0$ one 
computes 
\bea
\delta^2 / A \, = \, \delta_0^2 / A_0 (q) + {4 \over 3^{3/2} } \, \Bigl [
& \delta l^2_{01} 
\left ( q_{01} + q_{02} + q_{03} + q_{04} + q_{05} + q_{06}
- q_{12} - q_{23} - q_{34} - q_{45} - q_{56} - q_{16} \right ) &
\nonumber \\
+ & \delta l^2_{02} 
\left ( q_{01} + q_{02} + q_{03} + q_{04} + q_{05} + q_{06}
- q_{12} - q_{23} - q_{34} - q_{45} - q_{56} - q_{16} \right ) &
\nonumber \\
+ & \cdots \; \Bigl ] \; + \; O(\delta l^4 ) \;\; .
\;\;\;\;\;\;\;\;\;\;\;\;\;\;\;\;\;\;\;\;\;\;\;\;\;\;\;\;\;\;\;\;\;\; &
\nonumber \\
\eea
Adding up all seven contributions one gets
\beq
\Delta \left ( \sum_h \delta_h^2 / A_h \right ) \; = \;
\sum_{P_0 \dots P_6} \Delta ( \delta^2 / A ) \;\; ,
\eeq 
and therefore
\beq
\sum_{P_0 \dots P_6} \delta^2 / A \, = \, 
\sum_{P_0 \dots P_6} \left ( \delta^2 / A \right )_0 +
\vec v_{R^2} (q) \cdot \vec {\delta l^2} + O(\delta l^4 ) \;\; .
\eeq
In this case the curvature squared associated with the vertex $0$
will remain unchanged, provided the variations in the squared edge
lengths meeting at $0$ satisfy the constraint
\beq
\vec v_{R^2} (q) \cdot \vec {\delta l^2} \; = \; 0 \;\; .
\eeq
which provides a second constraint on the edge length variations
$\vec {\delta l^2}$ at the vertex $0$.
Again this constraint is the generalization to curved space of the condition
of Eq.~(\ref{eq:gauge_r}), which was valid for flat space.

\vskip 10pt
\subsection{Square Lattice}
\hspace*{\parindent}

A similar calculation can be performed for the square lattice 
with $l_{01}=l_{02}=1$ and $l_{03}=\sqrt{2}$.
The baricentric area associated with vertex $0$ is now given by
\bea
A \, = \, A_0 (q) + {1 \over 24} \, \Bigl [
& \delta l^2_{01} &
\left ( 4 - 4 q_{01} + q_{02} + q_{16} \right )
\nonumber \\
+ & \delta l^2_{02} &
\left ( q_{01} - 2 q_{02} + q_{03} + q_{12} + q_{23} \right )
\nonumber \\
+ & \delta l^2_{03} &
\left ( 4 + q_{02} - 4 q_{03} + q_{34} \right )
\nonumber \\
+ & \delta l^2_{04} &
\left ( 4 - 4 q_{04} + q_{05} + q_{34} \right )
\nonumber \\
+ & \delta l^2_{05} &
\left ( q_{04} - 2 q_{05} + q_{06} + q_{45} + q_{56} \right )
\nonumber \\
+ & \delta l^2_{06} &
\left ( 4 + q_{05} - 4 q_{06} + q_{16} \right ) \, \Bigr ]
\nonumber \\
+ & O(\delta l^4 ) \;\; . & 
\nonumber \\
\eea
Our normalization here is such that $A_0=1$ for $q_i=0$.
Summing up all relevant contributions, one can write
\beq
\sum_{P_0 \dots P_6} A \, = \,
\sum_{P_0 \dots P_6} A_0 (q) +
\vec {v'}_{A} (q) \cdot \vec {\delta l^2} + O(\delta l^4 ) \;\; ,
\eeq
leading to the invariance constraint on the edge length 
variations
\beq
\vec {v'}_{A} (q) \cdot \vec {\delta l^2} \; = \; 0 \;\; .
\label{eq:area_invar}
\eeq
This provides a first constraint on the edge length variations
$\vec {\delta l^2}$ at the vertex $0$, for the deformed square
lattice.

For the dual (Voronoi) area associated with vertex $0$ one has instead
\bea
A \, = \, A_0 (q) + {1 \over 16} \, \Bigl [
& \delta l^2_{01} &
\left ( 2 - 3 q_{01} + 2 q_{02} + 2 q_{06} - q_{16} - q_{12} \right )
\nonumber \\
+ & \delta l^2_{02} &
\left (-2 + 2 q_{01} - 3 q_{02} + 2 q_{03} + q_{12} + q_{23} \right )
\nonumber \\
+ & \delta l^2_{03} &
\left ( 2 + 2 q_{02} - 3 q_{03} + 2 q_{04} - q_{23} - q_{34} \right )
\nonumber \\
+ & \delta l^2_{04} &
\left ( 2 + 2 q_{03} - 3 q_{04} + 2 q_{05} - q_{34} - q_{45} \right )
\nonumber \\
+ & \delta l^2_{05} &
\left (-2 + 2 q_{04} - 3 q_{05} + 2 q_{06} + q_{45} + q_{56} \right )
\nonumber \\
+ & \delta l^2_{06} &
\left ( 2 + 2 q_{01} + 2 q_{05} - 3 q_{06} - q_{56} - q_{16} \right )
\, \Bigr ] \nonumber \\
+ & O(\delta l^4 ) \;\; , &
\nonumber \\
\eea
leading to a constraint similar to the one for the baricentric area.
For the curvature associated with vertex $0$ one has
\bea
\delta \, = \, \delta_0 (q) + {1 \over 4} \, \Bigl [
& \delta l^2_{01} &
\left ( 2 - 2 q_{06} - q_{01} - q_{02} + q_{16} + q_{12} \right )
\nonumber \\
+ & \delta l^2_{02} &
\left ( 2 - q_{01} - q_{03} \right )
\nonumber \\
+ & \delta l^2_{03} &
\left ( 2 - q_{02} - q_{03} - 2 q_{04} + q_{23} + q_{34} \right )
\nonumber \\
+ & \delta l^2_{04} &
\left ( 2 - 2 q_{03} - q_{04} - q_{05} + q_{34} + q_{45} \right )
\nonumber \\
+ & \delta l^2_{05} &
\left ( 2 - q_{04} - q_{06} \right )
\nonumber \\
+ & \delta l^2_{06} &
\left ( 2 - q_{05} - q_{06} - 2 q_{01} + q_{56} + q_{16} \right ) \, \Bigr ]
\nonumber \\
+ & O(\delta l^4 ) \;\; , &
\nonumber \\
\eea
which gives now
\beq
\sum_{P_0 \dots P_6} \delta \, = \, 
\sum_{P_0 \dots P_6} \delta_0 (q) +
\vec {v'}_{R} (q) \cdot \vec {\delta l^2} + O(\delta l^4 ) \;\; ,
\eeq
with $ \vec {v'}_{R} (q) \equiv 0 $.

Finally, the curvature squared associated with just the vertex $0$ is given by
\bea
\delta^2 / A \, = \, \delta_0^2 / A_0 (q) + \half \, \Bigl [
& \delta l^2_{01} 
\left ( q_{01} + q_{02} + q_{03} + q_{04} + q_{05} + q_{06}
- q_{12} - q_{23} - q_{34} - q_{45} - q_{56} - q_{16} \right ) &
\nonumber \\
+ & \delta l^2_{02} 
\left ( q_{01} + q_{02} + q_{03} + q_{04} + q_{05} + q_{06}
- q_{12} - q_{23} - q_{34} - q_{45} - q_{56} - q_{16} \right ) &
\nonumber \\
+ & \cdots \; \Bigl ] \; + \; O(\delta l^4 ) \;\; .
\;\;\;\;\;\;\;\;\;\;\;\;\;\;\;\;\;\;\;\;\;\;\;\;\;\;\;\;\;\;\;\;\;\; &
\nonumber \\
\eea
The labeling in the previous formulae is a bit clumsy in dealing
with nearest next-nearest neighbor interactions; for the labeling in
the following formula we refer to Figure 15.

\begin{center}
\leavevmode
\epsfysize=6cm
\epsfbox{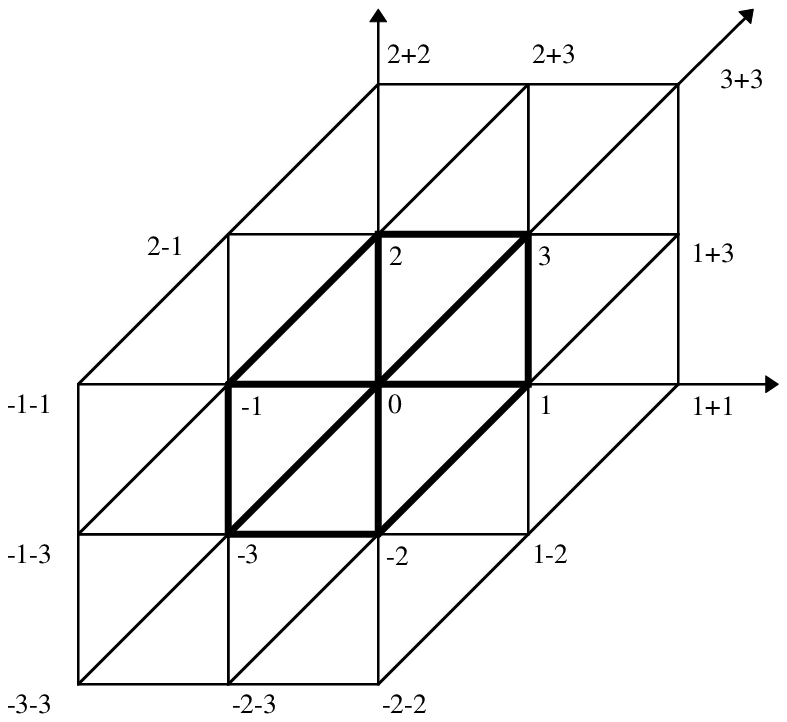}
\end{center}
\noindent
{\small{\it Fig.\ 15.
Labeling of lattice vertices for the expansion of the curvature squared
term around a minimally deformed square lattice.
\medskip}}

Adding up the contributions from the seven distinct vertices $0-6$
one obtains for the term linear in $ \delta l^2 $
\bea
\half \, \Bigl [
& \delta l^2_{0,1} ( 
& 4 q_{0,1} + 2 q_{0,2} + q_{0,-1} - q_{0,-2} - q_{0,3} + 2 q_{0,-3}
\nonumber \\
&& - q_{1,3} - 2 q_{2,3} - q_{-1,-3} - 2 q_{-2,-3} - q_{-1,2} - q_{1,-2}
\nonumber \\
&& + q_{1,1+1} + 2 q_{1,1-2} + 2 q_{1,1+3} - q_{1+1,1-2} - q_{1+1,1+3}
\nonumber \\
&& + q_{2,2+3} - 2 q_{-2,1-2} - q_{-2,-2-2} - q_{-2,-2-3} + q_{-2-2,-2+1}
+ q_{-2-2,-2-3}
\nonumber \\
&& - 2 q_{3,1+3} - q_{3,2+3} - q_{3,3+3} - q_{-3,-2-3} + q_{3+3,1+3}
+ q_{3+3,2+3} )
\nonumber \\
+ & \delta l^2_{0,3} ( 
& - q_{0,1} - q_{0,2} + 4 q_{0,3} + 2 q_{0,-1} + 2 q_{0,-2} + q_{0,-3}
\nonumber \\
&& - q_{1,3} - q_{2,3} - q_{-1,-3} - q_{-2,-3} - 2 q_{-1,2} - 2 q_{1,-2}
\nonumber \\
&& - q_{1,1+1} - q_{1,1-2} - 2 q_{1,1+3} + q_{-1,2-1} + q_{1+1,1-2}
+ q_{1+1,1+3}
\nonumber \\
&& - q_{2,2+2} - q_{2,2-1} - 2 q_{2,2+3} + q_{-2,1-2} + q_{2+2,2-1}
+ q_{2+2,2+3}
\nonumber \\
&& + 2 q_{3,1+3} + 2 q_{3,2+3} + q_{3,3+3} - q_{3+3,1+3} - q_{3+3,2+3} )
\nonumber \\
& + \cdots & \Bigr ] \; + \; O(\delta l^4 ) \;\; .
\nonumber \\
\eea
which can be written in short form as
\beq
\sum_{P_0 \dots P_6} \delta^2 / A (q) \, = \, 
\sum_{P_0 \dots P_6} \left ( \delta^2 / A \right )_0 +
\vec {v'}_{R^2} (q) \cdot \vec {\delta l^2} + O(\delta l^4 ) \;\; .
\eeq
Again the curvature squared associated with the vertex $0$
will remain unchanged provided the variations in the squared 
edge lengths meeting at $0$ satisfy the constraint
\beq
\vec {v'}_{R^2} (q) \cdot \vec {\delta l^2} \; = \; 0 \;\; .
\eeq
which provides the second constraint on the edge length variations
$\vec {\delta l^2}$ at the vertex $0$, for the deformed square
lattice (compare with Eq.~(\ref{eq:gauge_r})).

In conclusion, we have shown explicitly how gauge variations
of the edge lengths at each vertex can be defined by
requiring that the action contributions be locally invariant.
We have looked at small deformations, but large deformations
can be treated as well along the same lines, provided one is
careful not to violate the triangle inequalities, which impose
a non-perturbative cutoff in orbit space
The same approach can also be extended to higher dimensions, leading
to similar (but rather more complicated, when written out
explicitly!) results. The main conclusions do not change.

\vskip 10pt
\newsection{Scalar Field}
\hspace*{\parindent}

In the previous section we have discussed the invariance
properties of the lattice action for pure gravity.
Next a scalar field is introduced, as the simplest type of dynamical
matter that can be coupled to gravity.
The scalar lattice action in the continuum is
\beq
I [ g, \phi ] \; = \; \half \int d^2 x \, \sqrt g \, [ \,
g^{ \mu \nu } \, \partial_\mu \phi \, \partial_\nu \phi
+ ( m^2 + \xi R ) \phi^2 ] \;\; ,
\label{eq:scalar}
\eeq
The dimensionless coupling $\xi$ is arbitrary;
two special cases are the minimal ($\xi = 0$) and the conformal
($\xi = \sixth $) coupling case.
In the following we will mostly consider the case $\xi=0$.

\vskip 10pt
\subsection{Construction of the Lattice Action}
\hspace*{\parindent}

On the lattice consider a scalar $\phi_i$ and
define this field at the vertices of the simplices.
The corresponding lattice action can be obtained
through the usual procedure which replaces the original continuum metric
with the induced metric on the lattice, written in terms of the
edge lengths ~\cite{rowi,hw3d}. Here we shall consider only the
two-dimensional case; the generalization to higher dimensions is
straightforward.
It is convenient to use the notation of Figure 16, which will bring out
more readily the symmetries of the resulting lattice action.
Here coordinates will be picked in each triangle along the
(1,2) and (1,3) directions.

\begin{center}
\leavevmode
\epsfysize=5cm
\epsfbox{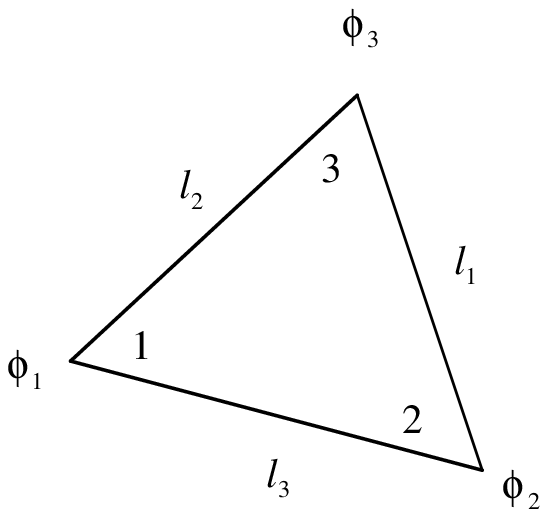}
\end{center}
\noindent
{\small{\it Fig.\ 16.
Labeling of edges and fields for the construction of the scalar
field action.
\medskip}}

To construct the scalar lattice action, one performs in two
dimensions the replacement
\beq
g_{\mu\nu} (x) \, \longrightarrow \, g_{ij} (\Delta) = \left( \begin{array}{cc}
l_{3}^2 & \half ( - l_1^2 + l_2^2 + l_{3}^2 ) \\
\half ( - l_1^2 + l_2^2 + l_{3}^2 ) & l_2^2 \\
\end{array} \right) \;\; ,
\eeq
which then gives
\beq
\det g_{\mu\nu} (x) \, \longrightarrow \, \det g_{ij} (\Delta) = 
\quarter \left \{ 
2 ( l_1^2 l_2^2 + l_2^2 l_{3}^2 + l_{3}^2 l_1^2 ) -
l_1^4 - l_2^4 - l_{3}^4 \right \} \, \equiv \, 4 A_{\Delta}^2 \;\; ,
\eeq
and also
\beq
g^{\mu\nu} (x) \, \longrightarrow \, g^{ij} (\Delta) = 
{ 1 \over \det g (\Delta) } \, \left( \begin{array}{cc}
l_2^2 & \half (l_1^2 - l_2^2 - l_{3}^2 ) \\
\half (l_1^2 - l_2^2 - l_{3}^2 ) & l_{3}^2 \\
\end{array} \right) \;\; .
\eeq
For the scalar field derivatives one writes ~\cite{bi,jns}
\beq
\partial_\mu \phi \, \partial_\nu \phi \, \longrightarrow \, 
\Delta_{i} \phi \Delta_{j} \phi = 
\left( \begin{array}{cc}
( \phi_2 - \phi_1 )^2 & ( \phi_2 - \phi_1 ) ( \phi_3 - \phi_1 ) \\
( \phi_2 - \phi_1 ) ( \phi_3 - \phi_1 ) & ( \phi_3 - \phi_1 )^2 \\
\end{array} \right) \;\; ,
\eeq
which corresponds to introducing finite lattice differences
defined in the usual way by
\beq
\partial_\mu \phi \, \longrightarrow \, ( \Delta_\mu \phi )_i \; = \;
\phi_{ i + \mu } - \phi_ i \;\; .
\eeq
Here the index $\mu$ labels the possible directions in which one can move from
a point in a given triangle.
The discrete scalar field action then takes the form
\beq
I_m (l^2, \phi ) \, = \, {\textstyle {1\over16} \displaystyle}
\sum_{\Delta} { 1 \over A_{\Delta} } \left [
l_1^2 ( \phi_2 - \phi_1 ) ( \phi_3 - \phi_1 ) +
l_2^2 ( \phi_3 - \phi_2 ) ( \phi_1 - \phi_2 ) +
l_3^2 ( \phi_1 - \phi_3 ) ( \phi_2 - \phi_3 ) \right ] \;\; .
\label{eq:jnac}
\eeq
Using the identity
\beq
( \phi_i - \phi_j ) ( \phi_i - \phi_k ) \; = \; \half \left [
( \phi_i - \phi_j )^2 + ( \phi_i - \phi_k )^2 -
( \phi_j - \phi_k )^2 \right ] \;\; ,
\eeq
one obtains after some re-arrangements the simpler expression ~\cite{bi}
\beq
I_m (l^2, \phi ) \; = \; \half \sum_{<ij>} A_{ij} \,
\Bigl ( { \phi_i - \phi_j \over l_{ij} } \Bigr )^2 \;\; ,
\label{eq:acdual} 
\eeq
where $A_{ij}$ is the dual (Voronoi) area associated with the edge $ij$.
In terms of the edge length $l_{ij}$ and the dual edge length $h_{ij}$,
connecting neighboring vertices in the dual lattice,
one has $A_{ij} = \half h_{ij} l_{ij}$ (see Figure 17).
Other choices for the lattice subdivision will lead to a similar
formula for the lattice action, but with the Voronoi dual volumes replaced
by their appropriate counterparts in the new lattice.

\begin{center}
\leavevmode
\epsfysize=6cm
\epsfbox{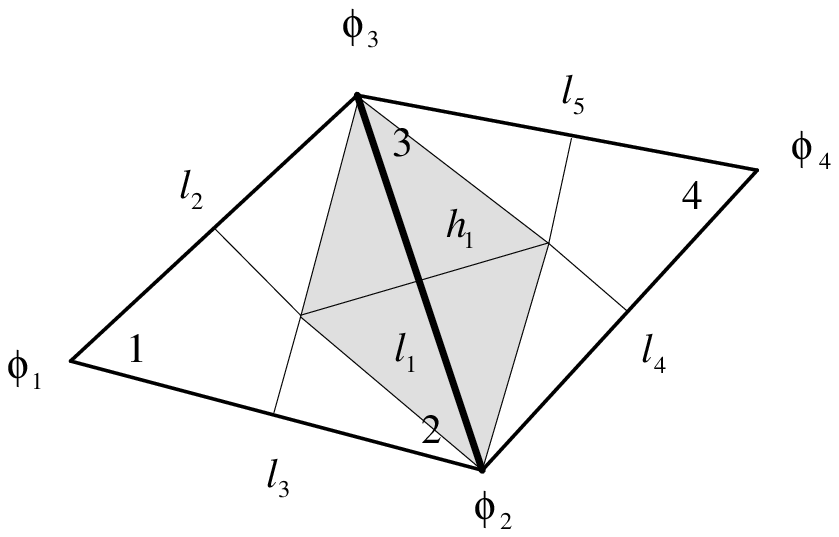}
\end{center}
\noindent
{\small{\it Fig.\ 17.
Dual area associated with the edge $l_1$ (shaded area),
and the corresponding dual link $h_1$.
% \label{fig:dualed1}
\medskip}}

For the edge of length $l_1$ the dihedral dual volume contribution is given by
\beq
A_{l_1} =
{ l_1^2 ( l_2^2 + l_{3}^2 - l_1^2 ) \over 16 A_{123} } +
{ l_1^2 ( l_{4}^2 + l_{5}^2 - l_1^2 ) \over 16 A_{234} }
\; = \; \half \, l_1 h_1 \;\; ,
\label{eq:dual_vol_edge}
\eeq
with $h_1$ is the length of the edge dual to $l_1$.
The baricentric dihedral volume for the same edge would be simply
\beq
A_{l_1} = ( A_{123} + A_{234} ) / 3 \;\; .
\eeq
It is well known that one of the disadvantages of the Voronoi construction
is the lack of positivity of the dual volumes, as already pointed out
in ~\cite{hw84}.
Thus some of the weights appearing in Eq.~(\ref{eq:acdual}) can be negative
for such an action. On the other hand, for the baricentric subdivision this
problem does not arise, as the areas $A_{ij}$ are always positive due
to the enforcement of the triangle inequalities.
It is immediate to generalize the action of Eq.~(\ref{eq:acdual}) to
higher dimensions, with the two-dimensional Voronoi volumes
replaced by their higher dimensional analogs.

Mass and curvature terms can be added to the action,
so that the total scalar action contribution becomes
\beq
I_m (l^2, \phi ) \; = \; \half \sum_{<ij>} A_{ij} \,
\Bigl ( { \phi_i - \phi_j \over l_{ij} } \Bigr )^2 \, +
\half \sum_{i} A_{i} \, (m^2 + \xi R_i ) \, \phi_i^2 \;\; .
\label{eq:acp} 
\eeq
The term containing the discrete analog of the scalar curvature involves
the quantity
\beq
A_{i} R_i \equiv \sum_{ h \supset i } \delta_h \; \sim \; \sqrt{g} \, R \;\; .
\eeq
In the above expression for the scalar action,
$A_{ij}$ is the area associated with the edge $l_{ij}$,
while $A_i$ is associated with the site $i$.
Again there is more than one way to define the volume element $A_i$,
~\cite{hw84}, but under reasonable assumptions, such as positivity,
one expects to get equivalent results in the lattice continuum limit.
In the following we shall mainly consider the simplest form for the scalar
action, with $m^2=\xi=0$.

One of the simplest problems which can be studied analytically in the
continuum as well as on the lattice
is the analysis of small fluctuations about some classical background solution.
In the continuum, the weak field expansion is often performed by expanding
the metric and the action about flat Euclidean space
\beq
g_{\mu\nu} \; = \; \delta_{\mu\nu} \; + \; \kappa \; h_{\mu\nu} \;\; .
\eeq
In four dimensions $\kappa=\sqrt{32 \pi G}$, which shows that the weak
field expansion there corresponds to an expansion in powers of $G$.
In two dimensions this is no longer the case and the relation between
$\kappa$ and $G$ is lost; instead one should regard $\kappa$ as
a dimensionless expansion parameter which is eventually set to one,
$\kappa = 1 $, at the end of the calculation. The procedure will be
sensible as long as wildly fluctuating geometries are not important
in two dimensions (on the lattice or in the continuum).
The influence of the latter configurations 
can only be studied by numerical simulations of the full
path integral ~\cite{hw2d,gh}.

In the continuum, the Feynman rules are obtained by expanding out the
action in the weak fields $ h_{\mu\nu} (x) $,
\beq
\half \int d^2 x \,
\sqrt{g} g^{\mu\nu} \, \partial_\mu \phi \, \partial_\nu \phi
\, = \, \half \int d^2 x \, ( \partial \phi )^2 +
\half \int d^2 x \, h_{\mu\nu} \left \{ 
\half \delta_{\mu\nu} ( \partial \phi )^2 -
\partial_\mu \phi \, \partial_\nu \phi \right \} + O (h^2) \;\; ,
\label{eq:wfesca}
\eeq
and by then transforming the resulting expressions to momentum space.

On the lattice the action is again expanded in the small fluctuation
fields $\epsilon_i$, which depend on the specific choice of parameterization
for the flat background lattice - a convenient starting point is (in
two dimensions) the square lattice with diagonals.
It is convenient to define the edge variables
at the midpoints of the links ~\cite{fey}.
For the edge lengths one then defines the lattice Fourier transforms as
\bea
\epsilon_1 (n) & = &
\int_{- \pi}^{\pi} \int_{- \pi}^{\pi} { d^2 k \over (2 \pi)^2 } \;
e^{ - i k \cdot n - i k_1 /2 } \; \epsilon_1 (k)
\nonumber \\
\epsilon_2 (n) & = &
\int_{- \pi}^{\pi} \int_{- \pi}^{\pi} { d^2 k \over (2 \pi)^2 } \;
e^{ - i k \cdot n - i k_2 /2 } \; \epsilon_2 (k)
\nonumber \\
\epsilon_3 (n) & = &
\int_{- \pi}^{\pi} \int_{- \pi}^{\pi} { d^2 k \over (2 \pi)^2 } \;
e^{ - i k \cdot n - i k_1 /2 - i k_2 /2 } \; \epsilon_3 (k) \;\; ,
\nonumber \\
\eea
while the scalar fields are still defined on the vertices, and
are Fourier transformed in the usual way, namely
\beq
\phi (n) = \int_{- \pi}^{\pi} \int_{- \pi}^{\pi} { d^2 p \over (2 \pi)^2 } \;
e^{ - i p \cdot n } \; \phi (p) \;\; .
\eeq
These formulae are completely analogous to the ones used in developing
the perturbative expansion for lattice gauge theories ~\cite{kns}.
They are easy to generalize to higher dimensions when a simplicial
subdivision of a hypercubic lattice is employed, as first
suggested in ~\cite{rowi}. Following this procedure,
the Feynman rules for the lattice scalar field
action were derived in ~\cite{fey}
(using the Voronoi dual volumes), and shown to agree
completely with the continuum Feynman rules for small momenta.
For more details, and the computation of the Feynman diagrams
relevant to the conformal anomaly,
we refer the interested reader to the cited work.

\vskip 10pt
\subsection{Equilateral Lattice}
\hspace*{\parindent}

In order to compare and analyze the difference between the two
volume discretizations (baricentric vs. dual)
for the scalar action, it will be useful
to look at their form for small deformations of the edges about
a regular lattice, such as an equilateral or a square one.
One would expect that the precise form of the discretization
should not matter, as long as the correct long distance
(small momentum) properties are preserved. Let us see how
this can come about.

The next step is therefore the expansion of the lattice scalar field action
of Eq.~(\ref{eq:acdual}) with volumes $A_{ij}$ defined via for example
a {\it baricentric} subdivision, starting from an equilateral
lattice with $l_i^2 = l_i^{0\,2} + \delta l^2_i $, and $l_i^0=1$.
It will be sufficient to limit oneself to the contributions coming
from one site (0) and its six neighbors (1-6), which for the kinetic
term is given by
\bea
( \phi_1 - \phi_0 )^2 \; \Bigl [ & {1 \over 4 \sqrt{3} } + 
{1 \over 24 \sqrt{3} } & ( 
\delta l^2_{06} -4 \delta l^2_{01} + \delta l^2_{02} + \delta l^2_{16}
+ \delta l^2_{12}) + O(\delta l^4 ) \, \Bigr ]
\nonumber \\
+ \, ( \phi_2 - \phi_0 )^2 \; \Bigl [ & {1 \over 4 \sqrt{3} } + 
{1 \over 24 \sqrt{3} } & ( 
\delta l^2_{01} -4 \delta l^2_{02} + \delta l^2_{03} + \delta l^2_{12}
+ \delta l^2_{23}) + O(\delta l^4 ) \, \Bigr ]
\nonumber \\
+ \, ( \phi_3 - \phi_0 )^2 \; \Bigl [ & {1 \over 4 \sqrt{3} } + 
{1 \over 24 \sqrt{3} } & ( 
\delta l^2_{02} -4 \delta l^2_{03} + \delta l^2_{04} + \delta l^2_{23}
+ \delta l^2_{34}) + O(\delta l^4 ) \, \Bigr ]
\nonumber \\
+ \, ( \phi_4 - \phi_0 )^2 \; \Bigl [ & {1 \over 4 \sqrt{3} } + 
{1 \over 24 \sqrt{3} } & ( 
\delta l^2_{03} -4 \delta l^2_{04} + \delta l^2_{05} + \delta l^2_{34}
+ \delta l^2_{45}) + O(\delta l^4 ) \, \Bigr ]
\nonumber \\
+ \, ( \phi_5 - \phi_0 )^2 \; \Bigl [ & {1 \over 4 \sqrt{3} } + 
{1 \over 24 \sqrt{3} } & ( 
\delta l^2_{04} -4 \delta l^2_{05} + \delta l^2_{06} + \delta l^2_{45}
+ \delta l^2_{56}) + O(\delta l^4 ) \, \Bigr ]
\nonumber \\
+ \, ( \phi_6 - \phi_0 )^2 \; \Bigl [ & {1 \over 4 \sqrt{3} } + 
{1 \over 24 \sqrt{3} } & ( 
\delta l^2_{05} -4 \delta l^2_{06} + \delta l^2_{01} + \delta l^2_{56}
+ \delta l^2_{16}) + O(\delta l^4 ) \, \Bigr ] \;\; .
\nonumber \\
\eea
For our notation and labeling of the edge lengths we refer
again to Figure 12.
In the case of the {\it dual} (Voronoi) subdivision, the results is
identical to this
order except for the replacement of the coefficient
$ {1 \over 24 \sqrt{3} } \rightarrow {1 \over 12 \sqrt{3} } $, which
can be interpreted as a rescaling of the
gravitational coupling between the metric field and the scalar.
For fields that are smoothly varying on the scale of the cutoff,
the scalar action contribution is
invariant under a gauge transformation on the edges acting
at the origin $0$, if the defining
condition for gauge transformations, analogous to
Eq.~(\ref{eq:gauge_r}) pertaining to the square lattice
in the weak field expansion, is satisfied
\beq
\sum_{i=1}^6 \; \delta l^2_i (n) \; \approx \; 0 \;\; .
\eeq
This result is also very similar to what happens in one dimensions,
where the exact invariance of the scalar action can be written down
explicitly with more ease ~\cite{oned}.

For the same (equilateral) lattice let us consider next the mass term;
it is given by
\beq
\half \, m^2 \, \int d^2 x \, \sqrt g \, \phi^2 
\; \sim \; \half \, m^2 \, \sum_{i} A_{i} \, \phi_i^2 \;\; .
\label{eq:acpm} 
\eeq
Again expanding about the equilateral lattice with
$l_i^2 = l_i^{0\,2} + \delta l^2_i $ one obtains for the baricentric
subdivision 
\beq
\half \; m^2 \; \phi_0^2 \; \Bigl [ \, { \sqrt{3} \over 2 } + {1 \over 12 \sqrt{3} } ( 
2 \delta l^2_{01} + 2 \delta l^2_{02} + \cdots +
\delta l^2_{12} + \delta l^2_{23} + \cdots ) +
O(\delta l^4) \Bigl ] \;\; ,
\eeq
while in the dual case one has.
\beq
\half \; m^2 \; \phi_0^2 \; \Bigl [ \, { \sqrt{3} \over 2 } + {1 \over 12 \sqrt{3} } ( 
\delta l^2_{01} + \delta l^2_{02} + \cdots +
2 \delta l^2_{12} + 2 \delta l^2_{23} + \cdots ) +
O(\delta l^4) \Bigl ] \;\; .
\eeq
Again to this order
the mass term is invariant under a gauge transformation on the edges acting
at the origin $0$, if the defining
condition for gauge transformations (compare to Eq.~(\ref{eq:gauge_v})
obtained in the weak field expansion) is satisfied
\beq
\sum_{i=1}^6 \; \delta l^2_i (n) \; = \; 0 \;\; .
\eeq
Finally one can consider the curvature term,
\beq
\half \; \xi \; \int d^2 x \, \sqrt g \, R \; \phi^2 
\; \sim \;
\half \; \xi \; \sum_{i} A_{i} \, ( 2 \delta_i / A_i ) \, \phi_i^2 \;\; .
\label{eq:acpr} 
\eeq
Its expansion is given by
\beq
\half \; \xi \; \phi_0^2 \; \Bigl [ { 2 \over \sqrt{3} } ( 
\delta l^2_{01} + \delta l^2_{02} + \cdots - \delta l^2_{12}
- \delta l^2_{23} - \cdots ) + O(\delta l^4) \Bigl ] \;\; ,
\eeq
with baricentric and dual forms identical to all orders in $\delta
l^2$, since $A_i$ does not appear in this term.
Again it is obvious that this term is invariant under gauge
transformations at $0$.

Further higher order terms involving the curvature, such as
\beq
\int d^2 x \, \sqrt g \, R \, 
g^{\mu\nu} \, \partial_\mu \phi \, \partial_\nu \phi
\; \sim \; \sum_{i} 2 { \delta_i \over A_{i} } \,
\sum_{ j \supset i } { A_{ij} \over l_{ij}^2 } \, ( \phi_i - \phi_j )^2 \;\; ,
\label{eq:acpkr} 
\eeq
become increasingly complicated in their structure, as they involve
neighbors which are further apart. They are strongly suppressed
for smooth manifolds due to the presence of the deficit angle.
In the baricentric case one finds
\bea
& ( \phi_1 - \phi_0 )^2 \; {1 \over 3 \sqrt{3} } \; ( 
& 2 \delta l^2_{01} + 0 \cdot \delta l^2_{02} + \delta l^2_{03} +
\delta l^2_{04} + \delta l^2_{05} + 0 \cdot \delta l^2_{06} + \cdots
\nonumber \\
& & - \, 0 \cdot \delta l^2_{12} - \delta l^2_{23} - \delta l^2_{34} -
\delta l^2_{45} - 0 \cdot \delta l^2_{16} + \cdots ) + O(\delta l^4 ) \;\; ,
\nonumber \\
\eea
and we have omitted the terms involving
$( \phi_2 - \phi_0 )^2 $ etc. since they can be obtained by symmetry.
Again for slowly varying scalar fields the above term is invariant
under gauge transformations at the origin.

\vskip 10pt
\subsection{Square Lattice}
\hspace*{\parindent}

A similar calculation can be performed for the square background
lattice with again $l_i^2 = l_i^{0\,2} + \delta l^2_i $, and
$l_{01}=l_{02}=1$ and $l_{03}=\sqrt{2}$.
Using the baricentric subdivision, the kinetic term gives
\bea
( \phi_1 - \phi_0 )^2 \; \Bigl [ & {1 \over 6 } + 
{1 \over 24 } & ( 
\delta l^2_{06} - 2 \delta l^2_{01} + \delta l^2_{12})
+ O(\delta l^4 ) \, \Bigr ]
\nonumber \\
+ \, ( \phi_2 - \phi_0 )^2 \; \Bigl [ & {1 \over 12 } + 
{1 \over 48 } & ( 
\delta l^2_{01} - 2 \delta l^2_{02} + \delta l^2_{03} +
\delta l^2_{12} + \delta l^2_{23})
+ O(\delta l^4 ) \, \Bigr ]
\nonumber \\
+ \, ( \phi_3 - \phi_0 )^2 \; \Bigl [ & {1 \over 6 } + 
{1 \over 24 } & ( 
- 2 \delta l^2_{03} + \delta l^2_{04} + \delta l^2_{23})
+ O(\delta l^4 ) \, \Bigr ]
\nonumber \\
+ \, ( \phi_4 - \phi_0 )^2 \; \Bigl [ & {1 \over 6 } + 
{1 \over 24 } & ( 
\delta l^2_{03} -2 \delta l^2_{04} + \delta l^2_{45})
+ O(\delta l^4 ) \, \Bigr ]
\nonumber \\
+ \, ( \phi_5 - \phi_0 )^2 \; \Bigl [ & {1 \over 12 } + 
{1 \over 48 } & ( 
\delta l^2_{04} - 2 \delta l^2_{05} + \delta l^2_{06} +
\delta l^2_{45} +  \delta l^2_{56})
+ O(\delta l^4 ) \, \Bigr ]
\nonumber \\
+ \, ( \phi_6 - \phi_0 )^2 \; \Bigl [ & {1 \over 6 } + 
{1 \over 24 } & ( 
- 2 \delta l^2_{06} + \delta l^2_{01} + \delta l^2_{56})
+ O(\delta l^4 ) \, \Bigr ] \;\; .
\nonumber \\
\eea
Again for smoothly varying scalar fields, the above action will be
invariant under gauge variations of the edge lengths at the vertex $0$,
provided one has
$\delta l^2_{01} + 2 \delta l^2_{02} + \delta l^2_{03} +
\delta l^2_{04} + 2 \delta l^2_{05} + \delta l^2_{06} =0 $
(which is the scalar field action analog of Eq.~(\ref{eq:area_invar})).

In the case of the dual subdivision, one obtains for the same lattice
\bea
( \phi_1 - \phi_0 )^2 \; \Bigl [ & {1 \over 4 } + 
{1 \over 16 } & ( 
- 4 \delta l^2_{01} + \delta l^2_{02} + \delta l^2_{16})
+ O(\delta l^4 ) \, \Bigr ]
\nonumber \\
+ \, ( \phi_2 - \phi_0 )^2 \; \Bigl [ & { 0 } + 
{1 \over 16 } & ( 
\delta l^2_{01} - 2 \delta l^2_{02} + \delta l^2_{03} +
\delta l^2_{12} + \delta l^2_{23})
+ O(\delta l^4 ) \, \Bigr ]
\nonumber \\
+ \, ( \phi_3 - \phi_0 )^2 \; \Bigl [ & {1 \over 4 } + 
{1 \over 16 } & ( 
- 4 \delta l^2_{03} + \delta l^2_{02} + \delta l^2_{34})
+ O(\delta l^4 ) \, \Bigr ]
\nonumber \\
+ \, ( \phi_4 - \phi_0 )^2 \; \Bigl [ & {1 \over 4 } + 
{1 \over 16 } & ( 
- 4 \delta l^2_{04} + \delta l^2_{05} + \delta l^2_{34})
+ O(\delta l^4 ) \, \Bigr ]
\nonumber \\
+ \, ( \phi_5 - \phi_0 )^2 \; \Bigl [ & { 0 } + 
{1 \over 16 } & ( 
\delta l^2_{04} - 2 \delta l^2_{05} + \delta l^2_{06} +
\delta l^2_{45} + \delta l^2_{56})
+ O(\delta l^4 ) \, \Bigr ]
\nonumber \\
+ \, ( \phi_6 - \phi_0 )^2 \; \Bigl [ & {1 \over 4 } + 
{1 \over 16 } & ( 
- 4 \delta l^2_{06} + \delta l^2_{05} + \delta l^2_{16})
+ O(\delta l^4 ) \, \Bigr ] \;\; ,
\nonumber \\
\eea
Due to the asymmetry of the lattice in this case, the difference
between the two discretizations is quite marked. We will argue
below that the dual volume form represents in fact an ``improved''
discretization over the baricentric form.

For the mass term one proceeds in the same way, and for the
baricentric subdivision one obtains
\beq
\half \; m^2 \; \phi_0^2 \; \Bigl [ 1 + {1 \over 12 } ( 
2 \delta l^2_{01} + 0 \cdot \delta l^2_{02} + 2 \delta l^2_{03} + \cdots +
\delta l^2_{12} + \delta l^2_{23} + 0 \cdot l^2_{34} + \cdots ) +
O(\delta l^4) \Bigr ] \;\; ,
\eeq
Again, this term is invariant under a gauge transformation
if Eq.~(\ref{eq:area_invar}) is satisfied.
For the dual (Voronoi) case one finds similarly
\beq
\half \; m^2 \; \phi_0^2 \; \Bigl [ 1 + {1 \over 8 } ( 
\delta l^2_{01} - \delta l^2_{02} + \delta l^2_{03} + \cdots +
\delta l^2_{12} + \delta l^2_{23} + \delta l^2_{34} + \cdots ) +
O(\delta l^4) \Bigr ] \;\; .
\eeq
which is invariant again if the area at $0$ is invariant.

Finally for the curvature term of Eq.~(\ref{eq:acpr}) one has
\bea
& \half \; \xi \; \phi_0^2 \; \Bigl [ &
\delta l^2_{01} + \delta l^2_{02} + \delta l^2_{03} + \delta l^2_{04} +
\delta l^2_{05} + \delta l^2_{06}
\nonumber \\
&& - \delta l^2_{12} - \delta l^2_{23} -
\delta l^2_{34} - \delta l^2_{45} - \delta l^2_{56} - \delta l^2_{16} +
O(\delta l^4) \Bigr ]
\nonumber \\
+ & \half \; \xi \; \phi_1^2 \; \Bigl [ &
\delta l^2_{01} - \delta l^2_{02} - \delta l^2_{06} +
\delta l^2_{12} + \delta l^2_{16} +
O(\delta l^4) \Bigl ] + \cdots \;\; ,
\eea
with the baricentric discretization equal to the dual discretization
to all orders in $\delta l^2$, since the local area $A_i$ does not appear.
Again it is obvious that this term is invariant under gauge
transformations at $0$.

Let us now return to the apparent discrepancy between the scalar
kinetic term in the baricentric and dual discretizations.
It is useful to look at the two discretized forms in momentum space.
Thus we assume that the edge length fluctuations
$\epsilon_i \equiv \delta l_i^2 / 2 \, l_i^{0\,2} $ and $\phi$
at the point $i$, $j$ steps in one coordinate direction and $k$ steps in the 
other coordinate direction from the origin, are related to the 
corresponding $\epsilon_i$ and $\phi$ at the origin by 
\beq
\epsilon_i^{(j+k)} \, = \, \omega_1^j \, \omega_2^k \, \epsilon_i^{(0)} \;\; ,
\eeq
where $\omega_i = e^{- i k_{i} }$, and $k_i$ is the momentum in the
direction $i$.
Similarly for $\phi$ one writes
\beq
\phi^{(j+k)} \, = \, {\omega '}_1^j \, {\omega '}_2^k \, \phi_i^{(0)} \;\; ,
\eeq
where ${\omega '}_i=e^{- i p_{i} }$.

After making the substitution $\epsilon_i \rightarrow h_{\mu\nu}$ 
of Eq.~(\ref{eq:epstoh}) (which is the same as the change of variables
in Eq.~(\ref{eq:change})) one obtains for the interaction in
momentum space, in the dual case
\bea
& & p_1^2 \; \phi^2 \;
\left [ \; \half + \quarter ( - h_{11} + h_{22} ) + O(h^2) \; \right ]
\nonumber \\
&+& p_2^2 \; \phi^2 \;
\left [ \; \half + \quarter ( + h_{11} - h_{22} ) + O(h^2) \; \right ]
\nonumber \\
&+& p_1 p_2 \; \phi^2 \;
\left [ \; - h_{12} + O(h^2) \; \right ] \;\; ,
\nonumber \\
\label{eq:wfedual}
\eea
which indeed coincides with the weak field expansion of the kinetic term
for the original scalar field action in the continuum, Eqs.~(\ref{eq:scalar})
and (\ref{eq:wfesca}).
In the baricentric case one obtains instead
\bea
& &p_1^2 \; \phi^2 \;
\left [ \; \half + \sixth (-h_{11}+h_{22}-h_{12})+O(h^2) \; \right ]
\nonumber \\
&+&p_2^2 \; \phi^2 \;
\left [ \; \half + \sixth (+h_{11}-h_{22}-h_{12})+O(h^2) \; \right ]
\nonumber \\
&+&p_1 p_2 \; \phi^2 \;
\left [ \; \third - \third h_{12} + O(h^2) \; \right ] \;\; ,
\nonumber \\
\label{eq:wfebari}
\eea
which at first looks quite different from the dual (Voronoi) volume case.
\begin{center}
\leavevmode
\epsfysize=7cm
\epsfbox{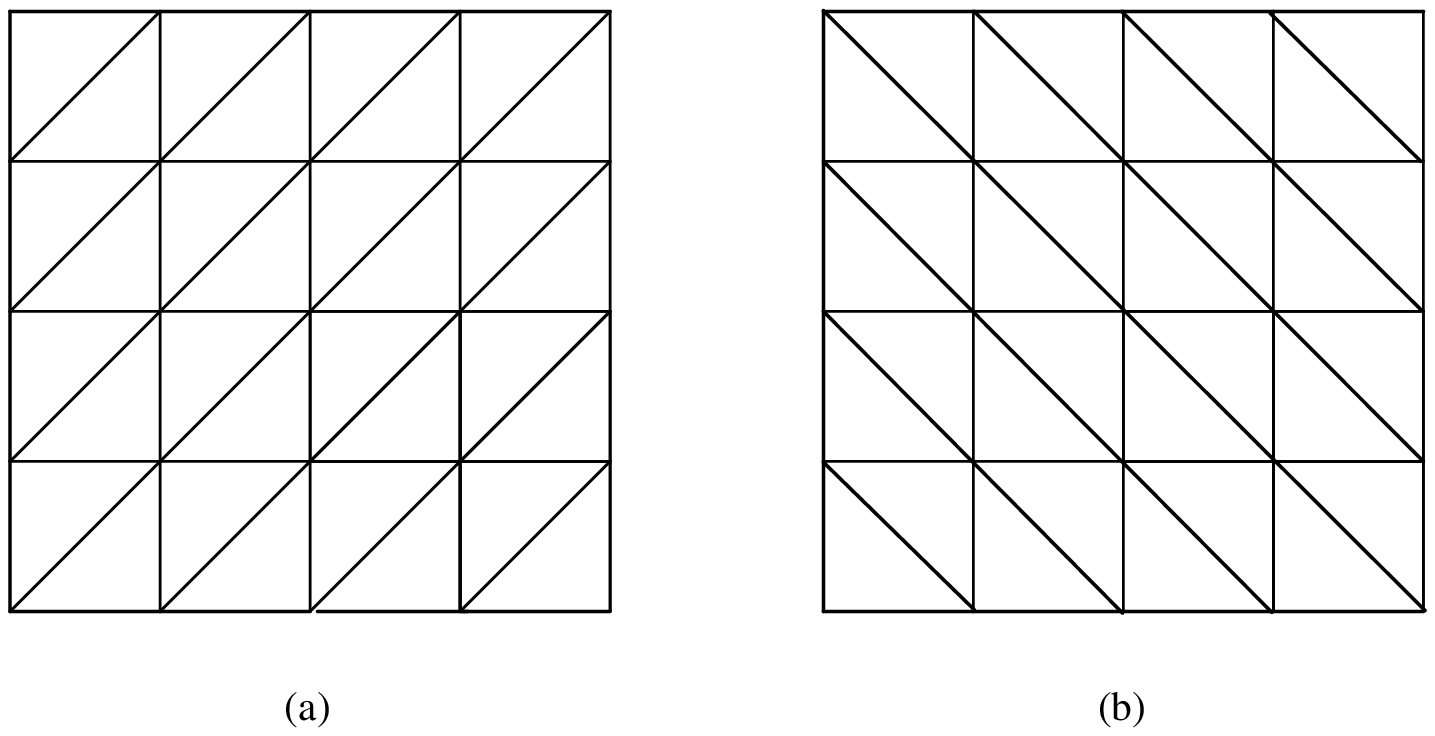}
\end{center}
\noindent
{\small{\it Fig.\ 18.
Two equivalent triangulation of flat space, based on different subdivisions
of the square lattice.
\medskip}}

On the other hand this result is not too surprising, as an
analogous situation occurs in flat space.
Consider for example the two lattices of Figure 18 (up to now we only
considered the one corresponding to $18a$), and compute in each
case the action for one momentum mode, which is just the inverse
scalar propagator in momentum space.
In the absence of any interaction terms along the lattice diagonals one has
\beq
2 \sum_{\mu} \; [ \; 1 - \cos p_\mu \; ] \; \sim \; p_1^2 + p_2^2 +
O(p^4) \;\; .
\eeq
The interaction terms along the lattice diagonals in Figure $18a$
contribute
\beq
2 [ \; 1 - \cos (p_1+p_2) \; ] \; \sim \; p_1^2 + p_2^2 + 2 p_1 p_2
+ O(p^4) \;\; ,
\eeq
while the interaction terms along the lattice diagonals in Figure $18b$
contribute
\beq
2 [ \; 1 - \cos (p_1-p_2) \; ] \; \sim \; p_1^2 + p_2^2 - 2 p_1 p_2 +
O(p^4) \;\; .
\eeq
Thus when one averages over the two equivalent contributions (since
the lattice is dynamical, and both contributions are equally probable)
one obtains
\beq
2 [ \; 1 - \cos p_1 \cos p_2 \; ] \; \sim \; p_1^2 + p_2^2 + O(p^4) \;\; ,
\eeq
which is now rotationally invariant to order $p^2$.
When the same procedure is applied to the lattice scalar action in the presence
of the gravitational field,
Eqs.~(\ref{eq:wfedual}) and (\ref{eq:wfebari}) can be shown to become
equivalent, after a rescaling of the gravitational coupling.
Still, the action based on dual Voronoi volumes
of Eq.~(\ref{eq:wfedual}) appears to lead to a more attractive
discretization, as the unwanted terms do not appear at all for the
choice of lattice of Figure $18a$, and no averaging over the
gravitational field
fluctuations is necessary to exhibit the correct correspondence with
the continuum action. How seriously one takes this class of problems depends
on how seriously one trusts low-order perturbation theory about a
fixed flat background as a tool for the study of fluctuating geometries.

\vskip 10pt
\subsection{Invariance Properties of the Scalar Action}
\hspace*{\parindent}

It is instructive to look at the invariance properties of the
scalar action under the continuous lattice 
gauge transformations defined in Eqs.~(\ref{eq:gauge_wfe}) and
(\ref{eq:gauge_guess}).
Physically, these local gauge transformations, which act on the vertices,
correspond to re-assignments of edge lengths which
leave the distance between two fixed points unchanged.
In the simplest case, only two neighboring edge lengths are changed, leaving
the total distance between the end points unchanged.
On physical grounds one would like to maintain such an invariance also
in the case of coupling to matter, just as is done in the continuum.

The scalar nature of the field requires that in the continuum under a 
change of coordinates $x \rightarrow x'$,
\beq
\phi' (x') = \phi (x) \;\; ,
\eeq
where $x$ and $x'$ refer to the same physical point in the
two coordinate systems.
Let us first look at the one-dimensional case, which is the simplest.
On the lattice, as discussed previously, gauge transformations
move the points around, and at the same vertex labeled by $n$ we expect
\beq
\phi_n \; \rightarrow \; \phi'_n \; \approx \; \phi_n + 
\left ( { \phi_{n+1} - \phi_n \over l_n } \right ) \epsilon_n \;\; ,
\eeq
One can determine the exact form of the change needed
in $\phi_n$ by requiring that the local variation of the scalar field action
\bea
& & { 1 \over l_{n-1} + \epsilon_{n} } \;
( \phi_{n} + \Delta \phi_n - \phi_{n-1} )^2 +
{ 1 \over l_{n} - \epsilon_{n} } \;
( \phi_{n+1} - \phi_n - \Delta \phi_n )^2 \nonumber \\
& & -
{ 1 \over l_{n-1} } \;
( \phi_{n} - \phi_{n-1} )^2 -
{ 1 \over l_{n} } \; ( \phi_{n+1} - \phi_n )^2 \; = \; 0
\eea
be identically zero.
Solving the resulting quadratic equation for $\Delta \phi_n$ one
obtains a rather unwieldy expression, which to lowest order is
given by ~\cite{oned}
\bea
\Delta \phi_n & = & { \epsilon_n \over 2 } \left [
{ \phi_{n} - \phi_{n-1} \over l_{n-1} } +
{ \phi_{n+1} - \phi_n \over l_{n} }
\right ] \nonumber \\
& & + { \epsilon_n^2 \over 8 } \left [ 
- { \phi_{n} - \phi_{n-1} \over l_{n-1}^2 } +
{ \phi_{n+1} - \phi_n \over l_{n}^2 } +
{ \phi_{n+1} - 2 \phi_n + \phi_{n-1} \over l_{n-1} l_{n} }
\right ] + O( \epsilon_n^3 ) \;\; ,
\label{eq:sgauge} 
\eea
and which is indeed of the expected form (as well as being
symmetric in the vertices $n-1$ and $n+1$).
For fields which are reasonably smooth, this correction
is suppressed if $ | \phi_{n+1} - \phi_n | / l_{n} \ll 1 $.
On the other hand it should be clear that
the functional measure $d \phi_n $ is no longer manifestly
invariant, due to the rather involved transformation property of the scalar
field.

A similar line of argument can be pursued in higher dimensions.
Thus in two dimensions one should require that locally
the variation of the action contribution be again zero for edge
length deformations $\delta l_{ij}$, associated with edges
meeting at the vertex $i$, and which
correspond to lattice gauge transformations
(in the weak field, for example, they have the explicit form given in
Eqs.~(\ref{eq:gauge_wfe_o}) to (\ref{eq:gauge_wfe})), namely
\beq
\sum_{ j } \; A_{ij} (l^2_{ij} + \delta l^2_{ij} ) \;
\left ( { \phi_i + \Delta \phi_i - \phi_j \over
\sqrt{ l^2_{ij} + \delta l^2_{ij} }  } \right )^2 \; - \;
\sum_{ j } \; A_{ij} (l^2_{ij} ) \;
\left ( { \phi_i - \phi_j \over
\sqrt{ l^2_{ij} }  } \right )^2 \; = \; 0 \;\; ,
\eeq
where $j$ labels the neighbors of the site $i$. The transformation 
law for $\phi_i$ is then determined by solving the above
equation for $\Delta \phi_i$, given an arbitrary 
{\it gauge} variation $\delta l^2 (\chi)$ at the vertex $i$.

\vskip 10pt
\subsection{Equations of Motion and Lattice Energy Momentum Tensor}
\hspace*{\parindent}

The equations of motion for $\phi_i$ are obtained from
\beq
{ \partial \over \partial \phi_k } \; \left \{ \sum_{<ij>} A_{ij} \,
\Bigl ( { \phi_i - \phi_j \over l_{ij} } \Bigr )^2 \, + \,
m^2 \sum_{i} A_{i} \, \phi_i^2 \right \} \; = \; 0 \;\; ,
\eeq
and read
\beq
\sum_{j(i)} \;
{ A_{ij} \over l_{ij} } \cdot { \phi_i - \phi_j \over l_{ij} } \, + \,
m^2 A_{i} \, \phi_i \; = \; 0 \;\; ,
\label{eq:phieq}
\eeq
where the notation $j(i)$ indicates that the site
$j$ is taken to be adjacent to $i$.
For the choice of indices in Figure 12 (see also Figure 17),
the equation with $i=0$ reads
\beq
- \; { 2 \over A_{01} + \dots + A_{06} } \; \left \{
{ A_{01} \over l_{01} } \cdot { \phi_1 - \phi_0 \over l_{01} } + \cdots +
{ A_{06} \over l_{06} } \cdot { \phi_6 - \phi_0 \over l_{06} } \right \}
\; + \; m^2 \phi_0 \; = \; 0 \;\; ,
\label{eq:eqmo_phi}
\eeq
and represents the discrete analog of
${ 1 \over \sqrt{g} } \, \partial_{\mu} g^{\mu\nu} g \partial_{\nu} \phi 
+ m^2 \phi = 0 $.
Eq.~(\ref{eq:phieq}) can be re-written, for $m^2=0$, as
\beq
\sum_{j(i)} \; \left ( { A_{ij} \over A_i \; d } \right )
\,{ \phi_i - \phi_j \over l_{ij}^2 }
\; \equiv \; 
\sum_{j(i)} \; P_{ij} \,{ \phi_i - \phi_j \over l_{ij}^2 } = 0 \;\; ,
\eeq
The weights $P_{ij} = A_{ij} / A_i d $ can be interpreted as
normalized hopping amplitudes, with a normalization condition
\beq
\sum_{j(i)} \; P_{ij} \; = \; 1 \;\; ,
\eeq
It is easy to verify that this last property does not depend
on the way the original lattice has been subdivided in order to
construct the dual lattice. Both the Voronoi construction 
of Eq.~(\ref{eq:voronoi}) and the baricentric one of
Eq.~(\ref{eq:baricentric}) lead to the same normalization for $P_{ij}$.
And indeed it is easy to show that this result holds in higher
dimensions as well.

Some very interesting properties regarding the spectrum of the Laplacian
on flat random lattices have been worked out in
\cite{gid,itzcar} and the reader is referred to these papers for
further details. Perhaps one of the most interesting results which
can be derived by the method of replicas in the weak disorder
limit is that, at least
in sufficiently low dimension, the spectrum of the Laplacian
coincides with the continuum result at low frequencies, with
subleading corrections which then reflect specific aspects
of the edge length distribution such as its Poissonian form
\cite{gid,oned}. In other words, quenched random lattices and regular
lattices give the same (continuum) result at low frequencies.

In the continuum the energy momentum tensor for matter described by
action $I_m$ is defined via the relationship 
\beq
\delta I_m \; = \; \half \; \int d^d x \,
\sqrt{g} \; T^{\mu\nu} \, \delta g_{\mu\nu}
\label{eq:emdef}
\eeq
For infinitesimal gauge variations, which have the form
\beq
\delta g_{\mu\nu} (x) \, = \, 
- g_{\mu\lambda} (x) \, \partial_\nu \chi^\lambda (x)
- g_{\lambda\nu} (x) \, \partial_\mu \chi^\lambda (x)
- \partial_{\lambda} g_{\mu\nu} (x) \, \chi^\lambda (x)
\; \equiv \; \Delta_{\chi} \; g_{\mu\nu} \;\; ,
\label{eq:varg}
\eeq
one expects $\delta I_m = 0$.
After an integration by parts in Eq.~(\ref{eq:emdef}), one then obtains
\beq
\left ( T^{\mu}_{\;\; \lambda} \right )_{;\mu} \; = \; 0 \;\; .
\eeq
The energy momentum tensor defined by equation 
Eq.~(\ref{eq:emdef}) will be conserved if and only if the matter action
is a scalar. More concretely, for one real scalar field of mass $m$, one has
\beq
T_{\mu\nu} \; = \; \partial_\mu \phi \; \partial_\nu \phi -
\half \, g_{\mu\nu} \, ( \partial_\lambda \phi \, \partial^\lambda \phi
+ m^2 \phi^2 ) \;\; .
\eeq
Taking the trace one obtains
\beq
T_{\;\;\mu}^\mu \; = \;
{\textstyle \left ( { 2 - d \over 2 } \right ) \displaystyle} \;
\partial_\mu \phi \; \partial^\mu \phi -
{\textstyle { d \over 2 } \displaystyle} \; m^2 \phi^2 \;\; .
\eeq
It is natural to proceed on the lattice in analogy with the continuum case.
For the scalar action contribution of Eq.~(\ref{eq:acdual}) one
computes the variation
\beq
\delta I_m \; = \; \sum_k \; \sum_{ij} \; ( \phi_i - \phi_j )^2 \;
{ \partial \over \partial l^2_k }
\left ( { A_{ij} \over 2 l_{ij}^2 } \right ) \; \delta l^2_k \;\; .
\eeq
\begin{center}
\leavevmode
\epsfysize=5cm
\epsfbox{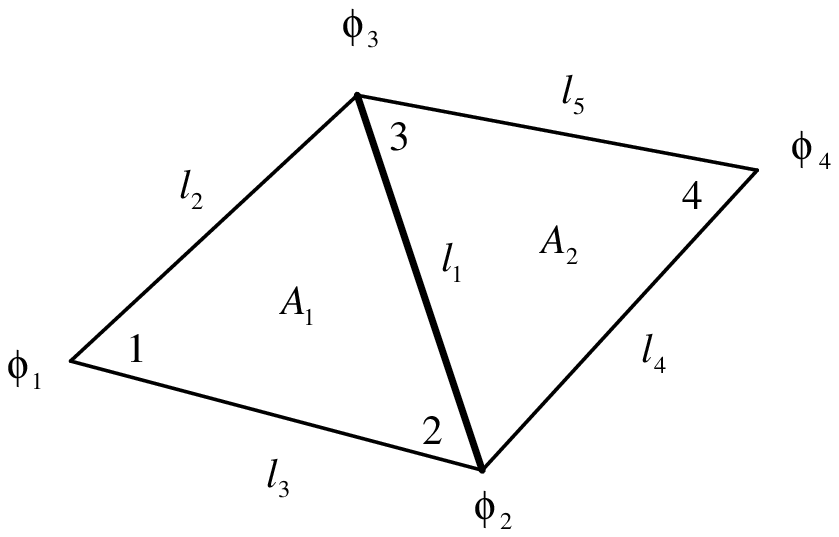}
\end{center}
\noindent
{\small{\it Fig.\ 19.
Notation for Eq.~(\ref{eq:variation}) describing the variation of the action.
\medskip}}

Since the derivative term inside the sum is non-zero only for edges
$ij$ which belong to triangles touching the edge $k$ (see Figure 19),
only four terms contribute to the sum over $<ij>$.
It is actually more convenient to start from the equivalent form
for the scalar action given in Eq.~(\ref{eq:jnac}),
and one obtains
\bea
\delta I_m \; = \; {1 \over 16} \; \Bigl [
& { 1 \over 16 A_1^3 } &
\left \{ l_1^2 ( l_2^2+l_3^2) - (l_2^2 - l_3^2 )^2 \right \}
\; ( \phi_2 - \phi_1 ) \; ( \phi_3 - \phi_1 )
\\ \nonumber
& + { 1 \over 16 A_2^3 } &
\left \{ l_1^2 ( l_4^2+l_5^2) - (l_4^2 - l_5^2 )^2 \right \}
\; ( \phi_2 - \phi_4 ) \; ( \phi_3 - \phi_4 ) \; \Bigr ] \;
\delta l^2_1 \; +  \; {1 \over 16} \; \left [ \; \dots \; \right ] \;
\delta l_2^2 + \cdots
\\ \nonumber
\label{eq:variation}
\eea
Using the definition for the dual volumes (see Eq.~(\ref{eq:voronoi}),
and Figure 20 for our notation here)
one can re-write the above expression more compactly as
\beq
\delta I_m \; = \; {1 \over 8} \; \Bigl [ \;
{ A_{11} \over A_1^2 } \; ( \phi_2 - \phi_1 ) \; ( \phi_3 - \phi_1 )
+ { A_{24} \over A_2^2 } \; ( \phi_2 - \phi_4 ) \; ( \phi_3 - \phi_4 )
\; \Bigr ] \;
\delta l^2_1 \; +  \; {1 \over 8} \; \left [ \; \dots \; \right ] \;
\delta l_2^2 \; + \; \cdots
\label{eq:vardual}
\eeq
Therefore one can introduce the quantities $T_k$ such that
\beq
\delta I_m \; = \; \half \sum_k \; T_k (l^2) \; \delta l^2_k \;\; ,
\eeq
with
\beq
T_1 \; = \; {1 \over 4} \; \Bigl \{ \;
{ A_{11} \over A_1^2 } \; ( \phi_2 - \phi_1 ) \; ( \phi_3 - \phi_1 )
+ { A_{24} \over A_2^2 } \; ( \phi_2 - \phi_4 ) \; ( \phi_3 - \phi_4)
\; \Bigr \} \;\; ,
\eeq
associated with the edge labeled by $1$ in Figure 19,
and similarly for all the other edges in the lattice.
It is clear that this equation defines the analog of the
energy-momentum tensor in the discrete case.

\begin{center}
\leavevmode
\epsfysize=5cm
\epsfbox{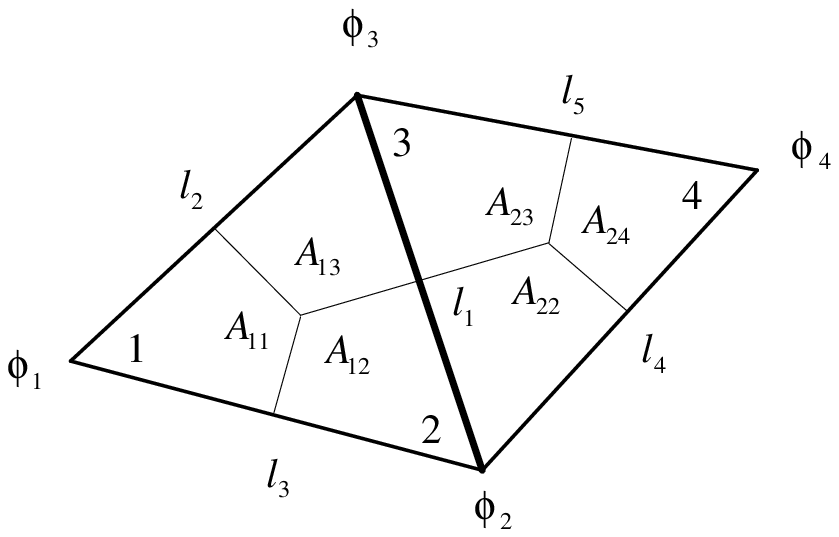}
\end{center}
\noindent
{\small{\it Fig.\ 20.
Labeling of the dual areas appearing in Eq.~(\ref{eq:vardual}).
\medskip}}

The coefficients in the above expansion in $\delta l_i^2$
can be regarded as the components
of the Regge lattice analog of the energy-momentum tensor, 
\beq
T_k \; = \; T_{\mu\nu} \; l_k^{\mu} \; l_k^{\nu} \;\; ,
\eeq
just as we can define for the simplicial
components of the metric tensor ~\cite{lund}
\beq
g_k \; = \; g_{\mu\nu} l_k^{\mu} l_k^{\nu} \; \equiv \; l_i^2 \;\; .
\eeq
From Eq.~(\ref{eq:gij_simplex}) and Eq.~(\ref{eq:gpert})
\beq
\delta g_{ij} (l^2) \; = \; \half \;
\Bigl [ \delta l_{0i}^2 + \delta l_{0j}^2 - \delta l_{ij}^2 \Bigr ] \;\; ,
\eeq
and therefore within triangle 1, 
(with vertices 1,2,3 and choosing coordinates along 23 and 21),
one has (see Figure 20)
\bea
\delta l_1^2 & = & \delta g_{11}
\nonumber \\ 
\delta l_2^2 & = & \delta g_{22}
\nonumber \\ 
\delta l_3^2 & = & \delta g_{11} + \delta g_{22} - 2 \delta g_{12}
\;\; .
\nonumber \\ 
\eea
For one triangle (for example, triangle 1 in Figure 19) one obtains
\beq
T^{\mu\nu} \, \delta g_{\mu\nu} \; \longrightarrow \;
( T^{11} + T^{12} ) \; \delta l_1^2 \; + \;
( T^{22} + T^{12} ) \; \delta l_2^2 \; - \;
T^{12} \; \delta l_3^2
\eeq
In the lattice case it is clear that, inserting for the variation
of the squared edge lengths corresponding to gauge variations,
as in Eq.~(\ref{eq:gauge_wfe_o}) and Eq.~(\ref{eq:gauge_wfe}),
and then equating the resulting coefficients of the arbitrary
gauge parameters $\chi_n^i$ to zero, gives the discrete equation
of conservation for the energy-momentum tensor.

On the other hand, the equations of motion for $l_i^2$
(as opposed to the equations of motion for $\phi$, which
are given in Eq.~(\ref{eq:eqmo_phi})) are
obtained directly from Eq.~(\ref{eq:vardual}), namely
\bea
{ \partial \; I_m [ l^2 ] \over \partial \; l^2_i } \; = \;
& { A_{11} \over 16 \; A_1^2 }  & \Bigl [
( \phi_2 - \phi_1 )^2 + ( \phi_3 - \phi_1 )^2 - ( \phi_3 - \phi_2 )^2
\; \Bigr ]
\\ \nonumber
+ & { A_{24} \over \; 16 A_2^2 } & \Bigl [
( \phi_2 - \phi_4 )^2 + ( \phi_3 - \phi_4 )^2 - ( \phi_3 - \phi_2 )^2
\; \Bigr ] \; = \; 0  \;\; .
\label{eq:eqmo_l}
\eea
It corresponds to the continuum equation $T_{\mu\nu}=0$.
In the presence of a cosmological constant term,
$\lambda \sum_{t} A_t$, an additional term appears in the 
equations of motion, which become, again with the notation of Figures 19 and 20,
\bea
& { A_{11} \over 16 \; A_1^2 } & \Bigl [
( \phi_2 - \phi_1 )^2 + ( \phi_3 - \phi_1 )^2 - ( \phi_3 - \phi_2 )^2
\; \Bigr ]
\\ \nonumber
+ & { A_{24} \over 16 \; A_2^2 } & \Bigl [
( \phi_2 - \phi_4 )^2 + ( \phi_3 - \phi_4 )^2 - ( \phi_3 - \phi_2 )^2
\; \Bigr ]
\\ \nonumber
& & + \; \lambda \; \Bigl [ \; 
{ 1 \over 16 A_1 } \; ( l_2^2 + l_3^2 -l_1^2 ) 
+ { 1 \over 16 A_2 } \; ( l_4^2 + l_5^2 -l_1^2 ) \Bigr ]
\; = \; 0  \;\; ,
\\ \nonumber
\label{eq:eqmo_l2}
\eea
and relates the squared edge lengths to the derivatives (or, more
properly here, finite differences) of the scalar field.

\vskip 10pt
\newsection{Gravitational Functional Measure}
\hspace*{\parindent}

In this section we will re-examine the issue of the gravitational
functional measure in light
of the results of the previous sections, and in particular
the local gauge invariance of the lattice gravitational action.
It is well know that in ordinary (lattice) gauge theories the
invariance of the action selects a unique measure (the group
invariant Haar measure).
A natural way to construct the gravitational functional measure in the
continuum is to introduce a metric over metrics (or super-metric),
and then compute the resulting functional volume element.
We shall see that what appears at first
as a rather straightforward procedure, is in fact affected by
a number of rather subtle ambiguities.

\vskip 10pt
\subsection{Continuum Case}
\hspace*{\parindent}

Following DeWitt ~\cite{dewitt}, one introduces a super-metric $G$
over metric deformations $\delta g_{\mu\nu}(x)$,
which in the simplest local form leads to the following
norm-squared for metric deformations
\beq
\Vert \delta g \Vert^2 \; \equiv \;
\int d^d x \; G^{\mu \nu, \alpha \beta} [g(x)] \;
\delta g_{\mu \nu}(x) \, \delta g_{\alpha \beta}(x) \;\; ,
\eeq
with the inverse of the DeWitt supermetric given by
\beq
G^{\mu \nu, \alpha \beta} [g(x)] \; = \;
\half {\textstyle \sqrt{g(x)} \displaystyle} \left [
g^{\mu \alpha}(x) g^{\nu \beta}(x) +
g^{\mu \beta}(x) g^{\nu \alpha}(x) + \lambda \,
g^{\mu \nu}(x) g^{\alpha \beta}(x) \right ] \;\; ,
\label{eq:dewittsuper}
\eeq
and $\lambda \neq - 2 / d $, to avoid the vanishing of the determinant
of $G$. It is easy to check that the above expression for
$\Vert \delta g \Vert^2 $ is invariant under diffeomorphisms.
\footnote{While it should be clear that the functional norm should
be invariant, it is less obvious that it should be {\it local}.
Ultimately the justification for locality lies in the fact that
the resulting functional {\it measure} is local in the continuum,
a rather desirable feature.}
The usual procedure is then to derive the functional measure in the form
\beq
\int d \mu [g] \; = \; \int \prod_x \left ( \det [ G(g(x)) ] \right )^{\half} 
\prod_{\mu \geq \nu} d g_{\mu \nu} (x) \;\; ,
\eeq
with the determinant of the super-metric $G^{\mu \nu, \alpha \beta} (g(x))$
given by
\beq
\det G (g(x)) \propto (1 + \half d \lambda ) \; [ g(x) ]^{ (d-4)(d+1)/4 }
\;\; .
\eeq
Up to irrelevant constants, in four dimensions it reduces to the very
simple expression
\beq
\int d \mu [g] \; = \; \int \prod_x \; [ g(x) ]^{ (d-4)(d+1)/8 } \;
\prod_{\mu \ge \nu} \, d g_{\mu \nu} (x)
\; \mathrel{\mathop\rightarrow_{ d = 4}} \;
\int \prod_x \prod_{\mu \geq \nu} d g_{\mu \nu} (x) \;\; .
\label{eq:dewitt}
\eeq
Unfortunately this measure is {\it not} gauge invariant, if 
the product over $x$ is interpreted
as one over `physical' points, and coordinate invariance is imposed
at one and the same `physical' point, as discussed in ~\cite{popov}.
Here this is seen as a consequence of the fact that
$\Vert \delta g \Vert^2 $ has
been split in two separately non-invariant parts.
The measure that {\it does} satisfies the invariance property is
\beq
\int d \mu [g] \; = \; \int \; \prod_x \;
\left ( \det \left [
G(g(x)) / {\textstyle \sqrt{g(x)} \displaystyle} \right ] \right )^{\half} 
\prod_{\mu \geq \nu} d g_{\mu \nu} (x) \;\; ,
\eeq
and was originally proposed by Misner ~\cite{misner}. Explicitly,
\beq
\int d \mu [ g] \; = \; \int \; \prod_x \; \left ( g(x) \right )^{-(d+1)/2}
\; \prod_{ \mu \ge \nu } \, d g_{ \mu \nu } (x) \;\; .
\label{eq:misner}
\eeq
It has the property of being scale invariant
in any dimension.
Unfortunately it is also singular, 
and needs therefore to be regulated in some way
for small $g$ (on the lattice this requires some cutoff for 
small local volumes).

Indeed both measures can be obtained as particular cases if one
introduces a real parameter $\omega$, and writes
\beq
\Vert \delta g \Vert^2 \; = \;
\int d^d x \left ( g(x) \right )^{\omega/2} \;
G^{\mu \nu, \alpha \beta} [g(x); \omega ] \;
\delta g_{\mu \nu}(x) \, \delta g_{\alpha \beta}(x) \;\; ,
\eeq
with
\beq
G^{\mu \nu, \alpha \beta} [g(x); \omega] \; = \;
\half \; \left ( g(x) \right )^{(1-\omega)/2} \left [
g^{\mu \alpha}(x) g^{\nu \beta}(x) +
g^{\mu \beta}(x) g^{\nu \alpha}(x) + \lambda \,
g^{\mu \nu}(x) g^{\alpha \beta}(x) \right ] \;\; .
\label{eq:dewittsuper1}
\eeq
The metric in function space is obviously left unchanged by this
rewriting, but the measure (obtained from $\det G$) depends
on $\omega$, and by the usual argument one obtains the
parameterized functional gravitational measure 
\beq
\int d \mu [ g] \; = \; \int \prod_x \;
\left [{\textstyle \sqrt{g(x)} \displaystyle} \right ]^{\sigma}
\; \prod_{ \mu \ge \nu } \, g_{ \mu \nu } (x) \;\; .
\label{eq:contmeas}
\eeq
with the {\it measure parameter} $\sigma$ related to the choice of $\omega$ via
\beq
\sigma \; = \; - (d+1) + ( \omega -1) \; { d(d+1) \over 4 } \;\; .
\eeq
For $\omega=0$ one obtains the DeWitt measure of Eq.~(\ref{eq:dewitt}),
while for $\omega=1$ one has the Misner measure of
Eq.~(\ref{eq:misner}). The close relationship between the DeWitt and Misner
measures was pointed out in ~\cite{ban}.
\footnote{The gravitational measure has to be 
modified in the presence of the matter fields.
For an $n$-component massless scalar field the invariant measure is
\beq
\int d \mu [ \phi ] \; = \; \int
\prod_x g^{n/4}(x) \; \prod_{a=1}^n d \phi^a (x) \;\; ,
\eeq
and therefore
$ \sigma \; = \; - (d+1) + ( \omega -1) \; { d(d+1) \over 4} + n/2 $.}

As there is no way of deciding between these two choices, or any
intermediate one for that matter, one is forced to consider
$\sigma$ as an {\it arbitrary}
(and hopefully ultimately irrelevant) {\it real parameter} of the
theory,
the only constraint being $\sigma > -(d+1)$.
In general the volume factor $g^{\sigma/2}$ in the functional measure
is absent in $d$ dimensions for the special choice $\omega = 1 + 4/d$.
It should be emphasized here that gauge invariance does not
select a specific value for $\sigma$, but does otherwise
completely constrain the form of the measure.
The criteria of simplicity and universality would suggest the above
to be the preferred choice for $\omega$.
The reason for the
ambiguity in the gravitational functional measure
appears to be a lack of a clear definition of what
is meant by $\prod_x$.
In spite of some recurrent claims to the contrary, 
such an ambiguity persists in {\it all}
lattice formulations.
Also, the volume term in the measure is 
completely {\it local} since it contains no derivatives.
It does not effect the propagation of
gravitons, as it contributes $\delta^d(0)$ terms to the effective
action. To some extent these can be regarded as renormalizations
of the cosmological constant, since they clearly only affect
the distribution of local volumes.
As such they are expected to only affect the short distance
behavior of the theory, leaving the more interesting universal
large distance properties unmodified (the recent work in
Ref. \cite{pot} gives a more concrete realization of these
ideas in the framework of continuum perturbation theory).
Support to this interpretation also comes from
a number of simple examples ~\cite{oned}.

\vskip 10pt
\subsection{Lattice Transcription}
\hspace*{\parindent}

Let us examine the consequences of this discussion in the discrete
case. As the edge lengths play the role of the metric in the
continuum, one expects the discrete measure to involve an
integration over edge lengths \cite{lesh,hartle}.
Indeed the induced metric at a simplex is related to the edge
lengths squared within that simplex, via the expression for the
invariant line element $ds^2 = g_{\mu \nu} dx^\mu dx^\nu$.
The relation between metric perturbations and squared edge
length variations for a given simplex in $d$ dimensions is
(see Eqs.~(\ref{eq:gij_simplex}) and ~(\ref{eq:gpert}), and Figure 3)
\beq
\delta g_{ij} (l^2) \; = \; \half \;
( \delta l_{0i}^2 + \delta l_{0j}^2 - \delta l_{ij}^2 ) \;\; .
\eeq
Thus for one simplex
the integration over the metric is equivalent to an 
integration over the edge lengths, and one has
\beq
\left ( {1 \over d ! } \sqrt { \det g_{ij}(s) } \right )^{\sigma} \;
\prod_{ i \geq j } \, d g_{i j} (s) \; = \; 
{\textstyle \left ( - { 1 \over 2 } \right ) \displaystyle}^{ d(d-1) \over 2 }
\; \left [ V_d (l^2) \right ]^{\sigma} \;
\prod_{ k = 1 }^{ d(d+1)/2 } \, dl_{k}^2 \;\; .
\label{eq:simpmeas}
\eeq
There are $d(d+1)/2$ edges for each
simplex, just as there are $d(d+1)/2$ independent components for the metric
tensor in $d$ dimensions.
(We are ignoring for the moment the triangle inequality constraints,
which further require all sub-determinants of $g_{ij}$ to be
positive as well, including the obvious restriction $l_k^2 >0$).
The extension to many simplices glued together at their common faces
is then immediate, and after summing over all simplices one obtains,
up to an irrelevant numerical constant,
\beq
\int d \mu [l^2] \; = \; 
\int_0^\infty \; \prod_s \; \left [ V_d (s) \right ]^{\sigma} \;
\prod_{ ij } \, dl_{ij}^2 \;
\;\; .
\label{eq:lattmeas}
\eeq
In four dimensions the lattice DeWitt measure ($\sigma=0$) is particularly
simple,
\beq
\int d \mu [l^2] \; = \; \int_0^\infty \prod_{ ij } \, dl_{ij}^2 
\; F [ l_{ij}^2 ] \;\; .
\label{eq:dewlattmeas}
\eeq
Here $ F_\epsilon [l] $ is
a (step) function of the edge lengths, with the property
that it is equal to one whenever the triangle inequalities and their
higher dimensional analogs are satisfied,
and zero otherwise.
\footnote{The functional measure over edge lengths in
Eq.~(\ref{eq:dewlattmeas}) does not have compact support, and a cosmological
term (with coefficient $\lambda > 0 $) is therefore essential in
obtaining convergence of the functional integral for large edge lengths.}

The above lattice measure over edge lengths
has recently been used extensively in numerical simulations of simplicial
gravity ~\cite{phases,lines,scalar,corr,berg,beirl,holm}
(Figure 21 gives an example of the edge length distribution
obtained in four dimensions with measure parameter $\sigma=0$).
One would expect that physical properties of the theory should not
depend on the arbitrary parameter $\sigma$, but for the moment
this universality argument remains largely an unproven conjecture.
The universality argument is in part tenuously supported by
some systematic numerical studies in two ~\cite{gh} and
three ~\cite{hw3d} dimensions,
although there is no solid evidence for it yet in four dimensions
~\cite{phases,beirl}.
The authors of Ref. ~\cite{beirl} have also independently
emphasized the importance of exploring the effects of different
choices for the measure parameter $\sigma$.
In the context of dynamical triangulations and their relationship
with simplicial gravity, the irrelevance
of the measure parameter has also been proposed in ~\cite{dt2regge}.
Finally let us mention that one can argue backwards ~\cite{kyoto} that the
Regge lattice measure of Eqs.~(\ref{eq:simpmeas}) and
(\ref{eq:dewlattmeas}) provides further support for the
correctness of the continuum DeWitt functional measure approach.

The above measure can also be obtained by considering
a simplicial analog of the DeWitt supermetric, as was suggested
in ~\cite{cms,cms1}. One writes
for the induced metric of Eq.~(\ref{eq:gij_simplex})
\beq
\Vert \, \delta g (s) \, \Vert^2 \; = \; \sum_{s} \;
G^{ i j k l } \left ( g(s) \right ) \; 
\delta g_{i j} (s) \, \delta g_{k l} (s) \;\; ,
\eeq
with the inverse of the lattice DeWitt supermetric given now by
\beq
G^{ i j k l } [ g(s) ] \; = \; 
\half \sqrt{g(s)} \; \left [ \,
g^{i k} (s) g^{j l} (s) +
g^{i l} (s) g^{j k} (s) + \lambda \,
g^{i j} (s) g^{k l} (s) \right ] \;\; ,
\label{eq:dewittsuperl}
\eeq
and $(\lambda \neq - 2 / d )$.
This defines a metric on the tangent space of positive real
symmetric matrices $g$. The resulting functional measure is the one
of Eq.~(\ref{eq:lattmeas}), with, by construction,
$\sigma=(d-4)(d+1)/4$, the DeWitt value. Thus
\beq
\int d \mu [l^2] \; = \; \int
\prod_{s} \, \left [ \; \det G(g(s)) \; \right ]^{\half}
\prod_{i \geq j} d g_{i j} (s) \;\; ,
\eeq
with the determinant of the super-metric $G^{i j k l} (g(s))$
given by
\beq
\det G (g(s)) \; \propto \; 
(1 + \half d \lambda ) \; \left [ g(s) \right ]^{ (d-4)(d+1)/4 } \;\; ,
\eeq
and therefore up to irrelevant constants
\beq
\int d \mu [l^2] \; = \; \int_0^\infty
\prod_{s} \, \left [ V (s) \right ]^{\sigma}
\prod_{ij} \, d l^2_{i j} \;\; ,
\eeq
with $\sigma = (d-4)(d+1)/4 $. The measure factor can be 
exponentiated and written equivalently as an effective action contribution
\beq
\int d \mu [l^2] \; = \; \int_0^\infty
\prod_{ij} \, dl_{ij}^2 \; \exp \left \{ \sigma
\sum_{s} \log V (s) \right \} \;\; .
\eeq
Its effect is to suppress or enhance, depending on the sign
of $\sigma$, contributions from small volumes. As such, it
acts much like a cosmological constant contribution
\beq
\exp \left \{ - \; \lambda \; \sum_{s} \; V (s) \right \} \;\; .
\eeq

\vskip 10pt
\subsection{Lund Regge Approach}
\hspace*{\parindent}

The previous approach to the functional measure is based on a direct
discretization of the continuum measure, and leads to a unique
local measure over the squared edge lengths (modulo the volume
factors), in close analogy to the continuum expression.
Alternatively, one can try to find a discrete form for the supermetric,
and then evaluate the resulting determinant.

In a paper Lund and Regge offered a slightly different approach
to the measure problem ~\cite{lund}, in connection with the
$3+1$ formulation of simplicial gravity.
The idea, recently re-analyzed by the authors of Ref. ~\cite{hmw},
was to obtain a lattice analog of the DeWitt supermetric, by
considering the quantity
\beq
\Vert \delta l^2 \Vert^2 \; = \; \sum_{ij} \; G_{ij} (l^2)
\; \delta l^2_i \; \delta l^2_j \; \; ,
\label{eq:lund}
\eeq
where $G_{ij} (l^2)$ plays a role analogous to the DeWitt supermetric,
but now on the space of squared edge lengths.
One way of constructing the explicit form for $G_{ij} (l^2)$ is to
write the squared volume of a given simplex in terms of the induced
metric within the same simplex,
\beq
V^2 ( s ) \; = \; {\textstyle \left ( { 1 \over d! } \right )^2 \displaystyle}
\det \left \{ g_{ij}(l^2(s)) \right \} \;\; .
\eeq
Then compare the expansion of the determinant of the metric in
the continuum,
\bea
\det ( g_{ij} + \delta g_{ij} ) & = & 
\exp \Tr \log ( g_{ij} + \delta g_{ij} )
\nonumber \\
& = & \det ( g_{ij} ) \left [ \, 1 + g^{ij} \delta g_{ij}
+ \half g^{ij} g^{kl} \delta g_{ij} \delta g_{kl}
- \half g^{ij} g^{kl} \delta g_{jk} \delta g_{li}
+ \cdots \; \right ] \;\; ,
\nonumber \\
\label{eq:gex}
\eea
to the analogous expansion for the square of the volume of a simplex
\beq
V^2 ( l^2 + \delta l^2 ) \; = \; V^2 (l^2)
+ \sum_i \; { \partial V^2 (l^2) \over \partial l^2_i } \; \delta l^2_i
+ \half \sum_{ij} \;
{ \partial^2 V^2 (l^2) \over \partial l^2_i \partial l^2_j } \;
\delta l^2_i \; \delta l^2_j + \cdots \;\; .
\label{eq:vex}
\eeq
Identifying terms of order $(\delta l_i^2)^n$ in Eq.~(\ref{eq:vex})
with terms of order $(\delta g_{ij})^n$ in Eq.~(\ref{eq:gex}) one obtains
to linear order
\beq
{1 \over V (l^2)} \; {\partial V^2 (l^2) \over \partial l^2_i} \; \delta l^2_i
\; = \; {\textstyle { 1 \over d! } \displaystyle} \sqrt{ \det ( g_{ij} ) }
\; g^{ij} \; \delta g_{ij} \;\; ,
\eeq
and to quadratic order
\beq
{1 \over V (l^2)} \; 
\sum_{ij} { \partial^2 V^2 (l^2) \over \partial l^2_i \partial l^2_j }
\; \delta l^2_i \; \delta l^2_j \; = \;
{\textstyle { 1 \over d! } \displaystyle} \sqrt{ \det ( g_{ij} ) }
\left [ \; g^{ij} g^{kl} \delta g_{ij} \delta g_{kl}
- g^{ij} g^{kl} \delta g_{jk} \delta g_{li} \; \right ] \;\; .
\eeq
Remarkably, the right hand side of this equation contains precisely
the expression appearing in the continuum supermetric
of Eq.~(\ref{eq:dewittsuper}), for the specific choice $\lambda = -2$.
After summing over simplices one obtains
\beq
\half \; \sum_{s} \sqrt{ \det ( g_{ij}(s) ) } \;
\left [ \; g^{ik}(s) \; g^{jl}(s) + g^{il}(s) \; g^{jk}(s)
- 2 g^{ij}(s) \; g^{kl}(s) \; \right ]
\delta g_{ij}(s) \; \delta g_{kl}(s)
\; = \; \sum_{ij} G_{ij} (l^2) \; \delta l^2_i \; \delta l^2_j \;\; ,
\eeq
with 
\beq
G_{ij} (l^2) \; = \; - \; d! \; \sum_{s} \;
{1 \over V (s)} \; 
{ \partial^2 \; V^2 (s) \over \partial l^2_i \; \partial l^2_j } \;\; ,
\eeq
which now determines the matrix $G_{ij}(l^2)$ 
appearing in the Lund-Regge metric for deformations in the space
of squared edge lengths, Eq.~(\ref{eq:lund}).
The analogy with the continuum expression is brought out more clearly
when one factors out the volume element, and writes
\beq
\Vert \delta l^2 \Vert^2 \; = \; \sum_{s} \; V (s) \;
\left \{
\; - \; {d! \over V^2 (s)} \; \sum_{ij} \;
{ \partial^2 \; V^2 (s) \over \partial l^2_i \; \partial l^2_j }
\; \delta l^2_i \; \delta l^2_j \; \right \} \;\; .
\eeq
The volume factor ambiguity present in the continuum measure
is not removed though.
As in the continuum, different measures on the edge
lengths are obtained, depending on whether the local volume
factor $V(s)$ is included in the supermetric or not.
In parallel with Eq.~(\ref{eq:dewittsuper1}) one writes therefore
\beq
\Vert \delta l^2 \Vert^2 \; = \; \sum_{s} \; [ V (s) ]^{\omega'} \;
\left \{
\; - \; {d! \over [ V(s) ]^{1+{\omega'}} } \; \sum_{ij} \;
{ \partial^2 \; V^2 (s) \over \partial l^2_i \; \partial l^2_j }
\; \delta l^2_i \; \delta l^2_j \; \right \} \;\; .
\eeq
The metric in edge length space is again left unchanged by this
rewriting, but the measure obtained from $\det G$ depends
on a parameter ${\omega'}$, and by the usual arguments one obtains a
parameterized functional gravitational measure 
\beq
\int \; d \mu [ l^2 ] \; = \; \int \; \prod_i \;
\sqrt{ \det G_{ij}^{( \omega ' )} (l^2) } \, dl^2_i \;\; .
\label{eq:lundmeas}
\eeq
with 
\beq
G^{(\omega')}_{ij} (l^2) \; = \; - \; d! \; \sum_{s} \;
{ 1 \over [ V(s) ]^{1+{\omega'}} } \;
{ \partial^2 \; V^2 (s) \over \partial l^2_i \; \partial l^2_j } \;\; ,
\eeq
Thus the matrix $ G^{({\omega'})}_{ij} $
should be thought of as defining a one-parameter family of measures.

A somewhat disturbing feature of the Lund-Regge lattice measure
of Eq.~(\ref{eq:lundmeas}) is that it does not give the correct
result already in one dimension ~\cite{oned}. The one-dimensional
action for pure gravity of Eq. ~(\ref{eq:length}), proportional to the length
of the curve, is invariant under the local gauge transformations of
Eq.~(\ref{eq:gauge_1d}),
\beq
\delta l_n \; = \; \chi_{n+1} - \chi_n \;\; ,
\eeq
where the $\chi_n$'s represent continuous parameters defined on
the lattice vertices.
Any local variations of the edges which have the above form (and, we should
add, also do not violate the constraint $l_n>0$)
obviously leave the physical length of the curve unchanged.
The invariant measure in one dimension is therefore
\beq
\int \; d \mu [ l^2 ] \; = \; \int_0^\infty \; \prod_{i=0}^N \, dl_i \;\; .
\eeq
which falls precisely in the class of measures encompassed
by Eq.~(\ref{eq:lattmeas}), with $\sigma=-1$.
On the other hand, the arguments leading to the measure
of Eq.~(\ref{eq:lundmeas}) give $l^2=g$ to zeroth order,
$\delta l^2/l^2 = \delta g / g $ to first order, and $0=0$
to second order, and therefore, since $G_{nm}=0$, $ \Vert \delta l^2
\Vert^2 = 0 $!
Incidentally, in one dimensions a physically motivated
invariant distance between
manifolds is $ d^2 (l,l') \equiv \left [ L(l) - L'(l') \right ]^2 $,
which is nonlocal.

Another somewhat undesirable feature of the Lund-Regge metric is that in 
general it is non-local, in spite of the fact that the original
continuum measure of Eq.~(\ref{eq:contmeas}) {\it is} completely local.
\footnote{After imposing a gauge condition, such as the conformal gauge, the
measure can become non-local as in ordinary gauge theories. But such
a gauge fixing term is only necessary in perturbation theory to remove
the zero modes of the action, discussed in Sections 3. and 4.
Nonperturbatively one would expect that no gauge fixing is necessary,
as the effects of the gauge zero modes are expected to cancel out in 
averages of physical quantities, as in ordinary lattice gauge theories
~\cite{hartle}.}
After all, metric perturbations and squared edge lengths are
linearly related to each other, and locality for one measure
should translate into locality for the other measure.
Non-local contribution to the original (non gauge-fixed) 
measure seem as unattractive as non-local action contributions.
A non-local measure makes it virtually impossible to study
the theory non-perturbatively.
On the other hand it is clear that, for some special choices
of $\omega'$ and $d$, one {\it does} recover a local measure.
Thus in two dimensions for ${\omega'} = -1$ one obtains again the simple
result,
\beq
\int \; d \mu [ l^2 ] \; = \; \int_0^\infty \; \prod_i \, dl^2_i \;\; ,
\eeq
(which incidentally is non-singular at small edge lengths and
represents therefore in this respect an acceptable measure).
This measure appears therefore, on the basis of purely theoretical arguments,
to be as good as any other measure with a different $\omega'$.
Given the possibility of a choice for ${\omega'}$,
it would seem natural that one should chose its
value in such a way that the measure has the simplest form, here the
local form without any volume factors: after all the physical theory
should not depend on the bare parameter $\omega'$.

We should also remark that the appearance of non-local
measure contributions (and in particular for some values of the 
parameter $\omega'$, but not for others) makes one question the
initial justification for starting in the first place with an expression,
such as the one in Eq.~(\ref{eq:dewittsuperl}), which is local.
In conclusion it appears that the local measure of Eq.~(\ref{eq:dewlattmeas})
provides the simplest theoretically justifiable
starting point, if not the only possible one.

\begin{figure}
  \begin{center}
    \input{plot1} 
  \end{center}  
\noindent
{\small{\it Fig.\ 24 . Typical edge length distribution ${\cal P}(l)$
in four dimensions, for the lattice analog of the DeWitt measure
$\sigma=0$ (see Eq.~(\ref{eq:lattmeas})),
close to the critical point at $G_c$ (from ref. ~\cite{phases}).
\medskip}}
\end{figure}

\vskip 10pt
\newsection{Gauge Fixing and Lattice Conformal Gauge}
\hspace*{\parindent}

Regge's simplicial quantum gravity does not require gauge
fixing ~\cite{lesh,hartle}, unless one intends to perform a diagrammatic
perturbative expansion on the lattice ~\cite{fey}.
In this respect, the situation is completely
analogous to ordinary gauge theories, and one expects the volume
of the gauge group, the diffeomorphism group in the case of gravity,
eventually to cancel out in the expression for physical averages,
\beq
\langle O \rangle \; = \;
{ \int d \mu [l^2] \; O (l^2) \; \exp \{ - I[l^2] \} \over
\int d \mu [l^2] \; \exp \{ - I[l^2] \} }
\eeq
The lattice diffeomorphism zero modes discussed here
(see Eq.~(\ref{eq:gauge_wfe_o})
and subsequent expressions) do not therefore in principle pose a problem
in nonperturbative studies of quantized gravity, such as the
ones presented in ~\cite{phases,lines}.
Zero modes are automatically taken into account as the measure
explores gauge-equivalent choices of metrics.
One important distinction with the formal continuum theory
is the presence of a cutoff in orbit space, due to the
enforcement of the triangle inequalities. As a result, the
gravitational functional measure is highly non-trivial.
Such a constraint is not seen to any order in the perturbative
weak field expansion, it is a genuinely non-perturbative
constraint.

On the other hand in two dimensions the continuum theory
can be studied by perturbative methods, which are most
suitably applied in the conformal gauge ~\cite{poly}.
It is the purpose of this section to elaborate on the
connection between the continuum and the lattice theory,
both being formulated here in a particular gauge.
Let us first summarize the results in the continuum.
The weak field expansion is used, and one sets as usual
\beq
g_{\mu\nu} (x) \; = \;
\delta_{\mu\nu} \; + \; \kappa \; h_{\mu\nu} (x) \;\; .
\eeq
As there is no small parameter in two dimension to play with,
one assumes $\kappa \ll 1$, expands, and then sets $\kappa=1$ at the end.
The easiest quantity to compute is the ``graviton'' vacuum polarization
due to one massless scalar particle, a one loop diagram here.
It is given by
\bea
\Pi_{\mu \nu,\alpha\beta} (q) & = & \half \,
\int \frac{d^2 p}{(2\,\pi)^2}
\frac{t_{\mu\nu}(p,q)\,t_{\alpha\beta}(p,q)}{p^2\,(p+q)^2}
\nonumber \\
t_{\mu\nu} (p,q) & = & \half \; [ \, \delta_{\mu\nu} \;
p \cdot (p + q) - p_{\mu} \, (p_{\nu} + q_{\nu}) - p_{\nu} \, 
(p_{\mu} + q_{\mu})\, ] \;\; .
\eea
The calculation of the integral is easily done using dimensional
regularization \cite{fey}, or by the methods of ~\cite{polybook}.
In either case one obtains
\beq
\Pi_{\mu \nu,\alpha\beta} (q) \; = \;
{1 \over {48 \, \pi}} \; (q^2 \delta_{\mu\nu} - q_{\mu} q_{\nu})
\, { 1 \over q^2 } \,
(q^2 \delta_{\alpha\beta} - q_{\alpha} q_{\beta}) \;\; .
\label{eq:vacpol}
\eeq
For a $D$-component scalar field, the above result is simply
multiplied by a factor of $D$, and the effective action, to lowest
order in the weak field expansion, is then 
\beq
I_{eff} \, = \, - \half \int { d^2 q \over (2 \pi)^2 } 
\, h_{\mu\nu} (q) \, \Pi_{\mu\nu\rho\sigma} (q) \, h_{\rho\sigma} (-q) \;\; .
\label{eq:ieff}
\eeq
In the conformal gauge, coordinates are chosen
which are locally orthogonal, so as to bring the
metric into the form
\beq
g_{\mu\nu} (x) \, = \, \delta_{\mu\nu} \, e^{ \varphi (x) } \;\; .
\label{eq:conf_gauge}
\eeq
Then one has for the scalar curvature
\beq
R(q) \; = \;
( q_{\mu} q_{\nu} - \delta_{\mu\nu} q^2 ) \; h_{\mu\nu} (q)
\; = \; q^2 \; \varphi (q) \;\; ,
\eeq
and one can therefore re-write the effective action in the form
\beq
I_{eff} (\varphi) \, = \,
- {D \over 96 \pi} \int { d^2 q \over (2 \pi)^2 } \;
\varphi (q) \; q^2 \; \varphi (-q) \; = \;
- {D \over 96 \pi} \int d^2 x \left [ ( \partial_{\mu} \varphi )^2
+ ( \lambda - \lambda_c ) \, e^{\varphi} \right ] \;\; .
\label{eq:vacpols}
\eeq
On the lattice one can perform a similar computation,
using again perturbation theory \cite{fey}.
The lattice Feynman rules are written down, the integration
over the scalar is performed, and an effective action results,
which can be expanded out in the weak field limit.
As the scalar couples invariantly to the gravitational degrees of
freedom, one would expect that the result should be eventually expressible
in terms of invariants. Indeed in the continuum one can re-write 
the effective action of Eq.~(\ref{eq:vacpols}) in an invariant form,
\beq
\half \int d^2 x \, d^2 y \, R \sqrt g (x) \,
\langle x | { 1 \over - \partial^2 } | y \rangle \,
R \sqrt g (y) \;\; ,
\eeq
where $\partial^2$ is the continuum covariant Laplacian,
$\partial^2 \equiv \partial_{\mu} \sqrt{g} g^{\mu\nu} \partial_{\nu}$.
On the lattice this expression has an obvious invariant counterpart
~\cite{hw2d},
\beq
\half
\sum_{{\rm hinges} h,h'} \, \delta_h \;
\Bigl [ { 1 \over - \Delta } \Bigr ]_{ h,h' } \;
\delta_{h'} \;\; ,
\eeq
which is obtained from the correspondence between lattice and continuum
curvatures derived in ~\cite{hw84}.
$\Delta$ here is the nearest-neighbor covariant lattice Laplacian,
as obtained from the discrete scalar action (see 
Eqs.~(\ref{eq:acdual}) and (\ref{eq:eqmo_phi})); 
for a recent discussion of the discretization of
this term see also ~\cite{peir}.
It introduces an effective long-range interaction between deficit
angles.
For this reason it is actually preferable
to study non-perturbative aspects of the model leaving
the scalar fields un-integrated, which keeps the action {\it local}
~\cite{gh}.
\footnote{It is encouraging that the conformal mode stays massless
in the full non-perturbative treatment of the two-dimensional
simplicial lattice theory, without the necessity of any sort of fine-tuning
of bare parameters ~\cite{gh,elba}.}

The previous discussion provides a background for motivating the
introduction of the conformal gauge on the lattice.
It is legitimate to ask therefore what is the simplicial lattice
analog of the gauge condition of Eq.~(\ref{eq:conf_gauge}).
The conformal gauge implies a local choice of
orthogonal coordinates. It is clear from the discussion
in Section 3.1 that there is a corresponding choice on the
lattice. Indeed in the development of the weak field expansion 
a uniform orthogonal set of coordinates was chosen,
with a diagonal background metric
(with edge length assignments $l_i^0 = 1 $ for the body principals
($i=1,2$) and $l_i^0 = \sqrt{2} $ for the diagonal ($i=3$)).
For these coordinates (see  Eq.~(\ref{eq:gij_square})) one has
for the background metric
\beq
g_{ij}^{(0)} \; = \; \delta_{ij} \;\; .
\eeq
The lattice conformal gauge choice corresponds to an assignment of edge
lengths such that locally
\beq
g_{ij} (n) \; = \; \left( \begin{array}{cc}
l_1^2 (n) & \half (l_3^2 (n) - l_1^2 (n) - l_2^2 (n) ) \\
\half (l_3^2 (n) - l_1^2 (n) - l_2^2 (n) ) & l_2^2 (n) \\
\end{array} \right) \; \approx \; \delta_{ij} \; e^{\varphi (n)} \;\; .
\eeq
Here the lattice fields $\varphi(n)$ have to be defined on the lattice
vertices, and so are the gauge degrees of freedom
$\chi_{\mu}(n)$, as can be inferred from Eq.~(\ref{eq:gauge_wfe}).
It is clear therefore that a choice of lattice conformal gauge corresponds
to a re-assignment of edge lengths about each vertex, in such a way
that the local curvature is left unchanged, but at the same time
the induced metric is brought into diagonal form;
it corresponds to a choice of approximately right-angle triangles at each
lattice vertex.

This result is further illustrated in Figures 22 to 24.
The surface shown in Figure
22 has been brought into the lattice analog of the conformal gauge,
by re-assigning edge lengths in such a way that individual triangles
look as close as possible to right-angle triangles. In going from
Figure 24 to Figure 23, repeated gauge transformations must be performed
on the vertices, by reassigning edge lengths in such a
way that local areas and volumes are kept unchanged.
It is easy to see that such a construction can always be done,
except in some rather pathological cases.

\begin{center}
\leavevmode
\epsfysize=7cm
\epsfbox{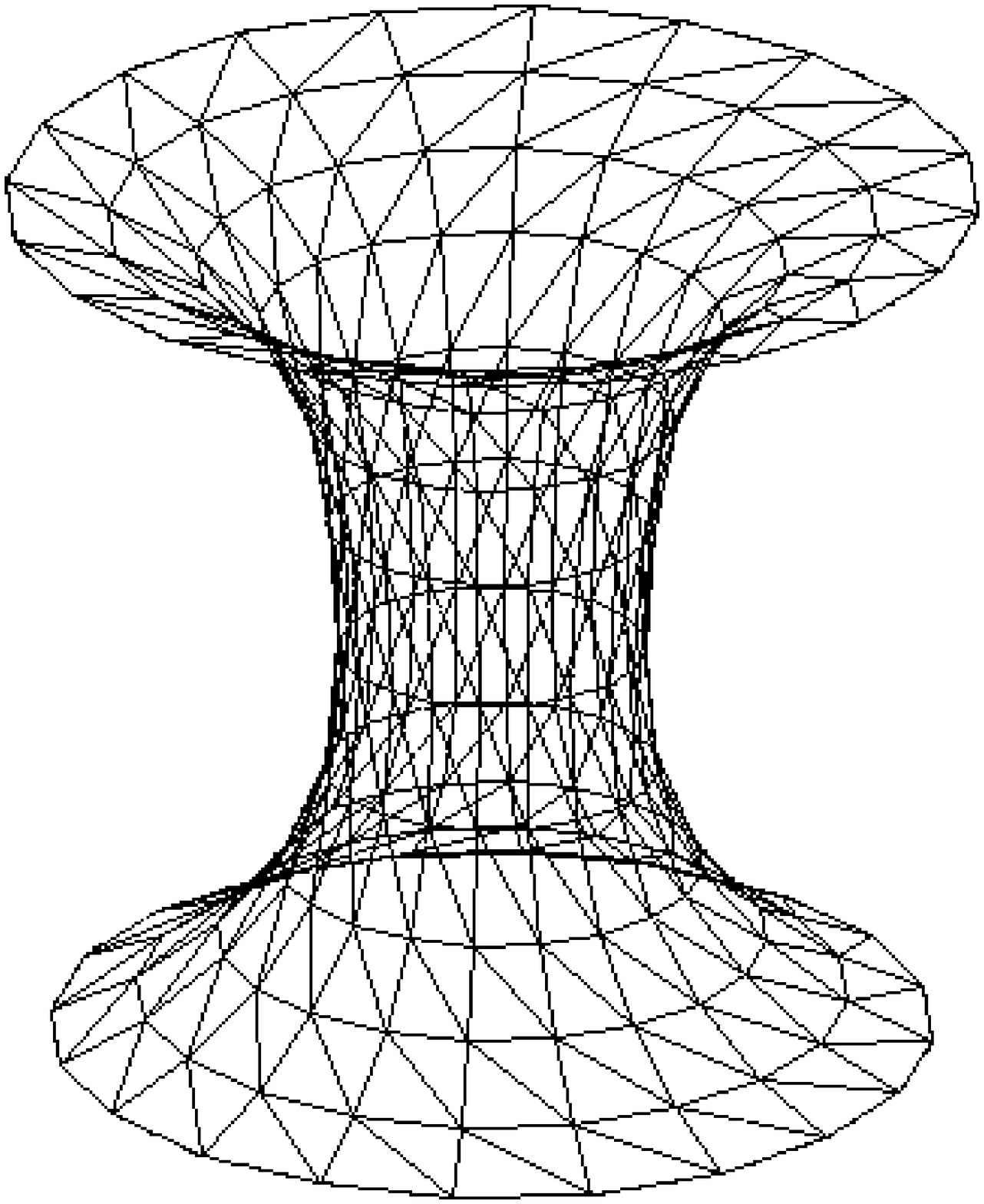}
\end{center}
\noindent
{\small{\it Fig.\ 21.
Description of a smooth surface in the lattice analog of the conformal
gauge.
\medskip}}

\begin{center}
\leavevmode
\epsfysize=5cm
\epsfbox{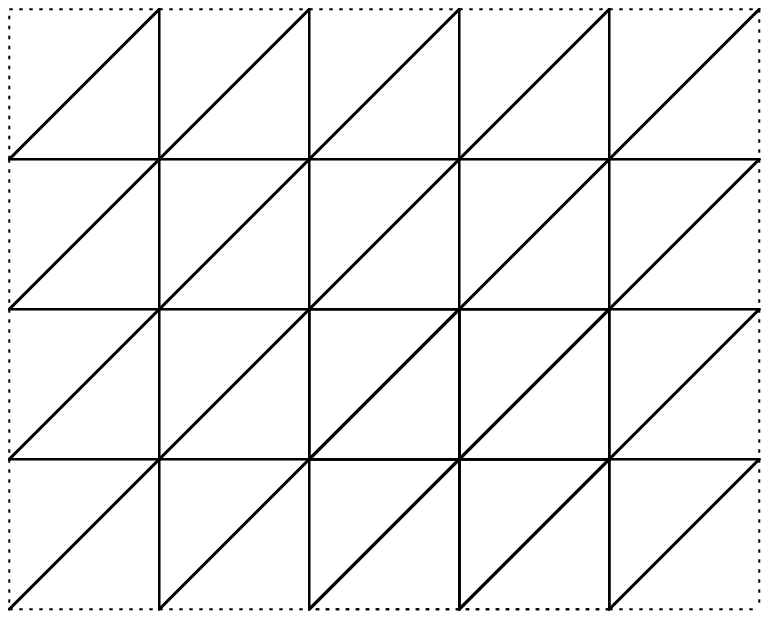}
\end{center}
\noindent
{\small{\it Fig.\ 22.
Enlarged view of a small region on the surface in Fig.\ 21.
\medskip}}

\begin{center}
\leavevmode
\epsfysize=5cm
\epsfbox{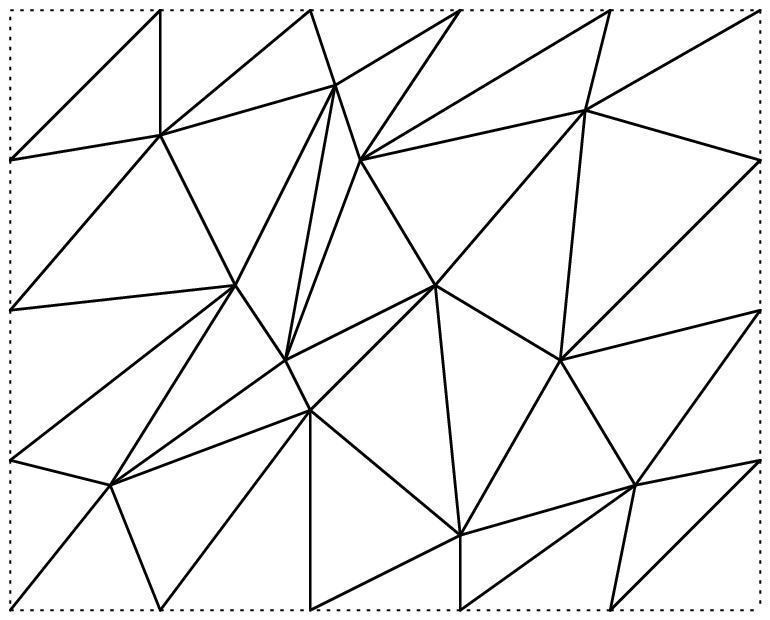}
\end{center}
\noindent
{\small{\it Fig.\ 23.
Gauge equivalent description of the enlarged view, of a small region
of the original surface, represented in Fig.\ 21.
\medskip}}

The gravitational contribution to the effective action in
the lattice conformal gauge can, at least in principle,
be computed in a similar way.
Let us sketch here how the analogous lattice calculation would
proceed; a more detailed discussion will be presented elsewhere.
In the continuum the metric perturbations are naturally decomposed
into orthogonal conformal and diffeomorphism parts,
\beq
\delta g_{\mu\nu} (x) = g_{\mu\nu} (x) \; \delta \varphi (x) +
\nabla_{\mu} \chi_{\nu} (x) + \nabla_{\nu} \chi_{\mu} (x) \;\; .
\label{eq:orth}
\eeq
where $\nabla_{\nu}$ denotes the covariant derivative.
It should be clear from the discussion in Section 3. that a rather similar
decomposition can be done for the lattice degrees of
freedom, by separating out the lattice gauge transformations
(which act on the vertices and change the edge lengths without
changing the local volumes and curvatures) from the conformal
transformations (which do change them) in Eq.~(\ref{eq:gpert}).
The explicit form for the lattice diffeomorphisms, to lowest
order in the lattice weak field expansion, is given in
Eq.~(\ref{eq:gauge_wfe_o}),
while the explicit form for the lattice conformal transformations
in given in Eq.~(\ref{eq:confdef}),
which makes it obvious that such a decomposition can indeed be
performed on the lattice.
In the continuum,
after rewriting the gravitational functional measure in terms of
conformal and diffeomorphism degrees of freedom,
\beq
\int \; d \mu [g] \; = \; \int \;
d \mu [\varphi] d \mu [\chi] \left [ \det ( L^+ L) \right ]^{\half} \;\; ,
\label{eq:fpdet}
\eeq
one has to compute Jacobian of the operator $L$.
It is determined, in the continuum, from
\beq
( L^+ L \; \chi )_{\mu} \; = \; \nabla^{\nu} (
\nabla_{\mu} \chi_{\nu} + \nabla_{\nu} \chi_{\mu} 
- g_{\mu\nu} \nabla^{\rho} \chi_{\rho} ) \;\; ,
\eeq
One then obtains for the effective action contribution 
in the conformal gauge,
\beq
\left [ \det ( L^+ L) \right ]^{-\half} \; \sim \;
\exp \left \{ - I_{eff} (\varphi) \right \} \;\; .
\eeq
with $I_{eff}$ of the form in Eq.~(\ref{eq:ieff}) to lowest
order in the weak field expansion.
A diagrammatic calculation, similar to the one for the scalar field
contribution, gives in the continuum the celebrated
result ~\cite{poly,polybook}
\beq
\Pi_{\mu\nu\rho\sigma} (q) \; = \;
{13 \over 48 \pi } \; ( q_{\mu} q_{\nu} - \delta_{\mu\nu} q^2 ) \,
{ 1 \over q^2 } \, ( q_{\rho} q_{\sigma} - \delta_{\rho\sigma} q^2 ) \;\; .
\label{eq:vacpolg}
\eeq

On the lattice the functional integration is performed over
the squared edge lengths, as in Eq.~(\ref{eq:lattmeas}).
But it seems a technically challenging task to
compute that Jacobian that maps the edge length variables
(which define the lattice metric) to the orthogonal lattice
diffeomorphism variables of Eq.~(\ref{eq:gauge_wfe}) and
the lattice conformal fields of Eq.~(\ref{eq:confdef}), with
the appropriate Jacobian included. It is also quite possible that
one might have to go beyond the lowest order in the lattice
perturbative expansion.

As a consequence the total Liouville action for the Liouville field
$ \varphi = { 1 \over \partial^2 } R $ becomes
\beq
I_{eff} (\varphi) \, = \,
{ 26 - D \over 96 \pi} \int d^2 x \left [ ( \partial_{\mu} \varphi )^2
+ ( \lambda - \lambda_c ) \; e^{\varphi} \right ]  \;\; ,
\label{eq:liouville}
\eeq
To lowest order in the weak field expansion the
critical value of $D$ for which the action vanishes is $D_c=26$, but
this number is modified by higher order quantum corrections.
In any case, for sufficiently large $D$ one expects an instability to develop.
Numerical nonperturbative studies of two-dimensional gravity
suggest that in the lattice theory the correction is large, and one finds
that the threshold of instability moves to values as low as
$D_c \approx 13 $ ~\cite{elba}. It is unclear if this critical value can
be regarded as truly universal, and independent for example on the detailed
choice of gravitational measure (we are referring here to
the choice of parameter $\sigma$).

\vskip 10pt
\newsection{Conclusions}
\hspace*{\parindent}

We have shown in this paper that Regge's formulation of simplicial
gravity is endowed with a remnant of the continuous local gauge
invariance of the original continuum theory. The appearance of zero
modes corresponding to the diffeomorphisms in the continuum is
particularly transparent in the weak field expansion. 
Nevertheless, the presence of a local invariance in the discrete
action can be exhibited,
via the detailed explicit calculations presented in this paper,
for almost any conceivable choice of background lattice.
It was shows in particular that the structure of the zero modes
corresponds precisely to the discretized form of the diffeomorphism
transformation law in the continuum.
We have underscored the fact that the squared edge lengths correspond
to the metric components in the continuum, and that such a
result is therefore hardly surprising (indeed it has been known
for some time in some special cases).
Explicit calculations also show that the gauge and conformal modes
are consistently defined as acting on the vertices of the lattice.
Although our derivations have mainly been restricted to the
two dimensional case, where they are more transparent and one
does not run the risk of drowning is a sea of indices,
wa have argued that they have general applicability, and in 
a number of cases we have indicated the structure of the general
result.

Our results have a bearing on the issue of the gravitational measure
in simplicial gravity. As the metric degrees of freedom in the
continuum correspond to the squared edge lengths in the lattice
theory, it is clear that functional integration in the latter should
be performed again over the edge lengths squared.
We have provided a number of arguments in support of this statement,
based on DeWitt's approach to the functional measure in the
continuum. We have argued that the lattice measure is essentially no
less unique than the original continuum (DeWitt) measure, with
ambiguities restricted to local volume factors, and which most likely
are not relevant in four dimensions. It is unlikely that further insight
into this issue can come from analytical work, and it is hoped that
future numerical simulations will support this conclusion, for which
there is already some partial and incomplete evidence.
At the end of the paper we have considered the introduction of gauge fixing
terms in the lattice action, which are needed in order to remove the
gauge zero modes of the gravitational action in perturbative
calculations.
Again the situation is similar to what happens in the continuum
when one performs perturbation theory, where one first separates out the
infinite gauge volume contribution.
As a specific example, we have discussed how one goes about
constructing the lattice conformal gauge.

A more practical motivation for our work has been to try to
understand the recently discovered discrepancy between the critical
exponents for matter coupled to gravity
in two dimensions as computed in the lattice regularized model for
gravity ~\cite{gh,holm},
and the corresponding conformal field theory predictions
for central charge $c=\half$ \cite{kpz,kazakov}.
Particularly significant in this respect appears to be the recent realization
that the conformal field theory exponents describe two-dimensional
random systems in {\it flat} space, and do not correspond to
``gravitational'' dressing of correlators ~\cite{ranis} (see Table I).
Recent independent calculations have to some extent confirmed
this result ~\cite{bb}.

\begin{table}
\caption{Critical exponents of random and non-random Ising models.}
\begin{center}
\begin{tabular}{|l|l|l|l|l|l|}
\hline
& $\gamma / \nu$ & $\beta / \nu$ & $\alpha / \nu$ & $\alpha$ & $\nu$
\\
\hline \hline
Onsager solution on regular flat lattice
& 1.75    & 0.125  & 0 & 0 & 1  \\ \hline
Ising spins coupled to gravity$^{\cite{gh,holm}}$
& 1.73(2) & 0.124(3)  & -0.06(11) & - & 0.98(1)  \\ \hline \hline
Matrix model and CFT $^{\cite{kazakov,kpz}}$
& 1.333...  & 0.333... & -0.666... & -1.0 & 1.5  \\ \hline
Random Ising spins in {\it flat} space$^{\cite{ranis}}$
& 1.32(3) & 0.31(4)  & -0.65(4) & -0.98(4) & 1.46(8) \\ \hline
\end{tabular}
\end{center}
\label{TabIsing}
\end{table}
\vskip 10pt

Previously the authors of Ref.~\cite{ranfer} had considered the Dirac equation
on a two-dimensional lattice where sites have been removed randomly
- a doped lattice. They argued that in this case
the fermions acquire a quartic interaction
and become Thirring fermions, thus changing the critical
exponents and the universality class.
In our opinion, the results for simplicial gravity found in ~\cite{gh},
and the conformal field theory exponents of ~\cite{kpz}, can simply be
made consistent with each other, if an additive {\it gravitational}
dressing of critical exponents is taken to be zero in both cases,
for both non-random and random matter (which are known to have
different critical exponents in flat space to begin with).

\vspace{12pt}

{\bf Acknowledgements}

The authors thank Gabriele Veneziano and the Theory Division at CERN
for hospitality during the completion of this paper.
The authors also acknowledge discussions with James Hartle, and
thank him for presenting his ideas on the supermetric.
The work of H.W.H. and R.M.W. was supported in part
by the UK Science and Engineering Research Counsel under grant GR/J64788.

\vfill
\newpage
\end{document}

%% file: plot1.tex
% GNUPLOT: LaTeX picture
\setlength{\unitlength}{0.240900pt}
\ifx\plotpoint\undefined\newsavebox{\plotpoint}\fi
\sbox{\plotpoint}{\rule[-0.175pt]{0.350pt}{0.350pt}}%
\begin{picture}(1650,1320)(0,0)
% rem by hwh \tenrm
\sbox{\plotpoint}{\rule[-0.175pt]{0.350pt}{0.350pt}}%
\put(264,158){\rule[-0.175pt]{4.818pt}{0.350pt}}
\put(242,158){\makebox(0,0)[r]{$0$}}
\put(1566,158){\rule[-0.175pt]{4.818pt}{0.350pt}}
\put(264,333){\rule[-0.175pt]{4.818pt}{0.350pt}}
\put(242,333){\makebox(0,0)[r]{$2000$}}
\put(1566,333){\rule[-0.175pt]{4.818pt}{0.350pt}}
\put(264,508){\rule[-0.175pt]{4.818pt}{0.350pt}}
\put(242,508){\makebox(0,0)[r]{$4000$}}
\put(1566,508){\rule[-0.175pt]{4.818pt}{0.350pt}}
\put(264,683){\rule[-0.175pt]{4.818pt}{0.350pt}}
\put(242,683){\makebox(0,0)[r]{$6000$}}
\put(1566,683){\rule[-0.175pt]{4.818pt}{0.350pt}}
\put(264,857){\rule[-0.175pt]{4.818pt}{0.350pt}}
\put(242,857){\makebox(0,0)[r]{$8000$}}
\put(1566,857){\rule[-0.175pt]{4.818pt}{0.350pt}}
\put(264,1032){\rule[-0.175pt]{4.818pt}{0.350pt}}
\put(242,1032){\makebox(0,0)[r]{$10000$}}
\put(1566,1032){\rule[-0.175pt]{4.818pt}{0.350pt}}
\put(264,1207){\rule[-0.175pt]{4.818pt}{0.350pt}}
\put(242,1207){\makebox(0,0)[r]{$12000$}}
\put(1566,1207){\rule[-0.175pt]{4.818pt}{0.350pt}}
\put(264,158){\rule[-0.175pt]{0.350pt}{4.818pt}}
\put(264,113){\makebox(0,0){$0$}}
\put(264,1187){\rule[-0.175pt]{0.350pt}{4.818pt}}
\put(484,158){\rule[-0.175pt]{0.350pt}{4.818pt}}
\put(484,113){\makebox(0,0){$1$}}
\put(484,1187){\rule[-0.175pt]{0.350pt}{4.818pt}}
\put(705,158){\rule[-0.175pt]{0.350pt}{4.818pt}}
\put(705,113){\makebox(0,0){$2$}}
\put(705,1187){\rule[-0.175pt]{0.350pt}{4.818pt}}
\put(925,158){\rule[-0.175pt]{0.350pt}{4.818pt}}
\put(925,113){\makebox(0,0){$3$}}
\put(925,1187){\rule[-0.175pt]{0.350pt}{4.818pt}}
\put(1145,158){\rule[-0.175pt]{0.350pt}{4.818pt}}
\put(1145,113){\makebox(0,0){$4$}}
\put(1145,1187){\rule[-0.175pt]{0.350pt}{4.818pt}}
\put(1366,158){\rule[-0.175pt]{0.350pt}{4.818pt}}
\put(1366,113){\makebox(0,0){$5$}}
\put(1366,1187){\rule[-0.175pt]{0.350pt}{4.818pt}}
\put(1586,158){\rule[-0.175pt]{0.350pt}{4.818pt}}
\put(1586,113){\makebox(0,0){$6$}}
\put(1586,1187){\rule[-0.175pt]{0.350pt}{4.818pt}}
\put(264,158){\rule[-0.175pt]{318.470pt}{0.350pt}}
\put(1586,158){\rule[-0.175pt]{0.350pt}{252.704pt}}
\put(264,1207){\rule[-0.175pt]{318.470pt}{0.350pt}}
\put(1,682){\makebox(0,0)[l]{\shortstack{ ${\cal P} (l)$ }}}
\put(925,68){\makebox(0,0){ $l$ }}
\put(925,1252){\makebox(0,0){ Edge Length Distribution }}
\put(264,158){\rule[-0.175pt]{0.350pt}{252.704pt}}
\put(265,158){\rule{.1pt}{.1pt}}
\put(268,158){\rule{.1pt}{.1pt}}
\put(271,158){\rule{.1pt}{.1pt}}
\put(273,158){\rule{.1pt}{.1pt}}
\put(276,158){\rule{.1pt}{.1pt}}
\put(279,158){\rule{.1pt}{.1pt}}
\put(281,158){\rule{.1pt}{.1pt}}
\put(284,158){\rule{.1pt}{.1pt}}
\put(286,158){\rule{.1pt}{.1pt}}
\put(289,158){\rule{.1pt}{.1pt}}
\put(292,158){\rule{.1pt}{.1pt}}
\put(294,158){\rule{.1pt}{.1pt}}
\put(297,158){\rule{.1pt}{.1pt}}
\put(300,158){\rule{.1pt}{.1pt}}
\put(302,158){\rule{.1pt}{.1pt}}
\put(305,158){\rule{.1pt}{.1pt}}
\put(308,158){\rule{.1pt}{.1pt}}
\put(310,158){\rule{.1pt}{.1pt}}
\put(313,158){\rule{.1pt}{.1pt}}
\put(316,158){\rule{.1pt}{.1pt}}
\put(318,158){\rule{.1pt}{.1pt}}
\put(321,158){\rule{.1pt}{.1pt}}
\put(323,158){\rule{.1pt}{.1pt}}
\put(326,158){\rule{.1pt}{.1pt}}
\put(329,158){\rule{.1pt}{.1pt}}
\put(331,158){\rule{.1pt}{.1pt}}
\put(334,158){\rule{.1pt}{.1pt}}
\put(337,158){\rule{.1pt}{.1pt}}
\put(339,158){\rule{.1pt}{.1pt}}
\put(342,158){\rule{.1pt}{.1pt}}
\put(345,158){\rule{.1pt}{.1pt}}
\put(347,158){\rule{.1pt}{.1pt}}
\put(350,158){\rule{.1pt}{.1pt}}
\put(353,158){\rule{.1pt}{.1pt}}
\put(355,158){\rule{.1pt}{.1pt}}
\put(358,158){\rule{.1pt}{.1pt}}
\put(361,158){\rule{.1pt}{.1pt}}
\put(363,158){\rule{.1pt}{.1pt}}
\put(366,158){\rule{.1pt}{.1pt}}
\put(368,158){\rule{.1pt}{.1pt}}
\put(371,158){\rule{.1pt}{.1pt}}
\put(374,158){\rule{.1pt}{.1pt}}
\put(376,158){\rule{.1pt}{.1pt}}
\put(379,158){\rule{.1pt}{.1pt}}
\put(382,158){\rule{.1pt}{.1pt}}
\put(384,158){\rule{.1pt}{.1pt}}
\put(387,158){\rule{.1pt}{.1pt}}
\put(390,158){\rule{.1pt}{.1pt}}
\put(392,158){\rule{.1pt}{.1pt}}
\put(395,158){\rule{.1pt}{.1pt}}
\put(398,158){\rule{.1pt}{.1pt}}
\put(400,158){\rule{.1pt}{.1pt}}
\put(403,158){\rule{.1pt}{.1pt}}
\put(405,158){\rule{.1pt}{.1pt}}
\put(408,158){\rule{.1pt}{.1pt}}
\put(411,158){\rule{.1pt}{.1pt}}
\put(413,158){\rule{.1pt}{.1pt}}
\put(416,158){\rule{.1pt}{.1pt}}
\put(419,158){\rule{.1pt}{.1pt}}
\put(421,158){\rule{.1pt}{.1pt}}
\put(424,159){\rule{.1pt}{.1pt}}
\put(427,159){\rule{.1pt}{.1pt}}
\put(429,159){\rule{.1pt}{.1pt}}
\put(432,159){\rule{.1pt}{.1pt}}
\put(435,159){\rule{.1pt}{.1pt}}
\put(437,159){\rule{.1pt}{.1pt}}
\put(440,159){\rule{.1pt}{.1pt}}
\put(442,160){\rule{.1pt}{.1pt}}
\put(445,160){\rule{.1pt}{.1pt}}
\put(448,160){\rule{.1pt}{.1pt}}
\put(450,161){\rule{.1pt}{.1pt}}
\put(453,161){\rule{.1pt}{.1pt}}
\put(456,162){\rule{.1pt}{.1pt}}
\put(458,161){\rule{.1pt}{.1pt}}
\put(461,163){\rule{.1pt}{.1pt}}
\put(464,162){\rule{.1pt}{.1pt}}
\put(466,164){\rule{.1pt}{.1pt}}
\put(469,164){\rule{.1pt}{.1pt}}
\put(472,165){\rule{.1pt}{.1pt}}
\put(474,165){\rule{.1pt}{.1pt}}
\put(477,168){\rule{.1pt}{.1pt}}
\put(479,168){\rule{.1pt}{.1pt}}
\put(482,171){\rule{.1pt}{.1pt}}
\put(485,174){\rule{.1pt}{.1pt}}
\put(487,175){\rule{.1pt}{.1pt}}
\put(490,176){\rule{.1pt}{.1pt}}
\put(493,177){\rule{.1pt}{.1pt}}
\put(495,181){\rule{.1pt}{.1pt}}
\put(498,181){\rule{.1pt}{.1pt}}
\put(501,186){\rule{.1pt}{.1pt}}
\put(503,191){\rule{.1pt}{.1pt}}
\put(506,193){\rule{.1pt}{.1pt}}
\put(509,196){\rule{.1pt}{.1pt}}
\put(511,200){\rule{.1pt}{.1pt}}
\put(514,200){\rule{.1pt}{.1pt}}
\put(517,207){\rule{.1pt}{.1pt}}
\put(519,215){\rule{.1pt}{.1pt}}
\put(522,215){\rule{.1pt}{.1pt}}
\put(524,221){\rule{.1pt}{.1pt}}
\put(527,222){\rule{.1pt}{.1pt}}
\put(530,232){\rule{.1pt}{.1pt}}
\put(532,239){\rule{.1pt}{.1pt}}
\put(535,245){\rule{.1pt}{.1pt}}
\put(538,254){\rule{.1pt}{.1pt}}
\put(540,255){\rule{.1pt}{.1pt}}
\put(543,264){\rule{.1pt}{.1pt}}
\put(546,269){\rule{.1pt}{.1pt}}
\put(548,279){\rule{.1pt}{.1pt}}
\put(551,285){\rule{.1pt}{.1pt}}
\put(554,299){\rule{.1pt}{.1pt}}
\put(556,310){\rule{.1pt}{.1pt}}
\put(559,319){\rule{.1pt}{.1pt}}
\put(561,323){\rule{.1pt}{.1pt}}
\put(564,331){\rule{.1pt}{.1pt}}
\put(567,340){\rule{.1pt}{.1pt}}
\put(569,356){\rule{.1pt}{.1pt}}
\put(572,368){\rule{.1pt}{.1pt}}
\put(575,375){\rule{.1pt}{.1pt}}
\put(577,396){\rule{.1pt}{.1pt}}
\put(580,405){\rule{.1pt}{.1pt}}
\put(583,418){\rule{.1pt}{.1pt}}
\put(585,433){\rule{.1pt}{.1pt}}
\put(588,441){\rule{.1pt}{.1pt}}
\put(591,456){\rule{.1pt}{.1pt}}
\put(593,475){\rule{.1pt}{.1pt}}
\put(596,475){\rule{.1pt}{.1pt}}
\put(598,504){\rule{.1pt}{.1pt}}
\put(601,512){\rule{.1pt}{.1pt}}
\put(604,531){\rule{.1pt}{.1pt}}
\put(606,546){\rule{.1pt}{.1pt}}
\put(609,562){\rule{.1pt}{.1pt}}
\put(612,571){\rule{.1pt}{.1pt}}
\put(614,576){\rule{.1pt}{.1pt}}
\put(617,603){\rule{.1pt}{.1pt}}
\put(620,616){\rule{.1pt}{.1pt}}
\put(622,628){\rule{.1pt}{.1pt}}
\put(625,649){\rule{.1pt}{.1pt}}
\put(628,655){\rule{.1pt}{.1pt}}
\put(630,668){\rule{.1pt}{.1pt}}
\put(633,693){\rule{.1pt}{.1pt}}
\put(635,700){\rule{.1pt}{.1pt}}
\put(638,723){\rule{.1pt}{.1pt}}
\put(641,738){\rule{.1pt}{.1pt}}
\put(643,747){\rule{.1pt}{.1pt}}
\put(646,770){\rule{.1pt}{.1pt}}
\put(649,774){\rule{.1pt}{.1pt}}
\put(651,786){\rule{.1pt}{.1pt}}
\put(654,793){\rule{.1pt}{.1pt}}
\put(657,817){\rule{.1pt}{.1pt}}
\put(659,840){\rule{.1pt}{.1pt}}
\put(662,853){\rule{.1pt}{.1pt}}
\put(665,863){\rule{.1pt}{.1pt}}
\put(667,874){\rule{.1pt}{.1pt}}
\put(670,882){\rule{.1pt}{.1pt}}
\put(672,897){\rule{.1pt}{.1pt}}
\put(675,911){\rule{.1pt}{.1pt}}
\put(678,914){\rule{.1pt}{.1pt}}
\put(680,917){\rule{.1pt}{.1pt}}
\put(683,936){\rule{.1pt}{.1pt}}
\put(686,944){\rule{.1pt}{.1pt}}
\put(688,942){\rule{.1pt}{.1pt}}
\put(691,955){\rule{.1pt}{.1pt}}
\put(694,974){\rule{.1pt}{.1pt}}
\put(696,970){\rule{.1pt}{.1pt}}
\put(699,965){\rule{.1pt}{.1pt}}
\put(702,996){\rule{.1pt}{.1pt}}
\put(704,987){\rule{.1pt}{.1pt}}
\put(707,1002){\rule{.1pt}{.1pt}}
\put(710,1006){\rule{.1pt}{.1pt}}
\put(712,1004){\rule{.1pt}{.1pt}}
\put(715,1018){\rule{.1pt}{.1pt}}
\put(717,1004){\rule{.1pt}{.1pt}}
\put(720,1020){\rule{.1pt}{.1pt}}
\put(723,1017){\rule{.1pt}{.1pt}}
\put(725,1014){\rule{.1pt}{.1pt}}
\put(728,1022){\rule{.1pt}{.1pt}}
\put(731,1019){\rule{.1pt}{.1pt}}
\put(733,1040){\rule{.1pt}{.1pt}}
\put(736,1012){\rule{.1pt}{.1pt}}
\put(739,1023){\rule{.1pt}{.1pt}}
\put(741,1027){\rule{.1pt}{.1pt}}
\put(744,1021){\rule{.1pt}{.1pt}}
\put(747,1021){\rule{.1pt}{.1pt}}
\put(749,1019){\rule{.1pt}{.1pt}}
\put(752,1023){\rule{.1pt}{.1pt}}
\put(754,996){\rule{.1pt}{.1pt}}
\put(757,993){\rule{.1pt}{.1pt}}
\put(760,992){\rule{.1pt}{.1pt}}
\put(762,970){\rule{.1pt}{.1pt}}
\put(765,998){\rule{.1pt}{.1pt}}
\put(768,968){\rule{.1pt}{.1pt}}
\put(770,976){\rule{.1pt}{.1pt}}
\put(773,972){\rule{.1pt}{.1pt}}
\put(776,957){\rule{.1pt}{.1pt}}
\put(778,937){\rule{.1pt}{.1pt}}
\put(781,936){\rule{.1pt}{.1pt}}
\put(784,918){\rule{.1pt}{.1pt}}
\put(786,921){\rule{.1pt}{.1pt}}
\put(789,918){\rule{.1pt}{.1pt}}
\put(791,899){\rule{.1pt}{.1pt}}
\put(794,886){\rule{.1pt}{.1pt}}
\put(797,904){\rule{.1pt}{.1pt}}
\put(799,881){\rule{.1pt}{.1pt}}
\put(802,857){\rule{.1pt}{.1pt}}
\put(805,841){\rule{.1pt}{.1pt}}
\put(807,836){\rule{.1pt}{.1pt}}
\put(810,833){\rule{.1pt}{.1pt}}
\put(813,815){\rule{.1pt}{.1pt}}
\put(815,804){\rule{.1pt}{.1pt}}
\put(818,803){\rule{.1pt}{.1pt}}
\put(821,780){\rule{.1pt}{.1pt}}
\put(823,763){\rule{.1pt}{.1pt}}
\put(826,769){\rule{.1pt}{.1pt}}
\put(828,747){\rule{.1pt}{.1pt}}
\put(831,754){\rule{.1pt}{.1pt}}
\put(834,725){\rule{.1pt}{.1pt}}
\put(836,703){\rule{.1pt}{.1pt}}
\put(839,695){\rule{.1pt}{.1pt}}
\put(842,684){\rule{.1pt}{.1pt}}
\put(844,664){\rule{.1pt}{.1pt}}
\put(847,648){\rule{.1pt}{.1pt}}
\put(850,646){\rule{.1pt}{.1pt}}
\put(852,643){\rule{.1pt}{.1pt}}
\put(855,614){\rule{.1pt}{.1pt}}
\put(858,615){\rule{.1pt}{.1pt}}
\put(860,613){\rule{.1pt}{.1pt}}
\put(863,586){\rule{.1pt}{.1pt}}
\put(866,566){\rule{.1pt}{.1pt}}
\put(868,574){\rule{.1pt}{.1pt}}
\put(871,560){\rule{.1pt}{.1pt}}
\put(873,550){\rule{.1pt}{.1pt}}
\put(876,544){\rule{.1pt}{.1pt}}
\put(879,528){\rule{.1pt}{.1pt}}
\put(881,520){\rule{.1pt}{.1pt}}
\put(884,502){\rule{.1pt}{.1pt}}
\put(887,492){\rule{.1pt}{.1pt}}
\put(889,482){\rule{.1pt}{.1pt}}
\put(892,480){\rule{.1pt}{.1pt}}
\put(895,459){\rule{.1pt}{.1pt}}
\put(897,447){\rule{.1pt}{.1pt}}
\put(900,447){\rule{.1pt}{.1pt}}
\put(903,431){\rule{.1pt}{.1pt}}
\put(905,425){\rule{.1pt}{.1pt}}
\put(908,413){\rule{.1pt}{.1pt}}
\put(910,404){\rule{.1pt}{.1pt}}
\put(913,393){\rule{.1pt}{.1pt}}
\put(916,398){\rule{.1pt}{.1pt}}
\put(918,383){\rule{.1pt}{.1pt}}
\put(921,368){\rule{.1pt}{.1pt}}
\put(924,368){\rule{.1pt}{.1pt}}
\put(926,361){\rule{.1pt}{.1pt}}
\put(929,359){\rule{.1pt}{.1pt}}
\put(932,350){\rule{.1pt}{.1pt}}
\put(934,337){\rule{.1pt}{.1pt}}
\put(937,339){\rule{.1pt}{.1pt}}
\put(940,332){\rule{.1pt}{.1pt}}
\put(942,319){\rule{.1pt}{.1pt}}
\put(945,311){\rule{.1pt}{.1pt}}
\put(947,306){\rule{.1pt}{.1pt}}
\put(950,309){\rule{.1pt}{.1pt}}
\put(953,292){\rule{.1pt}{.1pt}}
\put(955,288){\rule{.1pt}{.1pt}}
\put(958,289){\rule{.1pt}{.1pt}}
\put(961,280){\rule{.1pt}{.1pt}}
\put(963,276){\rule{.1pt}{.1pt}}
\put(966,268){\rule{.1pt}{.1pt}}
\put(969,263){\rule{.1pt}{.1pt}}
\put(971,258){\rule{.1pt}{.1pt}}
\put(974,258){\rule{.1pt}{.1pt}}
\put(977,251){\rule{.1pt}{.1pt}}
\put(979,241){\rule{.1pt}{.1pt}}
\put(982,245){\rule{.1pt}{.1pt}}
\put(984,239){\rule{.1pt}{.1pt}}
\put(987,239){\rule{.1pt}{.1pt}}
\put(990,234){\rule{.1pt}{.1pt}}
\put(992,226){\rule{.1pt}{.1pt}}
\put(995,231){\rule{.1pt}{.1pt}}
\put(998,223){\rule{.1pt}{.1pt}}
\put(1000,223){\rule{.1pt}{.1pt}}
\put(1003,222){\rule{.1pt}{.1pt}}
\put(1006,209){\rule{.1pt}{.1pt}}
\put(1008,212){\rule{.1pt}{.1pt}}
\put(1011,210){\rule{.1pt}{.1pt}}
\put(1014,211){\rule{.1pt}{.1pt}}
\put(1016,209){\rule{.1pt}{.1pt}}
\put(1019,204){\rule{.1pt}{.1pt}}
\put(1022,203){\rule{.1pt}{.1pt}}
\put(1024,200){\rule{.1pt}{.1pt}}
\put(1027,198){\rule{.1pt}{.1pt}}
\put(1029,198){\rule{.1pt}{.1pt}}
\put(1032,195){\rule{.1pt}{.1pt}}
\put(1035,191){\rule{.1pt}{.1pt}}
\put(1037,191){\rule{.1pt}{.1pt}}
\put(1040,186){\rule{.1pt}{.1pt}}
\put(1043,188){\rule{.1pt}{.1pt}}
\put(1045,187){\rule{.1pt}{.1pt}}
\put(1048,181){\rule{.1pt}{.1pt}}
\put(1051,185){\rule{.1pt}{.1pt}}
\put(1053,180){\rule{.1pt}{.1pt}}
\put(1056,181){\rule{.1pt}{.1pt}}
\put(1059,179){\rule{.1pt}{.1pt}}
\put(1061,177){\rule{.1pt}{.1pt}}
\put(1064,177){\rule{.1pt}{.1pt}}
\put(1066,179){\rule{.1pt}{.1pt}}
\put(1069,175){\rule{.1pt}{.1pt}}
\put(1072,173){\rule{.1pt}{.1pt}}
\put(1074,173){\rule{.1pt}{.1pt}}
\put(1077,173){\rule{.1pt}{.1pt}}
\put(1080,174){\rule{.1pt}{.1pt}}
\put(1082,173){\rule{.1pt}{.1pt}}
\put(1085,172){\rule{.1pt}{.1pt}}
\put(1088,171){\rule{.1pt}{.1pt}}
\put(1090,168){\rule{.1pt}{.1pt}}
\put(1093,169){\rule{.1pt}{.1pt}}
\put(1096,168){\rule{.1pt}{.1pt}}
\put(1098,167){\rule{.1pt}{.1pt}}
\put(1101,166){\rule{.1pt}{.1pt}}
\put(1103,167){\rule{.1pt}{.1pt}}
\put(1106,166){\rule{.1pt}{.1pt}}
\put(1109,165){\rule{.1pt}{.1pt}}
\put(1111,166){\rule{.1pt}{.1pt}}
\put(1114,165){\rule{.1pt}{.1pt}}
\put(1117,165){\rule{.1pt}{.1pt}}
\put(1119,164){\rule{.1pt}{.1pt}}
\put(1122,162){\rule{.1pt}{.1pt}}
\put(1125,164){\rule{.1pt}{.1pt}}
\put(1127,162){\rule{.1pt}{.1pt}}
\put(1130,162){\rule{.1pt}{.1pt}}
\put(1133,162){\rule{.1pt}{.1pt}}
\put(1135,163){\rule{.1pt}{.1pt}}
\put(1138,162){\rule{.1pt}{.1pt}}
\put(1140,162){\rule{.1pt}{.1pt}}
\put(1143,161){\rule{.1pt}{.1pt}}
\put(1146,161){\rule{.1pt}{.1pt}}
\put(1148,160){\rule{.1pt}{.1pt}}
\put(1151,161){\rule{.1pt}{.1pt}}
\put(1154,161){\rule{.1pt}{.1pt}}
\put(1156,161){\rule{.1pt}{.1pt}}
\put(1159,160){\rule{.1pt}{.1pt}}
\put(1162,161){\rule{.1pt}{.1pt}}
\put(1164,161){\rule{.1pt}{.1pt}}
\put(1167,160){\rule{.1pt}{.1pt}}
\put(1170,160){\rule{.1pt}{.1pt}}
\put(1172,160){\rule{.1pt}{.1pt}}
\put(1175,160){\rule{.1pt}{.1pt}}
\put(1178,159){\rule{.1pt}{.1pt}}
\put(1180,160){\rule{.1pt}{.1pt}}
\put(1183,159){\rule{.1pt}{.1pt}}
\put(1185,160){\rule{.1pt}{.1pt}}
\put(1188,159){\rule{.1pt}{.1pt}}
\put(1191,159){\rule{.1pt}{.1pt}}
\put(1193,160){\rule{.1pt}{.1pt}}
\put(1196,158){\rule{.1pt}{.1pt}}
\put(1199,159){\rule{.1pt}{.1pt}}
\put(1201,159){\rule{.1pt}{.1pt}}
\put(1204,159){\rule{.1pt}{.1pt}}
\put(1207,159){\rule{.1pt}{.1pt}}
\put(1209,159){\rule{.1pt}{.1pt}}
\put(1212,159){\rule{.1pt}{.1pt}}
\put(1215,159){\rule{.1pt}{.1pt}}
\put(1217,159){\rule{.1pt}{.1pt}}
\put(1220,159){\rule{.1pt}{.1pt}}
\put(1222,158){\rule{.1pt}{.1pt}}
\put(1225,159){\rule{.1pt}{.1pt}}
\put(1228,159){\rule{.1pt}{.1pt}}
\put(1230,158){\rule{.1pt}{.1pt}}
\put(1233,159){\rule{.1pt}{.1pt}}
\put(1236,159){\rule{.1pt}{.1pt}}
\put(1238,159){\rule{.1pt}{.1pt}}
\put(1241,159){\rule{.1pt}{.1pt}}
\put(1244,158){\rule{.1pt}{.1pt}}
\put(1246,158){\rule{.1pt}{.1pt}}
\put(1249,158){\rule{.1pt}{.1pt}}
\put(1252,158){\rule{.1pt}{.1pt}}
\put(1254,158){\rule{.1pt}{.1pt}}
\put(1257,158){\rule{.1pt}{.1pt}}
\put(1259,158){\rule{.1pt}{.1pt}}
\put(1262,158){\rule{.1pt}{.1pt}}
\put(1265,158){\rule{.1pt}{.1pt}}
\put(1267,158){\rule{.1pt}{.1pt}}
\put(1270,158){\rule{.1pt}{.1pt}}
\put(1273,158){\rule{.1pt}{.1pt}}
\put(1275,158){\rule{.1pt}{.1pt}}
\put(1278,158){\rule{.1pt}{.1pt}}
\put(1281,158){\rule{.1pt}{.1pt}}
\put(1283,158){\rule{.1pt}{.1pt}}
\put(1286,158){\rule{.1pt}{.1pt}}
\put(1289,158){\rule{.1pt}{.1pt}}
\put(1291,158){\rule{.1pt}{.1pt}}
\put(1294,158){\rule{.1pt}{.1pt}}
\put(1296,158){\rule{.1pt}{.1pt}}
\put(1299,158){\rule{.1pt}{.1pt}}
\put(1302,158){\rule{.1pt}{.1pt}}
\put(1304,158){\rule{.1pt}{.1pt}}
\put(1307,158){\rule{.1pt}{.1pt}}
\put(1310,158){\rule{.1pt}{.1pt}}
\put(1312,158){\rule{.1pt}{.1pt}}
\put(1315,158){\rule{.1pt}{.1pt}}
\put(1318,158){\rule{.1pt}{.1pt}}
\put(1320,158){\rule{.1pt}{.1pt}}
\put(1323,158){\rule{.1pt}{.1pt}}
\put(1326,158){\rule{.1pt}{.1pt}}
\put(1328,158){\rule{.1pt}{.1pt}}
\put(1331,158){\rule{.1pt}{.1pt}}
\put(1333,158){\rule{.1pt}{.1pt}}
\put(1336,158){\rule{.1pt}{.1pt}}
\put(1339,158){\rule{.1pt}{.1pt}}
\put(1341,158){\rule{.1pt}{.1pt}}
\put(1344,158){\rule{.1pt}{.1pt}}
\put(1347,158){\rule{.1pt}{.1pt}}
\put(1349,158){\rule{.1pt}{.1pt}}
\put(1352,158){\rule{.1pt}{.1pt}}
\put(1355,158){\rule{.1pt}{.1pt}}
\put(1357,158){\rule{.1pt}{.1pt}}
\put(1360,158){\rule{.1pt}{.1pt}}
\put(1363,158){\rule{.1pt}{.1pt}}
\put(1365,158){\rule{.1pt}{.1pt}}
\put(1368,158){\rule{.1pt}{.1pt}}
\put(1371,158){\rule{.1pt}{.1pt}}
\put(1373,158){\rule{.1pt}{.1pt}}
\put(1376,158){\rule{.1pt}{.1pt}}
\put(1378,158){\rule{.1pt}{.1pt}}
\put(1381,158){\rule{.1pt}{.1pt}}
\put(1384,158){\rule{.1pt}{.1pt}}
\put(1386,158){\rule{.1pt}{.1pt}}
\put(1389,158){\rule{.1pt}{.1pt}}
\put(1392,158){\rule{.1pt}{.1pt}}
\put(1394,158){\rule{.1pt}{.1pt}}
\put(1397,158){\rule{.1pt}{.1pt}}
\put(1400,158){\rule{.1pt}{.1pt}}
\put(1402,158){\rule{.1pt}{.1pt}}
\put(1405,158){\rule{.1pt}{.1pt}}
\put(1408,158){\rule{.1pt}{.1pt}}
\put(1410,158){\rule{.1pt}{.1pt}}
\put(1413,158){\rule{.1pt}{.1pt}}
\put(1415,158){\rule{.1pt}{.1pt}}
\put(1418,158){\rule{.1pt}{.1pt}}
\put(1421,158){\rule{.1pt}{.1pt}}
\put(1423,158){\rule{.1pt}{.1pt}}
\put(1426,158){\rule{.1pt}{.1pt}}
\put(1429,158){\rule{.1pt}{.1pt}}
\put(1431,158){\rule{.1pt}{.1pt}}
\put(1434,158){\rule{.1pt}{.1pt}}
\put(1437,158){\rule{.1pt}{.1pt}}
\put(1439,158){\rule{.1pt}{.1pt}}
\put(1442,158){\rule{.1pt}{.1pt}}
\put(1445,158){\rule{.1pt}{.1pt}}
\put(1447,158){\rule{.1pt}{.1pt}}
\put(1450,158){\rule{.1pt}{.1pt}}
\put(1452,158){\rule{.1pt}{.1pt}}
\put(1455,158){\rule{.1pt}{.1pt}}
\put(1458,158){\rule{.1pt}{.1pt}}
\put(1460,158){\rule{.1pt}{.1pt}}
\put(1463,158){\rule{.1pt}{.1pt}}
\put(1466,158){\rule{.1pt}{.1pt}}
\put(1468,158){\rule{.1pt}{.1pt}}
\put(1471,158){\rule{.1pt}{.1pt}}
\put(1474,158){\rule{.1pt}{.1pt}}
\put(1476,158){\rule{.1pt}{.1pt}}
\put(1479,158){\rule{.1pt}{.1pt}}
\put(1482,158){\rule{.1pt}{.1pt}}
\put(1484,158){\rule{.1pt}{.1pt}}
\put(1487,158){\rule{.1pt}{.1pt}}
\put(1489,158){\rule{.1pt}{.1pt}}
\put(1492,158){\rule{.1pt}{.1pt}}
\put(1495,158){\rule{.1pt}{.1pt}}
\put(1497,158){\rule{.1pt}{.1pt}}
\put(1500,158){\rule{.1pt}{.1pt}}
\put(1503,158){\rule{.1pt}{.1pt}}
\put(1505,158){\rule{.1pt}{.1pt}}
\put(1508,158){\rule{.1pt}{.1pt}}
\put(1511,158){\rule{.1pt}{.1pt}}
\put(1513,158){\rule{.1pt}{.1pt}}
\put(1516,158){\rule{.1pt}{.1pt}}
\put(1519,158){\rule{.1pt}{.1pt}}
\put(1521,158){\rule{.1pt}{.1pt}}
\put(1524,158){\rule{.1pt}{.1pt}}
\put(1527,158){\rule{.1pt}{.1pt}}
\put(1529,158){\rule{.1pt}{.1pt}}
\put(1532,158){\rule{.1pt}{.1pt}}
\put(1534,158){\rule{.1pt}{.1pt}}
\put(1537,158){\rule{.1pt}{.1pt}}
\put(1540,158){\rule{.1pt}{.1pt}}
\put(1542,158){\rule{.1pt}{.1pt}}
\put(1545,158){\rule{.1pt}{.1pt}}
\put(1548,158){\rule{.1pt}{.1pt}}
\put(1550,158){\rule{.1pt}{.1pt}}
\put(1553,158){\rule{.1pt}{.1pt}}
\put(1556,158){\rule{.1pt}{.1pt}}
\put(1558,158){\rule{.1pt}{.1pt}}
\put(1561,158){\rule{.1pt}{.1pt}}
\put(1564,158){\rule{.1pt}{.1pt}}
\put(1566,158){\rule{.1pt}{.1pt}}
\put(1569,158){\rule{.1pt}{.1pt}}
\put(1571,158){\rule{.1pt}{.1pt}}
\put(1574,158){\rule{.1pt}{.1pt}}
\put(1577,158){\rule{.1pt}{.1pt}}
\put(1579,158){\rule{.1pt}{.1pt}}
\put(1582,158){\rule{.1pt}{.1pt}}
\put(1585,158){\rule{.1pt}{.1pt}}
\sbox{\plotpoint}{\rule[-0.350pt]{0.700pt}{0.700pt}}%
\put(271,158){\usebox{\plotpoint}}
\put(271,158){\rule[-0.350pt]{35.171pt}{0.700pt}}
\put(417,159){\rule[-0.350pt]{4.818pt}{0.700pt}}
\put(437,160){\rule[-0.350pt]{3.132pt}{0.700pt}}
\put(450,161){\rule[-0.350pt]{1.686pt}{0.700pt}}
\put(457,162){\rule[-0.350pt]{0.723pt}{0.700pt}}
\put(460,163){\rule[-0.350pt]{0.723pt}{0.700pt}}
\put(463,164){\rule[-0.350pt]{0.843pt}{0.700pt}}
\put(466,165){\rule[-0.350pt]{0.843pt}{0.700pt}}
\put(470,166){\rule[-0.350pt]{0.843pt}{0.700pt}}
\put(473,167){\rule[-0.350pt]{0.843pt}{0.700pt}}
\put(477,168){\usebox{\plotpoint}}
\put(479,169){\usebox{\plotpoint}}
\put(481,170){\usebox{\plotpoint}}
\put(483,171){\usebox{\plotpoint}}
\put(484,172){\usebox{\plotpoint}}
\put(485,173){\usebox{\plotpoint}}
\put(487,174){\usebox{\plotpoint}}
\put(488,175){\usebox{\plotpoint}}
\put(489,176){\usebox{\plotpoint}}
\put(491,177){\usebox{\plotpoint}}
\put(492,178){\usebox{\plotpoint}}
\put(494,179){\usebox{\plotpoint}}
\put(495,180){\usebox{\plotpoint}}
\put(496,181){\usebox{\plotpoint}}
\put(498,182){\usebox{\plotpoint}}
\put(499,183){\usebox{\plotpoint}}
\put(500,184){\usebox{\plotpoint}}
\put(501,185){\usebox{\plotpoint}}
\put(502,186){\usebox{\plotpoint}}
\put(503,187){\usebox{\plotpoint}}
\put(504,188){\usebox{\plotpoint}}
\put(505,189){\usebox{\plotpoint}}
\put(506,190){\usebox{\plotpoint}}
\put(507,191){\usebox{\plotpoint}}
\put(508,192){\usebox{\plotpoint}}
\put(509,193){\usebox{\plotpoint}}
\put(510,194){\usebox{\plotpoint}}
\put(511,195){\usebox{\plotpoint}}
\put(512,197){\usebox{\plotpoint}}
\put(513,199){\usebox{\plotpoint}}
\put(514,200){\usebox{\plotpoint}}
\put(515,202){\usebox{\plotpoint}}
\put(516,204){\usebox{\plotpoint}}
\put(517,205){\usebox{\plotpoint}}
\put(518,207){\usebox{\plotpoint}}
\put(519,208){\usebox{\plotpoint}}
\put(520,210){\usebox{\plotpoint}}
\put(521,211){\usebox{\plotpoint}}
\put(522,213){\usebox{\plotpoint}}
\put(523,214){\usebox{\plotpoint}}
\put(524,216){\usebox{\plotpoint}}
\put(525,218){\usebox{\plotpoint}}
\put(526,220){\usebox{\plotpoint}}
\put(527,222){\usebox{\plotpoint}}
\put(528,224){\usebox{\plotpoint}}
\put(529,226){\usebox{\plotpoint}}
\put(530,228){\usebox{\plotpoint}}
\put(531,230){\usebox{\plotpoint}}
\put(532,233){\usebox{\plotpoint}}
\put(533,236){\usebox{\plotpoint}}
\put(534,238){\usebox{\plotpoint}}
\put(535,241){\usebox{\plotpoint}}
\put(536,244){\usebox{\plotpoint}}
\put(537,246){\usebox{\plotpoint}}
\put(538,249){\usebox{\plotpoint}}
\put(539,251){\usebox{\plotpoint}}
\put(540,254){\usebox{\plotpoint}}
\put(541,256){\usebox{\plotpoint}}
\put(542,259){\usebox{\plotpoint}}
\put(543,262){\rule[-0.350pt]{0.700pt}{0.723pt}}
\put(544,265){\rule[-0.350pt]{0.700pt}{0.723pt}}
\put(545,268){\rule[-0.350pt]{0.700pt}{0.723pt}}
\put(546,271){\rule[-0.350pt]{0.700pt}{0.723pt}}
\put(547,274){\rule[-0.350pt]{0.700pt}{0.723pt}}
\put(548,277){\rule[-0.350pt]{0.700pt}{0.723pt}}
\put(549,280){\rule[-0.350pt]{0.700pt}{0.723pt}}
\put(550,283){\rule[-0.350pt]{0.700pt}{0.964pt}}
\put(551,287){\rule[-0.350pt]{0.700pt}{0.964pt}}
\put(552,291){\rule[-0.350pt]{0.700pt}{0.964pt}}
\put(553,295){\rule[-0.350pt]{0.700pt}{0.964pt}}
\put(554,299){\rule[-0.350pt]{0.700pt}{0.964pt}}
\put(555,303){\rule[-0.350pt]{0.700pt}{0.964pt}}
\put(556,307){\rule[-0.350pt]{0.700pt}{0.929pt}}
\put(557,310){\rule[-0.350pt]{0.700pt}{0.929pt}}
\put(558,314){\rule[-0.350pt]{0.700pt}{0.929pt}}
\put(559,318){\rule[-0.350pt]{0.700pt}{0.929pt}}
\put(560,322){\rule[-0.350pt]{0.700pt}{0.929pt}}
\put(561,326){\rule[-0.350pt]{0.700pt}{0.929pt}}
\put(562,330){\rule[-0.350pt]{0.700pt}{0.929pt}}
\put(563,334){\rule[-0.350pt]{0.700pt}{0.964pt}}
\put(564,338){\rule[-0.350pt]{0.700pt}{0.964pt}}
\put(565,342){\rule[-0.350pt]{0.700pt}{0.964pt}}
\put(566,346){\rule[-0.350pt]{0.700pt}{0.964pt}}
\put(567,350){\rule[-0.350pt]{0.700pt}{0.964pt}}
\put(568,354){\rule[-0.350pt]{0.700pt}{0.964pt}}
\put(569,358){\rule[-0.350pt]{0.700pt}{0.964pt}}
\put(570,362){\rule[-0.350pt]{0.700pt}{1.164pt}}
\put(571,366){\rule[-0.350pt]{0.700pt}{1.164pt}}
\put(572,371){\rule[-0.350pt]{0.700pt}{1.164pt}}
\put(573,376){\rule[-0.350pt]{0.700pt}{1.164pt}}
\put(574,381){\rule[-0.350pt]{0.700pt}{1.164pt}}
\put(575,386){\rule[-0.350pt]{0.700pt}{1.164pt}}
\put(576,391){\rule[-0.350pt]{0.700pt}{1.067pt}}
\put(577,395){\rule[-0.350pt]{0.700pt}{1.067pt}}
\put(578,399){\rule[-0.350pt]{0.700pt}{1.067pt}}
\put(579,404){\rule[-0.350pt]{0.700pt}{1.067pt}}
\put(580,408){\rule[-0.350pt]{0.700pt}{1.067pt}}
\put(581,413){\rule[-0.350pt]{0.700pt}{1.067pt}}
\put(582,417){\rule[-0.350pt]{0.700pt}{1.067pt}}
\put(583,421){\rule[-0.350pt]{0.700pt}{1.136pt}}
\put(584,426){\rule[-0.350pt]{0.700pt}{1.136pt}}
\put(585,431){\rule[-0.350pt]{0.700pt}{1.136pt}}
\put(586,436){\rule[-0.350pt]{0.700pt}{1.136pt}}
\put(587,440){\rule[-0.350pt]{0.700pt}{1.136pt}}
\put(588,445){\rule[-0.350pt]{0.700pt}{1.136pt}}
\put(589,450){\rule[-0.350pt]{0.700pt}{1.136pt}}
\put(590,455){\rule[-0.350pt]{0.700pt}{1.365pt}}
\put(591,460){\rule[-0.350pt]{0.700pt}{1.365pt}}
\put(592,466){\rule[-0.350pt]{0.700pt}{1.365pt}}
\put(593,471){\rule[-0.350pt]{0.700pt}{1.365pt}}
\put(594,477){\rule[-0.350pt]{0.700pt}{1.365pt}}
\put(595,483){\rule[-0.350pt]{0.700pt}{1.365pt}}
\put(596,488){\rule[-0.350pt]{0.700pt}{1.205pt}}
\put(597,494){\rule[-0.350pt]{0.700pt}{1.204pt}}
\put(598,499){\rule[-0.350pt]{0.700pt}{1.204pt}}
\put(599,504){\rule[-0.350pt]{0.700pt}{1.204pt}}
\put(600,509){\rule[-0.350pt]{0.700pt}{1.204pt}}
\put(601,514){\rule[-0.350pt]{0.700pt}{1.204pt}}
\put(602,519){\rule[-0.350pt]{0.700pt}{1.204pt}}
\put(603,524){\rule[-0.350pt]{0.700pt}{1.445pt}}
\put(604,530){\rule[-0.350pt]{0.700pt}{1.445pt}}
\put(605,536){\rule[-0.350pt]{0.700pt}{1.445pt}}
\put(606,542){\rule[-0.350pt]{0.700pt}{1.445pt}}
\put(607,548){\rule[-0.350pt]{0.700pt}{1.445pt}}
\put(608,554){\rule[-0.350pt]{0.700pt}{1.445pt}}
\put(609,560){\rule[-0.350pt]{0.700pt}{1.273pt}}
\put(610,565){\rule[-0.350pt]{0.700pt}{1.273pt}}
\put(611,570){\rule[-0.350pt]{0.700pt}{1.273pt}}
\put(612,575){\rule[-0.350pt]{0.700pt}{1.273pt}}
\put(613,581){\rule[-0.350pt]{0.700pt}{1.273pt}}
\put(614,586){\rule[-0.350pt]{0.700pt}{1.273pt}}
\put(615,591){\rule[-0.350pt]{0.700pt}{1.273pt}}
\put(616,596){\rule[-0.350pt]{0.700pt}{1.273pt}}
\put(617,602){\rule[-0.350pt]{0.700pt}{1.273pt}}
\put(618,607){\rule[-0.350pt]{0.700pt}{1.273pt}}
\put(619,612){\rule[-0.350pt]{0.700pt}{1.273pt}}
\put(620,618){\rule[-0.350pt]{0.700pt}{1.273pt}}
\put(621,623){\rule[-0.350pt]{0.700pt}{1.273pt}}
\put(622,628){\rule[-0.350pt]{0.700pt}{1.273pt}}
\put(623,633){\rule[-0.350pt]{0.700pt}{1.486pt}}
\put(624,640){\rule[-0.350pt]{0.700pt}{1.486pt}}
\put(625,646){\rule[-0.350pt]{0.700pt}{1.486pt}}
\put(626,652){\rule[-0.350pt]{0.700pt}{1.486pt}}
\put(627,658){\rule[-0.350pt]{0.700pt}{1.486pt}}
\put(628,664){\rule[-0.350pt]{0.700pt}{1.486pt}}
\put(629,671){\rule[-0.350pt]{0.700pt}{1.308pt}}
\put(630,676){\rule[-0.350pt]{0.700pt}{1.308pt}}
\put(631,681){\rule[-0.350pt]{0.700pt}{1.308pt}}
\put(632,687){\rule[-0.350pt]{0.700pt}{1.308pt}}
\put(633,692){\rule[-0.350pt]{0.700pt}{1.308pt}}
\put(634,698){\rule[-0.350pt]{0.700pt}{1.308pt}}
\put(635,703){\rule[-0.350pt]{0.700pt}{1.308pt}}
\put(636,709){\rule[-0.350pt]{0.700pt}{1.239pt}}
\put(637,714){\rule[-0.350pt]{0.700pt}{1.239pt}}
\put(638,719){\rule[-0.350pt]{0.700pt}{1.239pt}}
\put(639,724){\rule[-0.350pt]{0.700pt}{1.239pt}}
\put(640,729){\rule[-0.350pt]{0.700pt}{1.239pt}}
\put(641,734){\rule[-0.350pt]{0.700pt}{1.239pt}}
\put(642,739){\rule[-0.350pt]{0.700pt}{1.239pt}}
\put(643,745){\rule[-0.350pt]{0.700pt}{1.405pt}}
\put(644,750){\rule[-0.350pt]{0.700pt}{1.405pt}}
\put(645,756){\rule[-0.350pt]{0.700pt}{1.405pt}}
\put(646,762){\rule[-0.350pt]{0.700pt}{1.405pt}}
\put(647,768){\rule[-0.350pt]{0.700pt}{1.405pt}}
\put(648,774){\rule[-0.350pt]{0.700pt}{1.405pt}}
\put(649,779){\rule[-0.350pt]{0.700pt}{1.170pt}}
\put(650,784){\rule[-0.350pt]{0.700pt}{1.170pt}}
\put(651,789){\rule[-0.350pt]{0.700pt}{1.170pt}}
\put(652,794){\rule[-0.350pt]{0.700pt}{1.170pt}}
\put(653,799){\rule[-0.350pt]{0.700pt}{1.170pt}}
\put(654,804){\rule[-0.350pt]{0.700pt}{1.170pt}}
\put(655,809){\rule[-0.350pt]{0.700pt}{1.170pt}}
\put(656,813){\rule[-0.350pt]{0.700pt}{1.136pt}}
\put(657,818){\rule[-0.350pt]{0.700pt}{1.136pt}}
\put(658,823){\rule[-0.350pt]{0.700pt}{1.136pt}}
\put(659,828){\rule[-0.350pt]{0.700pt}{1.136pt}}
\put(660,832){\rule[-0.350pt]{0.700pt}{1.136pt}}
\put(661,837){\rule[-0.350pt]{0.700pt}{1.136pt}}
\put(662,842){\rule[-0.350pt]{0.700pt}{1.136pt}}
\put(663,847){\rule[-0.350pt]{0.700pt}{1.204pt}}
\put(664,852){\rule[-0.350pt]{0.700pt}{1.204pt}}
\put(665,857){\rule[-0.350pt]{0.700pt}{1.204pt}}
\put(666,862){\rule[-0.350pt]{0.700pt}{1.204pt}}
\put(667,867){\rule[-0.350pt]{0.700pt}{1.204pt}}
\put(668,872){\rule[-0.350pt]{0.700pt}{1.204pt}}
\put(669,877){\rule[-0.350pt]{0.700pt}{0.998pt}}
\put(670,881){\rule[-0.350pt]{0.700pt}{0.998pt}}
\put(671,885){\rule[-0.350pt]{0.700pt}{0.998pt}}
\put(672,889){\rule[-0.350pt]{0.700pt}{0.998pt}}
\put(673,893){\rule[-0.350pt]{0.700pt}{0.998pt}}
\put(674,897){\rule[-0.350pt]{0.700pt}{0.998pt}}
\put(675,901){\rule[-0.350pt]{0.700pt}{0.998pt}}
\put(676,906){\rule[-0.350pt]{0.700pt}{0.860pt}}
\put(677,909){\rule[-0.350pt]{0.700pt}{0.860pt}}
\put(678,913){\rule[-0.350pt]{0.700pt}{0.860pt}}
\put(679,916){\rule[-0.350pt]{0.700pt}{0.860pt}}
\put(680,920){\rule[-0.350pt]{0.700pt}{0.860pt}}
\put(681,923){\rule[-0.350pt]{0.700pt}{0.860pt}}
\put(682,927){\rule[-0.350pt]{0.700pt}{0.860pt}}
\put(683,930){\rule[-0.350pt]{0.700pt}{0.923pt}}
\put(684,934){\rule[-0.350pt]{0.700pt}{0.923pt}}
\put(685,938){\rule[-0.350pt]{0.700pt}{0.923pt}}
\put(686,942){\rule[-0.350pt]{0.700pt}{0.923pt}}
\put(687,946){\rule[-0.350pt]{0.700pt}{0.923pt}}
\put(688,950){\rule[-0.350pt]{0.700pt}{0.923pt}}
\put(689,953){\usebox{\plotpoint}}
\put(690,956){\usebox{\plotpoint}}
\put(691,959){\usebox{\plotpoint}}
\put(692,962){\usebox{\plotpoint}}
\put(693,965){\usebox{\plotpoint}}
\put(694,968){\usebox{\plotpoint}}
\put(695,971){\usebox{\plotpoint}}
\put(696,973){\usebox{\plotpoint}}
\put(697,976){\usebox{\plotpoint}}
\put(698,979){\usebox{\plotpoint}}
\put(699,982){\usebox{\plotpoint}}
\put(700,985){\usebox{\plotpoint}}
\put(701,988){\usebox{\plotpoint}}
\put(702,990){\usebox{\plotpoint}}
\put(703,992){\usebox{\plotpoint}}
\put(704,994){\usebox{\plotpoint}}
\put(705,996){\usebox{\plotpoint}}
\put(706,998){\usebox{\plotpoint}}
\put(707,1000){\usebox{\plotpoint}}
\put(708,1002){\usebox{\plotpoint}}
\put(709,1003){\usebox{\plotpoint}}
\put(710,1005){\usebox{\plotpoint}}
\put(711,1006){\usebox{\plotpoint}}
\put(712,1008){\usebox{\plotpoint}}
\put(713,1009){\usebox{\plotpoint}}
\put(714,1011){\usebox{\plotpoint}}
\put(715,1012){\usebox{\plotpoint}}
\put(716,1014){\usebox{\plotpoint}}
\put(717,1015){\usebox{\plotpoint}}
\put(718,1016){\usebox{\plotpoint}}
\put(719,1017){\usebox{\plotpoint}}
\put(720,1018){\usebox{\plotpoint}}
\put(721,1019){\usebox{\plotpoint}}
\put(722,1021){\usebox{\plotpoint}}
\put(724,1022){\usebox{\plotpoint}}
\put(726,1023){\usebox{\plotpoint}}
\put(728,1024){\rule[-0.350pt]{2.048pt}{0.700pt}}
\put(737,1023){\usebox{\plotpoint}}
\put(739,1022){\usebox{\plotpoint}}
\put(740,1021){\usebox{\plotpoint}}
\put(742,1020){\usebox{\plotpoint}}
\put(743,1019){\usebox{\plotpoint}}
\put(744,1018){\usebox{\plotpoint}}
\put(745,1017){\usebox{\plotpoint}}
\put(746,1016){\usebox{\plotpoint}}
\put(747,1015){\usebox{\plotpoint}}
\put(748,1014){\usebox{\plotpoint}}
\put(749,1011){\usebox{\plotpoint}}
\put(750,1010){\usebox{\plotpoint}}
\put(751,1008){\usebox{\plotpoint}}
\put(752,1007){\usebox{\plotpoint}}
\put(753,1005){\usebox{\plotpoint}}
\put(754,1004){\usebox{\plotpoint}}
\put(755,1003){\usebox{\plotpoint}}
\put(756,1000){\usebox{\plotpoint}}
\put(757,998){\usebox{\plotpoint}}
\put(758,996){\usebox{\plotpoint}}
\put(759,994){\usebox{\plotpoint}}
\put(760,992){\usebox{\plotpoint}}
\put(761,990){\usebox{\plotpoint}}
\put(762,987){\usebox{\plotpoint}}
\put(763,985){\usebox{\plotpoint}}
\put(764,983){\usebox{\plotpoint}}
\put(765,980){\usebox{\plotpoint}}
\put(766,978){\usebox{\plotpoint}}
\put(767,976){\usebox{\plotpoint}}
\put(768,974){\usebox{\plotpoint}}
\put(769,971){\usebox{\plotpoint}}
\put(770,968){\usebox{\plotpoint}}
\put(771,966){\usebox{\plotpoint}}
\put(772,963){\usebox{\plotpoint}}
\put(773,961){\usebox{\plotpoint}}
\put(774,958){\usebox{\plotpoint}}
\put(775,956){\usebox{\plotpoint}}
\put(776,952){\rule[-0.350pt]{0.700pt}{0.803pt}}
\put(777,949){\rule[-0.350pt]{0.700pt}{0.803pt}}
\put(778,946){\rule[-0.350pt]{0.700pt}{0.803pt}}
\put(779,942){\rule[-0.350pt]{0.700pt}{0.803pt}}
\put(780,939){\rule[-0.350pt]{0.700pt}{0.803pt}}
\put(781,936){\rule[-0.350pt]{0.700pt}{0.803pt}}
\put(782,932){\rule[-0.350pt]{0.700pt}{0.792pt}}
\put(783,929){\rule[-0.350pt]{0.700pt}{0.792pt}}
\put(784,926){\rule[-0.350pt]{0.700pt}{0.792pt}}
\put(785,922){\rule[-0.350pt]{0.700pt}{0.792pt}}
\put(786,919){\rule[-0.350pt]{0.700pt}{0.792pt}}
\put(787,916){\rule[-0.350pt]{0.700pt}{0.792pt}}
\put(788,913){\rule[-0.350pt]{0.700pt}{0.792pt}}
\put(789,908){\rule[-0.350pt]{0.700pt}{1.004pt}}
\put(790,904){\rule[-0.350pt]{0.700pt}{1.004pt}}
\put(791,900){\rule[-0.350pt]{0.700pt}{1.004pt}}
\put(792,896){\rule[-0.350pt]{0.700pt}{1.004pt}}
\put(793,892){\rule[-0.350pt]{0.700pt}{1.004pt}}
\put(794,888){\rule[-0.350pt]{0.700pt}{1.004pt}}
\put(795,884){\rule[-0.350pt]{0.700pt}{0.895pt}}
\put(796,880){\rule[-0.350pt]{0.700pt}{0.895pt}}
\put(797,876){\rule[-0.350pt]{0.700pt}{0.895pt}}
\put(798,873){\rule[-0.350pt]{0.700pt}{0.895pt}}
\put(799,869){\rule[-0.350pt]{0.700pt}{0.895pt}}
\put(800,865){\rule[-0.350pt]{0.700pt}{0.895pt}}
\put(801,862){\rule[-0.350pt]{0.700pt}{0.895pt}}
\put(802,858){\rule[-0.350pt]{0.700pt}{0.929pt}}
\put(803,854){\rule[-0.350pt]{0.700pt}{0.929pt}}
\put(804,850){\rule[-0.350pt]{0.700pt}{0.929pt}}
\put(805,846){\rule[-0.350pt]{0.700pt}{0.929pt}}
\put(806,842){\rule[-0.350pt]{0.700pt}{0.929pt}}
\put(807,838){\rule[-0.350pt]{0.700pt}{0.929pt}}
\put(808,835){\rule[-0.350pt]{0.700pt}{0.929pt}}
\put(809,830){\rule[-0.350pt]{0.700pt}{1.164pt}}
\put(810,825){\rule[-0.350pt]{0.700pt}{1.164pt}}
\put(811,820){\rule[-0.350pt]{0.700pt}{1.164pt}}
\put(812,815){\rule[-0.350pt]{0.700pt}{1.164pt}}
\put(813,810){\rule[-0.350pt]{0.700pt}{1.164pt}}
\put(814,806){\rule[-0.350pt]{0.700pt}{1.164pt}}
\put(815,801){\rule[-0.350pt]{0.700pt}{0.998pt}}
\put(816,797){\rule[-0.350pt]{0.700pt}{0.998pt}}
\put(817,793){\rule[-0.350pt]{0.700pt}{0.998pt}}
\put(818,789){\rule[-0.350pt]{0.700pt}{0.998pt}}
\put(819,785){\rule[-0.350pt]{0.700pt}{0.998pt}}
\put(820,781){\rule[-0.350pt]{0.700pt}{0.998pt}}
\put(821,777){\rule[-0.350pt]{0.700pt}{0.998pt}}
\put(822,772){\rule[-0.350pt]{0.700pt}{1.032pt}}
\put(823,768){\rule[-0.350pt]{0.700pt}{1.032pt}}
\put(824,764){\rule[-0.350pt]{0.700pt}{1.032pt}}
\put(825,759){\rule[-0.350pt]{0.700pt}{1.032pt}}
\put(826,755){\rule[-0.350pt]{0.700pt}{1.032pt}}
\put(827,751){\rule[-0.350pt]{0.700pt}{1.032pt}}
\put(828,747){\rule[-0.350pt]{0.700pt}{1.032pt}}
\put(829,742){\rule[-0.350pt]{0.700pt}{1.205pt}}
\put(830,737){\rule[-0.350pt]{0.700pt}{1.204pt}}
\put(831,732){\rule[-0.350pt]{0.700pt}{1.204pt}}
\put(832,727){\rule[-0.350pt]{0.700pt}{1.204pt}}
\put(833,722){\rule[-0.350pt]{0.700pt}{1.204pt}}
\put(834,717){\rule[-0.350pt]{0.700pt}{1.204pt}}
\put(835,712){\rule[-0.350pt]{0.700pt}{1.032pt}}
\put(836,708){\rule[-0.350pt]{0.700pt}{1.032pt}}
\put(837,704){\rule[-0.350pt]{0.700pt}{1.032pt}}
\put(838,699){\rule[-0.350pt]{0.700pt}{1.032pt}}
\put(839,695){\rule[-0.350pt]{0.700pt}{1.032pt}}
\put(840,691){\rule[-0.350pt]{0.700pt}{1.032pt}}
\put(841,687){\rule[-0.350pt]{0.700pt}{1.032pt}}
\put(842,682){\rule[-0.350pt]{0.700pt}{1.032pt}}
\put(843,678){\rule[-0.350pt]{0.700pt}{1.032pt}}
\put(844,674){\rule[-0.350pt]{0.700pt}{1.032pt}}
\put(845,669){\rule[-0.350pt]{0.700pt}{1.032pt}}
\put(846,665){\rule[-0.350pt]{0.700pt}{1.032pt}}
\put(847,661){\rule[-0.350pt]{0.700pt}{1.032pt}}
\put(848,657){\rule[-0.350pt]{0.700pt}{1.032pt}}
\put(849,652){\rule[-0.350pt]{0.700pt}{1.205pt}}
\put(850,647){\rule[-0.350pt]{0.700pt}{1.204pt}}
\put(851,642){\rule[-0.350pt]{0.700pt}{1.204pt}}
\put(852,637){\rule[-0.350pt]{0.700pt}{1.204pt}}
\put(853,632){\rule[-0.350pt]{0.700pt}{1.204pt}}
\put(854,627){\rule[-0.350pt]{0.700pt}{1.204pt}}
\put(855,622){\rule[-0.350pt]{0.700pt}{0.998pt}}
\put(856,618){\rule[-0.350pt]{0.700pt}{0.998pt}}
\put(857,614){\rule[-0.350pt]{0.700pt}{0.998pt}}
\put(858,610){\rule[-0.350pt]{0.700pt}{0.998pt}}
\put(859,606){\rule[-0.350pt]{0.700pt}{0.998pt}}
\put(860,602){\rule[-0.350pt]{0.700pt}{0.998pt}}
\put(861,598){\rule[-0.350pt]{0.700pt}{0.998pt}}
\put(862,593){\rule[-0.350pt]{0.700pt}{0.998pt}}
\put(863,589){\rule[-0.350pt]{0.700pt}{0.998pt}}
\put(864,585){\rule[-0.350pt]{0.700pt}{0.998pt}}
\put(865,581){\rule[-0.350pt]{0.700pt}{0.998pt}}
\put(866,577){\rule[-0.350pt]{0.700pt}{0.998pt}}
\put(867,573){\rule[-0.350pt]{0.700pt}{0.998pt}}
\put(868,569){\rule[-0.350pt]{0.700pt}{0.998pt}}
\put(869,564){\rule[-0.350pt]{0.700pt}{1.124pt}}
\put(870,559){\rule[-0.350pt]{0.700pt}{1.124pt}}
\put(871,554){\rule[-0.350pt]{0.700pt}{1.124pt}}
\put(872,550){\rule[-0.350pt]{0.700pt}{1.124pt}}
\put(873,545){\rule[-0.350pt]{0.700pt}{1.124pt}}
\put(874,541){\rule[-0.350pt]{0.700pt}{1.124pt}}
\put(875,537){\rule[-0.350pt]{0.700pt}{0.929pt}}
\put(876,533){\rule[-0.350pt]{0.700pt}{0.929pt}}
\put(877,529){\rule[-0.350pt]{0.700pt}{0.929pt}}
\put(878,525){\rule[-0.350pt]{0.700pt}{0.929pt}}
\put(879,521){\rule[-0.350pt]{0.700pt}{0.929pt}}
\put(880,517){\rule[-0.350pt]{0.700pt}{0.929pt}}
\put(881,514){\rule[-0.350pt]{0.700pt}{0.929pt}}
\put(882,509){\rule[-0.350pt]{0.700pt}{1.044pt}}
\put(883,505){\rule[-0.350pt]{0.700pt}{1.044pt}}
\put(884,500){\rule[-0.350pt]{0.700pt}{1.044pt}}
\put(885,496){\rule[-0.350pt]{0.700pt}{1.044pt}}
\put(886,492){\rule[-0.350pt]{0.700pt}{1.044pt}}
\put(887,488){\rule[-0.350pt]{0.700pt}{1.044pt}}
\put(888,484){\rule[-0.350pt]{0.700pt}{0.860pt}}
\put(889,480){\rule[-0.350pt]{0.700pt}{0.860pt}}
\put(890,477){\rule[-0.350pt]{0.700pt}{0.860pt}}
\put(891,473){\rule[-0.350pt]{0.700pt}{0.860pt}}
\put(892,470){\rule[-0.350pt]{0.700pt}{0.860pt}}
\put(893,466){\rule[-0.350pt]{0.700pt}{0.860pt}}
\put(894,463){\rule[-0.350pt]{0.700pt}{0.860pt}}
\put(895,459){\rule[-0.350pt]{0.700pt}{0.826pt}}
\put(896,456){\rule[-0.350pt]{0.700pt}{0.826pt}}
\put(897,452){\rule[-0.350pt]{0.700pt}{0.826pt}}
\put(898,449){\rule[-0.350pt]{0.700pt}{0.826pt}}
\put(899,445){\rule[-0.350pt]{0.700pt}{0.826pt}}
\put(900,442){\rule[-0.350pt]{0.700pt}{0.826pt}}
\put(901,439){\rule[-0.350pt]{0.700pt}{0.826pt}}
\put(902,435){\rule[-0.350pt]{0.700pt}{0.923pt}}
\put(903,431){\rule[-0.350pt]{0.700pt}{0.923pt}}
\put(904,427){\rule[-0.350pt]{0.700pt}{0.923pt}}
\put(905,423){\rule[-0.350pt]{0.700pt}{0.923pt}}
\put(906,419){\rule[-0.350pt]{0.700pt}{0.923pt}}
\put(907,416){\rule[-0.350pt]{0.700pt}{0.923pt}}
\put(908,413){\rule[-0.350pt]{0.700pt}{0.723pt}}
\put(909,410){\rule[-0.350pt]{0.700pt}{0.723pt}}
\put(910,407){\rule[-0.350pt]{0.700pt}{0.723pt}}
\put(911,404){\rule[-0.350pt]{0.700pt}{0.723pt}}
\put(912,401){\rule[-0.350pt]{0.700pt}{0.723pt}}
\put(913,398){\rule[-0.350pt]{0.700pt}{0.723pt}}
\put(914,395){\rule[-0.350pt]{0.700pt}{0.723pt}}
\put(915,392){\rule[-0.350pt]{0.700pt}{0.723pt}}
\put(916,389){\rule[-0.350pt]{0.700pt}{0.723pt}}
\put(917,386){\rule[-0.350pt]{0.700pt}{0.723pt}}
\put(918,383){\rule[-0.350pt]{0.700pt}{0.723pt}}
\put(919,380){\rule[-0.350pt]{0.700pt}{0.723pt}}
\put(920,377){\rule[-0.350pt]{0.700pt}{0.723pt}}
\put(921,374){\rule[-0.350pt]{0.700pt}{0.723pt}}
\put(922,370){\rule[-0.350pt]{0.700pt}{0.763pt}}
\put(923,367){\rule[-0.350pt]{0.700pt}{0.763pt}}
\put(924,364){\rule[-0.350pt]{0.700pt}{0.763pt}}
\put(925,361){\rule[-0.350pt]{0.700pt}{0.763pt}}
\put(926,358){\rule[-0.350pt]{0.700pt}{0.763pt}}
\put(927,355){\rule[-0.350pt]{0.700pt}{0.763pt}}
\put(928,352){\usebox{\plotpoint}}
\put(929,349){\usebox{\plotpoint}}
\put(930,347){\usebox{\plotpoint}}
\put(931,344){\usebox{\plotpoint}}
\put(932,342){\usebox{\plotpoint}}
\put(933,339){\usebox{\plotpoint}}
\put(934,337){\usebox{\plotpoint}}
\put(935,334){\usebox{\plotpoint}}
\put(936,332){\usebox{\plotpoint}}
\put(937,330){\usebox{\plotpoint}}
\put(938,327){\usebox{\plotpoint}}
\put(939,325){\usebox{\plotpoint}}
\put(940,323){\usebox{\plotpoint}}
\put(941,321){\usebox{\plotpoint}}
\put(942,318){\usebox{\plotpoint}}
\put(943,315){\usebox{\plotpoint}}
\put(944,313){\usebox{\plotpoint}}
\put(945,310){\usebox{\plotpoint}}
\put(946,307){\usebox{\plotpoint}}
\put(947,305){\usebox{\plotpoint}}
\put(948,303){\usebox{\plotpoint}}
\put(949,301){\usebox{\plotpoint}}
\put(950,299){\usebox{\plotpoint}}
\put(951,297){\usebox{\plotpoint}}
\put(952,295){\usebox{\plotpoint}}
\put(953,293){\usebox{\plotpoint}}
\put(954,291){\usebox{\plotpoint}}
\put(955,289){\usebox{\plotpoint}}
\put(956,287){\usebox{\plotpoint}}
\put(957,285){\usebox{\plotpoint}}
\put(958,283){\usebox{\plotpoint}}
\put(959,281){\usebox{\plotpoint}}
\put(960,279){\usebox{\plotpoint}}
\put(961,277){\usebox{\plotpoint}}
\put(962,275){\usebox{\plotpoint}}
\put(963,273){\usebox{\plotpoint}}
\put(964,271){\usebox{\plotpoint}}
\put(965,269){\usebox{\plotpoint}}
\put(966,267){\usebox{\plotpoint}}
\put(967,265){\usebox{\plotpoint}}
\put(968,263){\usebox{\plotpoint}}
\put(969,261){\usebox{\plotpoint}}
\put(970,260){\usebox{\plotpoint}}
\put(971,258){\usebox{\plotpoint}}
\put(972,257){\usebox{\plotpoint}}
\put(973,255){\usebox{\plotpoint}}
\put(974,254){\usebox{\plotpoint}}
\put(975,252){\usebox{\plotpoint}}
\put(976,250){\usebox{\plotpoint}}
\put(977,248){\usebox{\plotpoint}}
\put(978,247){\usebox{\plotpoint}}
\put(979,245){\usebox{\plotpoint}}
\put(980,244){\usebox{\plotpoint}}
\put(981,242){\usebox{\plotpoint}}
\put(982,241){\usebox{\plotpoint}}
\put(983,239){\usebox{\plotpoint}}
\put(984,238){\usebox{\plotpoint}}
\put(985,236){\usebox{\plotpoint}}
\put(986,235){\usebox{\plotpoint}}
\put(987,234){\usebox{\plotpoint}}
\put(988,232){\usebox{\plotpoint}}
\put(989,231){\usebox{\plotpoint}}
\put(990,230){\usebox{\plotpoint}}
\put(991,229){\usebox{\plotpoint}}
\put(992,228){\usebox{\plotpoint}}
\put(993,227){\usebox{\plotpoint}}
\put(994,226){\usebox{\plotpoint}}
\put(995,224){\usebox{\plotpoint}}
\put(996,223){\usebox{\plotpoint}}
\put(997,222){\usebox{\plotpoint}}
\put(998,220){\usebox{\plotpoint}}
\put(999,219){\usebox{\plotpoint}}
\put(1000,218){\usebox{\plotpoint}}
\put(1001,218){\usebox{\plotpoint}}
\put(1001,218){\usebox{\plotpoint}}
\put(1002,217){\usebox{\plotpoint}}
\put(1003,216){\usebox{\plotpoint}}
\put(1004,215){\usebox{\plotpoint}}
\put(1005,214){\usebox{\plotpoint}}
\put(1006,213){\usebox{\plotpoint}}
\put(1007,212){\usebox{\plotpoint}}
\put(1008,211){\usebox{\plotpoint}}
\put(1009,210){\usebox{\plotpoint}}
\put(1010,209){\usebox{\plotpoint}}
\put(1011,208){\usebox{\plotpoint}}
\put(1012,207){\usebox{\plotpoint}}
\put(1013,206){\usebox{\plotpoint}}
\put(1015,205){\usebox{\plotpoint}}
\put(1016,204){\usebox{\plotpoint}}
\put(1017,203){\usebox{\plotpoint}}
\put(1018,202){\usebox{\plotpoint}}
\put(1019,201){\usebox{\plotpoint}}
\put(1020,200){\usebox{\plotpoint}}
\put(1021,199){\usebox{\plotpoint}}
\put(1022,198){\usebox{\plotpoint}}
\put(1023,197){\usebox{\plotpoint}}
\put(1025,196){\usebox{\plotpoint}}
\put(1026,195){\usebox{\plotpoint}}
\put(1028,194){\usebox{\plotpoint}}
\put(1029,193){\usebox{\plotpoint}}
\put(1031,192){\usebox{\plotpoint}}
\put(1033,191){\usebox{\plotpoint}}
\put(1035,190){\usebox{\plotpoint}}
\put(1036,189){\usebox{\plotpoint}}
\put(1038,188){\usebox{\plotpoint}}
\put(1039,187){\usebox{\plotpoint}}
\put(1041,186){\usebox{\plotpoint}}
\put(1042,185){\usebox{\plotpoint}}
\put(1044,184){\usebox{\plotpoint}}
\put(1046,183){\usebox{\plotpoint}}
\put(1048,182){\usebox{\plotpoint}}
\put(1050,181){\usebox{\plotpoint}}
\put(1052,180){\usebox{\plotpoint}}
\put(1055,179){\usebox{\plotpoint}}
\put(1057,178){\usebox{\plotpoint}}
\put(1059,177){\usebox{\plotpoint}}
\put(1061,176){\rule[-0.350pt]{0.843pt}{0.700pt}}
\put(1064,175){\rule[-0.350pt]{0.843pt}{0.700pt}}
\put(1068,174){\rule[-0.350pt]{0.723pt}{0.700pt}}
\put(1071,173){\rule[-0.350pt]{0.723pt}{0.700pt}}
\put(1074,172){\rule[-0.350pt]{0.843pt}{0.700pt}}
\put(1077,171){\rule[-0.350pt]{0.843pt}{0.700pt}}
\put(1081,170){\rule[-0.350pt]{0.843pt}{0.700pt}}
\put(1084,169){\rule[-0.350pt]{0.843pt}{0.700pt}}
\put(1088,168){\rule[-0.350pt]{1.445pt}{0.700pt}}
\put(1094,167){\rule[-0.350pt]{1.686pt}{0.700pt}}
\put(1101,166){\rule[-0.350pt]{0.843pt}{0.700pt}}
\put(1104,165){\rule[-0.350pt]{0.843pt}{0.700pt}}
\put(1108,164){\rule[-0.350pt]{1.445pt}{0.700pt}}
\put(1114,163){\rule[-0.350pt]{3.373pt}{0.700pt}}
\put(1128,162){\rule[-0.350pt]{1.445pt}{0.700pt}}
\put(1134,161){\rule[-0.350pt]{3.373pt}{0.700pt}}
\put(1148,160){\rule[-0.350pt]{4.577pt}{0.700pt}}
\put(1167,159){\rule[-0.350pt]{9.636pt}{0.700pt}}
\put(1207,158){\rule[-0.350pt]{91.301pt}{0.700pt}}
\end{picture}